\documentclass[longauth]{aa}
\usepackage[utf8]{inputenc}
\usepackage[varg]{txfonts}
\usepackage[breaklinks, colorlinks, citecolor=blue, linkcolor=blue,draft]{hyperref}
\usepackage{amsmath}
\usepackage{graphicx}
\usepackage[english]{babel}
\usepackage{siunitx}
\usepackage{float}
\usepackage{placeins}
\usepackage{url}
\usepackage{longtable}
\usepackage{threeparttable}
\usepackage{fancyhdr}  
\usepackage{natbib}
\usepackage[normalem]{ulem}
\usepackage[version=4]{mhchem}
\usepackage{caption}
\usepackage{subcaption}
\makeatletter  

\def\instrefs#1{{\def\scsep{\def\scsep{,}}\@for\w:=#1\do{\scsep\ref{inst:\w}}}}

\renewcommand{\inst}[1]{\unskip$^{\instrefs{#1}}$}

\let\orgautoref\autoref
\renewcommand{\autoref}
        {\def\equationautorefname{Eq.}
         \def\figureautorefname{Fig.}
         \def\sectionautorefname{Sect.}
         \def\subsectionautorefname{Sect.}
         \def\subsubsectionautorefname{Sect.}
         \orgautoref}

\renewcommand*\aa@pageof{, page \thepage{} of \pageref*{LastPage}}

\newcommand{\target}{TOI-1416}
\newcommand{\pname}{TOI-1416 $b$}  
\newcommand{\planet}{\pname}
\newcommand{\toib}{\pname}
\newcommand{\toic}{TOI-1416 $c$}

\newcommand{\mps}{$\mathrm{m\,s^{-1}}$}

\newcommand{\microns}{$\mu$m}

\newcommand{\tess}{\emph{{\it TESS}}}

\newcommand{\hn}{\emph{{HARPS-N}}}

\newcommand{\tw}{This Work}

\newcommand{\steffsme}[1][]{$4884 \pm 70$} 
\newcommand{\sloggsme}{$4.52 \pm 0.05$} 
\newcommand{\scahsme}[1][]{$+0.08 \pm 0.05$} 
\newcommand{\sfehsme}[1][]{$+0.08 \pm 0.05$} 
\newcommand{\svsinisme}[1][]{$2.0 \pm 0.7$} 
\newcommand{\svmicsme}[1][]{$0.8$} 
\newcommand{\svmacsme}[1][]{$1.5 \pm 1.0$} 
 
\newcommand{\steffspecm}[1][]{$4966 \pm 110$} 
\newcommand{\sfehspecm}[1][]{$+0.19 \pm 0.09$} 
\newcommand{\sradiusspecm}[1][]{$0.788 \pm 0.079$} 

\newcommand{\steffgaia}[1][]{$4909^{+97}_{-58}$} 
\newcommand{\sradiusgaia}[1][]{$0.819^{+0.020}_{-0.031}$} 

\newcommand{\smassisochrones}{$0.813\pm 0.013$}  
\newcommand{\sradiusisochrones}{$0.786\pm 0.007$}  
\newcommand{\srhoisochrones}{$2.36 \pm 0.09$} 
\newcommand{\Avisochrones}{$0.05\pm 0.04$}  
\newcommand{\Lisochrones}{$0.34\pm 0.03$}  

\newcommand{\smassparam}{$0.778^{+0.020}_{-0.018}$}  
\newcommand{\sradiusparam}{$0.785^{+0.009}_{-0.041}$}  
\newcommand{\srhoparam}{$2.27 \pm 0.22$}  

\newcommand{\sradiusSED}{$0.798\pm 0.008$}       
\newcommand{\smassSED}{$0.770^{+0.078}_{-0.065}$}         
\newcommand{\srhoSED}{$2.14 \pm 0.21$}         

\newcommand{\smasstorres}{$0.812\pm 0.055$}  
\newcommand{\sradiustorres}{$0.807\pm 0.056$}  
\newcommand{\srhotorres}{$2.18\pm 0.48$}  

\newcommand{\Tzerob}[1][days]   {$8739.4621 \pm 0.0008$~#1}
\newcommand{\Pb}[1][days]   {1.0697568$\pm$ 2.8e-06~#1}   
\newcommand{\bb}[1][ ]   {$0.39 _{ - 0.14 } ^ { + 0.10 }$~#1} 
 
\newcommand{\rrb}[1][ ]   {$0.01873 \pm 0.00054$~#1}
\newcommand{\kb}[1][${\rm m\,s^{-1}}$]   {$2.52 \pm 0.32$~#1} 
\newcommand{\mpb}[1][$M_{\oplus}$]   {$3.48 \pm 0.47$~#1} 
\newcommand{\rpb}[1][$R_{\oplus}$]   {$1.62 \pm 0.08$~#1}

\newcommand{\ib}[1][deg]   {$85.7 _{ - 1.4 } ^ { + 1.7 }$~#1} 
\newcommand{\insolationb}[1][${\rm F_{\oplus}}$]   {$883 \pm 96$~#1} 
 
\newcommand{\denstrb}[1][${\rm g\,cm^{-3}}$]   {$2.31 \pm 0.28$~#1} 

\newcommand{\Teqb}[1][K]   {$1517 \pm 39$~#1} 
\newcommand{\ttotb}[1][hours]   {$1.50 \pm 0.035$~#1}

\newcommand{\denpb}[1][${\rm g\,cm^{-3}}$]   {$4.50 _{ - 0.83 } ^ { + 0.99 }$~#1} 
\newcommand{\grapparsb}[1][${\rm cm\,s^{-2}}$]   {$1300 \pm 220$~#1} 

\newcommand{\Tzeroc}[1][days]   {$8876.78 \pm 0.69 $~#1} 
\newcommand{\Pc}[1][days]   {$29.5306$~#1} 
\newcommand{\kc}[1][${\rm m\,s^{-1}}$]   {$5.20 _{ - 0.65 } ^ { + 0.71 }$~#1} 
\newcommand{\mpc}[1][$M_{\oplus}$]   {$21.6 _{ - 2.8 } ^ { + 3.1 }$~#1}

\newcommand{\qone}[1][]   {$0.429 \pm 0.087$~#1} 
\newcommand{\qtwo}[1][]   {$0.355 \pm 0.030$~#1}

\newcommand{\HN}[1][${\rm m\,s^{-1}}$]   {$0.79 \pm 0.52$~#1} 
\newcommand{\dLW}[1][${\rm km\,s^{-1}}$]   {$-0.063 \pm 0.033$~#1} 
\newcommand{\jHN}[1][${\rm m\,s^{-1}}$]   {$0.30 \pm 0.25 $~#1} 

\newcommand{\jdLW}[1][${\rm m\,s^{-1}}$]   {$26.9 \pm 3.6 $~#1} 
\newcommand{\jtr}[1][]   {$788.2 \pm 2.8$~#1} 
\newcommand{\jArvc}[1][]   {$0.96 _{ - 0.54 } ^ { + 0.68 }$~#1} 
\newcommand{\jArvr}[1][]   {$15.7 _{ - 2.7 } ^ { + 3.6 }$~#1} 
\newcommand{\jAdlw}[1][]   {$0.121 _{ - 0.019 } ^ { + 0.026 }$~#1} 
\newcommand{\jlambdae}[1][]   {$24.0 \pm 6.2$~#1} 
\newcommand{\jlambdap}[1][]   {$0.62 \pm 0.10$~#1} 
\newcommand{\jPGP}[1][]   {$20.6 _{ - 1.0 } ^ { + 1.9 }$~#1}

\newcommand{\kbPcuni}[1][${\rm m\,s^{-1}}$]   {$2.5 \pm 0.32$~#1} 
\newcommand{\Pcuni}[1][days]   {$29.509 _{ - 0.065 } ^ { + 0.070 }$~#1} 
\newcommand{\eb}[1][ ]   {$0.034 _{ - 0.022 } ^ { + 0.038 }$~#1} 
\newcommand{\ec}[1][ ]   {$0.34 _{ - 0.21 } ^ { + 0.18 }$~#1} 

\newcommand{\UTzerob}[1][days]   {8739.4620$\pm0.0008$~#1} 
\newcommand{\UPb}[1][days]   {1.0697564 $\pm$ 2.8e-06~#1} 
\newcommand{\Ubb}[1][ ]   {$0.35 _{ - 0.15 } ^ { + 0.11 }$~#1} 
\newcommand{\Udenstrb}[1][${\rm g^{1/3}\,cm^{-1}}$]   {$2.40 \pm 0.27$~#1} 
\newcommand{\Urrb}[1][ ]   {$0.01963 \pm 0.00059$~#1} 
\newcommand{\Uqone}[1][]   {$0.430 \pm 0.089$~#1} 
\newcommand{\Uqtwo}[1][]   {$0.355 \pm 0.031$~#1} 
\newcommand{\Ukb}[1][${\rm m\,s^{-1}}$]   {$2.14 \pm 0.35$~#1} 
\newcommand{\UHN}[1][${\rm m\,s^{-1}}$]   {$-0.05 \pm 0.87$~#1}

\newcommand{\pyan}{\texttt{pyaneti}}
\newcommand{\IGkbfco}[1][${\rm m\,s^{-1}}$] {$2.12\pm0.36$~#1}
\newcommand{\IGkbonepl}[1][${\rm m\,s^{-1}}$]   {$2.28 \pm 0.33$~#1}

\newcommand{\utm}{\texttt{UTM}}
\newcommand{\ufit}{\texttt{UFIT}}
\makeatother

 \title{TOI-1416: A system with a super-Earth planet with a 1.07d period}

\titlerunning{\target: TOI-1416: A system with a super-Earth planet with a 1.07d period}

\author{H.\,J.~Deeg\inst{iac,ull}
\and I.\,Y.~Georgieva\inst{chalm}
\and G.~Nowak\inst{iac,ull,torun}
\and C.\,M.~Persson\inst{chalm}
\and B.\,L.~Cale\inst{nasaexo} 
\and F.~Murgas\inst{iac,ull}
\and E. Pall\'e\inst{iac,ull}
\and D.~Godoy Rivera\inst{iac,ull}
\and F.~Dai\inst{caltec_geo}
\and D.\,R.~Ciardi\inst{nasaexo}
\and J.\,M.~Akana Murphy\inst{ucsc}
\and P.\,G.~Beck\inst{ull,iac,graz} 
\and C.\,J.~Burke\inst{kavli} 
\and J.~Cabrera\inst{dlr} 
\and I. Carleo\inst{iac,ull,wes}
\and W.\,D. Cochran\inst{utexas}
\and K.\,A. Collins\inst{cfa}  
\and Sz.~Csizmadia\inst{dlr}
\and M.~El\,Mufti\inst{gmu} 
\and M.~Fridlund\inst{leiden,chalm}
\and A.~Fukui\inst{komaba,iac} 
\and D.~Gandolfi\inst{torino}
\and R.\,A.~Garc\'ia\inst{saclay}  
\and E.\,W.~Guenther\inst{tls}
\and P.~Guerra\inst{girona}
\and S.~Grziwa\inst{koln}
\and H.~Isaacson\inst{berkeley,queensland}
\and K.~Isogai\inst{okayama,multtokyo} 
\and J.\,M.~Jenkins\inst{ames} 
\and P.~K\'abath\inst{chec}
\and J.~Korth\inst{chalm}
\and K.W.F.~Lam\inst{dlr}
\and D.\,W.~Latham\inst{cfa}
\and R.~Luque\inst{chicago,iac,ull}  
\and M.\,B.~Lund \inst{nasaexo}
\and J.\,H.~Livingston\inst{osawa,naoj,sokendai}
\and S.~Mathis\inst{saclay}
\and S.~Mathur\inst{iac,ull}
\and N.~Narita\inst{komaba,osawa,iac}
\and J.~Orell-Miquel\inst{iac,ull} 
\and H.L.M.~Osborne\inst{dacp} 
\and H.~Parviainen\inst{iac,ull}
\and P.\,P.~Plavchan\inst{gmu} 
\and S.~Redfield\inst{wes}   
\and D.\,R.~Rodriguez\inst{stsci}  
\and R.\,P.~Schwarz\inst{cfa}
\and S.~Seager\inst{kavli,MIT1,MIT2} 
\and A.M.S.~Smith\inst{dlr}
\and V.~Van~Eylen\inst{dacp}
\and J.~Van~Zandt\inst{ucla}  
\and J.\,N~Winn\inst{princeton} 
\and C.~Ziegler\inst{austin} 
}

\institute{
\label{inst:iac}Instituto de Astrof\'isica de Canarias, 38205 La Laguna, Tenerife, Spain
\email{hdeeg@iac.es}
\and
\label{inst:ull}Departamento de Astrof\'isica, Universidad de La Laguna, 38206 La Laguna, Tenerife, Spain
\and
\label{inst:chalm}Department of Space, Earth and Environment,
Chalmers University of Technology, Onsala Space Observatory, 439 92 Onsala, Sweden
\and
\label{inst:torun}Institute of Astronomy, Faculty of Physics, Astronomy and Informatics, Nicolaus Copernicus University, Grudzi\c{a}dzka 5, 87-100 Toru\'n, Poland
\and
\label{inst:nasaexo}NASA Exoplanet Science Institute, 770 South Wilson Ave., Pasadena, CA 91125, USA
\and
\label{inst:caltec_geo} Division of Geological and Planetary Sciences, California Institute of Technology, 1200 E California Boulevard, Pasadena, CA 91125, USA
\and
\label{inst:ucsc}Department of Astronomy and Astrophysics, University of California, Santa Cruz, CA 95064, USA
\and
\label{inst:graz}Institut f\"ur Physik, Karl-Franzens Universit\"at Graz, Universit\"atsplatz 5/II, NAWI Graz, 8010 Graz, Austria 
\and
\label{inst:kavli}Department of Physics and Kavli Institute for Astrophysics and Space Research, Massachusetts Institute of Technology, Cambridge, MA 02139, USA
\and
\label{inst:dlr}Deutsches Zentrum f\"ur Luft- und Raumfahrt, Institut f\"ur Planetenforschung, 12489 Berlin, Rutherfordstrasse 2., Germany
\and
\label{inst:gmu}Department of Physics and Astronomy, George Mason University, 4400 University Drive, Fairfax, VA 22030, USA
\and 
\label{inst:stsci}Space Telescope Science Institute, 3700 San Martin Drive, Baltimore, MD 21218, USA
\and
\label{inst:dacp}Mullard Space Science Laboratory, University College London, Holmbury St. Mary, Dorking, Surrey, RH5 6NT, UK
\and 
\label{inst:tls}Th\"uringer Landessternwarte Tautenburg, Sternwarte 5, 07778 Tautenburg, Germany
\and
\label{inst:utexas}McDonald Observatory and Center for Planetary Systems Habitability, The University of Texas, Austin, TX 78730, USA
\and
\label{inst:cfa}Center for Astrophysics \textbar \ Harvard \& Smithsonian, 60 Garden Street, Cambridge, MA 02138, USA
\and
\label{inst:leiden}Leiden Observatory, Leiden University, 2333CA Leiden, The Netherlands
\and
\label{inst:ames}NASA Ames Research Center, Moffett Field, CA 94035, USA
\and
\label{inst:chec}Astronomical Institute, Czech Academy of Sciences, Fri\v{c}ova 298, 25165,
Ond\v{r}ejov, Czech Republic
\and
\label{inst:koln}Rheinisches Institut f\"ur Umweltforschung an der Universit\"at zu K\"oln, Aachener Strasse 209, 50931 K\"oln, Germany
\and
\label{inst:komaba}Komaba Institute for Science, The University of Tokyo, 3-8-1 Komaba, Meguro, Tokyo 153-8902, Japan
\and
\label{inst:torino}Dipartimento di Fisica, Universit\`a degli Studi di Torino, Torino, Italy 
\and
\label{inst:osawa}Astrobiology Center, 2-21-1 Osawa, Mitaka, Tokyo 181-8588, Japan
\and
\label{inst:wes}Astronomy Department and Van Vleck Observatory, Wesleyan University, Middletown, CT 06459, USA
\and
\label{inst:MIT1}Department of Earth, Atmospheric and Planetary Sciences, Massachusetts Institute of Technology, Cambridge, MA 02139, USA
\and
\label{inst:MIT2}Department of Aeronautics and Astronautics, Massachusetts Institute of Technology, 77 Massachusetts Avenue, Cambridge, MA 02139, USA
\and
\label{inst:chicago}Department of Astronomy \& Astrophysics, University of Chicago, Chicago, IL 60637, USA
\and
\label{inst:ucla}Department of Physics \& Astronomy, University of California Los Angeles, Los Angeles, CA 90095, USA
\and
\label{inst:okayama}Okayama Observatory, Kyoto University, 3037-5 Honjo, Kamogatacho, Asakuchi, Okayama 719-0232, Japan
\and
\label{inst:multtokyo}Department of Multi-Disciplinary Sciences, Graduate School of Arts and Sciences, The University of Tokyo, 3-8-1 Komaba, Meguro, Tokyo 153-8902, Japan
\and
\label{inst:saclay}D\'epartement d'Astrophysique, IRFU/DRF/CEA Saclay, L'Orme des Merisiers, bat. 709, 91191 Gif-sur-Yvette Cedex, France
\and
\label{inst:princeton}Department of Astrophysical Sciences, Princeton University, Princeton, NJ 08544, USA
\and
\label{inst:berkeley}University of California at Berkeley, 501 Campbell Hall, Berkeley, CA 94720, USA
\and
\label{inst:girona}Observatori Astron\`omic Albany\`a, Cam\'i de Bassegoda S/N, Albany\`a 17733, Girona, Spain
\and
\label{inst:austin}Department of Physics, Engineering and Astronomy, Stephen F. Austin State University, 1936 North St, Nacogdoches, TX 75962, USA
\and
\label{inst:naoj}National Astronomical Observatory of Japan, 2-21-1 Osawa, Mitaka, Tokyo 181-8588, Japan
\and
\label{inst:sokendai}Department of Astronomical Science, School of Physical Sciences, The Graduate University for Advanced Studies (SOKENDAI), 2-21-1, Osawa, Mitaka, Tokyo, 181-8588, Japan
\and
\label{inst:queensland}Centre for Astrophysics, University of Southern Queensland, Toowoomba, QLD, Australia
}

\date{Received 10 March 2023 / Accepted 10 May 2023}

\abstract{TOI 1416 (BD+42 2504, HIP 70705) is a V=10 late G or early K-type dwarf star. TESS detected transits in its Sectors 16, 23 and 50 with a 
depth of about 455 ppm and a period of 1.07 days. Radial velocities taken with 
the  \hn, CARMENES, Automated Planet Finder (APF)  and iSHELL instruments verify the presence of the transiting planet \pname, with a mass of $3.48 \pm 0.47 M_{\oplus}$ and a radius of $1.62 \pm 0.08 R_{\oplus}$, implying a slightly sub-Earth density of  \denpb[] ${\rm g\,cm^{-3}}$. The RV data also further indicate a tentative planet $c$ with a period of 27.4 or 29.5 days, whose nature cannot be verified due to strong suspicions about contamination by a signal related to the Moon's synodic period of 29.53 days. The near-USP (Ultra Short Period) planet \pname\ is a typical representative of a short-period and hot ($T_{eq} \approx$ 1570 K) super-Earth like planet. A planet model of an interior of molten magma containing a significant fraction of dissolved water provides a plausible explanation for its composition, and its atmosphere could be suitable for transmission spectroscopy with JWST. The position of \pname\ within the radius-period distribution corroborates that USPs with periods of less than one day do not form any special group of planets. Rather, this implies that USPs belong to a continuous distribution of super-Earth like planets with periods ranging from the shortest known ones up to $\approx$ 30 days, whose period-radius distribution is delimitated against larger radii by the Neptune desert and by the period-radius valley that separates super-Earths from sub-Neptune planets. {In the abundance of small-short periodic planets against period, a plateau between periods of 0.6 to 1.4 days has however become notable that is compatible with the low-eccentricity formation channel.} For the Neptune desert, its lower limits required a revision due to the increasing population of short period planets; for periods shorter then 2 days, we establish a radius of 1.6  $R_{\oplus}$ and a mass of 0.028 $M_\mathrm{jup}$ respectively 8.9 $M_{\oplus}$ as the desert's lower limits. {We also provide corresponding limits to the Neptune Desert against the planets' insulation and effective temperature}}

\keywords{planetary systems --
planets and satellites: detection --
    techniques: photometric --
    techniques: radial velocities --
    techniques: spectroscopic --
    stars: individual (\object{HIP 70705} TIC\,158025009, TOI 1416)
    stars: late-type
    }
             
\begin{document}

\maketitle

\section{Introduction}\label{sec:intro}

Small-sized exoplanets ($R \lesssim2.5 R_\oplus$) constitute  currently the most numerous group among the known exoplanets. Their population properties were first studied by \citet{2014ApJ...787...47S}, who identified several tens of  planets (or planet candidates) with periods of less than 1 day in data from the Kepler mission \citep{KEPLER}, and called them  Ultra Short Period Planets (USP). Nearly all of these planets were smaller than 2 R$\_{Earth}$ and a preference for the presence of further planets with periods of up to 50 days was identified. The upper limit of 1 day for USPs  - besides being a convenient number -  was due to the lower period limit of the Kepler planet detection pipeline \citep{2010ApJ...713L..87J}, which had missed out on these planets, but not due to any physical limit. However, the term 'USP' with that period limit has remained with the community, and currently there are 126 such planets known, albeit there are only 34 for which both masses and radii have been determined\footnote{Retrieved from the NASA Exoplanet Archive in February 2023.}. For overviews over this population and for theories for their development we refer to \citet{2018NewAR..83...37W} and \citet{2022A&A...668A.158M} and references therein. 

In this work, we describe the detection of a planet around the late G or early K star \object{TOI-1416} (see Table~\ref{table:star}), with a period of 1.067 days, just outside of the conventional definition of USPs, and place it in context with the population comprised of USP planets and of planets with slightly longer orbits. 
\object{TOI-1416 $b$} was found in lightcurves by the TESS mission \citep{Ricker2015}, and whose all-sky transit survey with relatively short coverages is well-suited of the detection of short-periodic planets. The TESS observations and their processing is described in Sect. 2. A ground-based follow-up campaign involving imaging and radial velocity observations is described in Sect. 3; where the analysis of the data by stellar modelling (Sect. 4) and planet system modelling (Sect. 5) led also to the detection of a potential second planet \target $c$,  with a period of 27 - 29 days, for which strong doubts remain about the origin of its RV signal from Moon-reflected solar light (with details about this Appendix A). The implications of these findings, in particular with regard to the planet's composition and its placement relative to the short-period planet population are provided in Sect. 6, with conclusions in Sect. 7.


\begin{table}
\centering
\small
\caption{Parameters of \target\ from catalogues} \label{table:star}
\resizebox{\linewidth}{!}{%
\begin{tabular}{lcr}
\hline\hline
\noalign{\smallskip}
Parameter   & Value             & Reference \\ 
\hline
\noalign{\smallskip}
\multicolumn{3}{c}{\em Identifiers}\\
\noalign{\smallskip}
 \multicolumn{2}{l}    {TOI 1416} &                ExoFOP\\  
 \multicolumn{2}{l}    {TIC 158025009}            & TIC \\  
 \multicolumn{2}{l} {BD+42 2504}&\\
\multicolumn{2}{l} {HIP   70705}  &  \\
\multicolumn{2}{l} {WISE J142741.68+415711.2}\\
\multicolumn{2}{l} {2MASS J14274177+4157124}&\\
\multicolumn{2}{l} {TYC 3039-00749-1}&\\
\multicolumn{2}{l} {Gaia DR3 1491634483976350720 }&\\
\noalign{\smallskip}
\multicolumn{3}{c}{\em Coordinates and kinematics}\\
\noalign{\smallskip}
ICRS coord (J2000) & 14 27 41.766 \  +41 57 12.32 & {Gaia} EDR3 \\

Pr. motion [$\mathrm{mas/yr}$]  & $-92.254\pm0.010, -101.233 \pm 0.012$ & {Gaia} EDR3 \\
Parallax [mas]  & $18.1671\pm0.0126$ & {Gaia} EDR3 \\
$d$ [pc]        & $55.044\pm0.038$ & {Gaia} EDR3 \\
Systemic velocity [km/s]&$1.1712\pm0.0010$&\tw\\

\noalign{\smallskip}
\multicolumn{3}{c}{\em Magnitudes and spectral type}\\
\noalign{\smallskip}
B [mag]        & $10.93\pm0.05$ & Tycho-2 \\
V [mag]        & $9.98\pm0.03$ & Tycho-2 \\
Gaia [mag]      & $9.6588\pm0.0028$ & {Gaia} EDR3 \\
TESS [mag]       &$9.0739\pm0.006$ &  TIC v.8.2\\ 
J [mag]       & $8.266\pm0.024$ & 2MASS\\
H [mag]       & $7.815\pm0.017$ & 2MASS\\
K [mag]       & $7.708\pm0.024$ & 2MASS\\
Extinction $A_\mathrm{v}$[mag]  & $<0.024 $&  IRSA\\
Spectral type & G9V    &This work \\
\noalign{\smallskip}
\hline
\end{tabular}
}  
\tablebib{
ExoFOP: TESS Exoplanet Follow-up Observing Program (ExoFOP) website (DOI: 10.26134/ExoFOP5) 
    {\it Gaia} EDR3: \citet{GaiaEDR3};
    Tycho-2: \citet{2000A&A...355L..27H};
    2MASS: \citet{2MASS};
    TIC: Tess Input Catalogue, \citet{2018AJ....156..102S,2019AJ....158..138S};
    IRSA: Upper limit from total Galactic extinction in target direction. Value from IRSA Galactic Reddening and Extinction Calculator, based on \citet{2011ApJ...737..103S}.
}
\end{table}

\section{Photometry by TESS} \label{sec:tessphot}

TESS observed \target\  in its Sectors 16, 23, and 50, with more detailed information given in Table~\ref{table:tessobs}. Planet $b$ was initially detected as a Tess Object of Interest (TOI) by the SPOC pipeline \citep{SPOC} in data from S16, as  TOI 1416.01. A subsequent analysis of the combined S16, S23 {and S50 data by the same pipeline specified a transit-like signal with a period of P=1.06975[1] d and an amplitude of 391.5$\pm$ 24.0 ppm} \footnote{These values are from the Data Validation Report Summary of TOI 1416.01 for the combined S14, S23 and S50 data, available at MAST (\url{https://mast.stsci.edu}) as file \texttt{tess2019199201929-s0014-s0050-0000000158025009-00611\_} \texttt{dvm.pdf}.}, indicating a candidate for a small planet of $\approx 1.6R{_{\oplus}}$. {The difference imaging test \citep{2018PASP..130f4502T} also revealed that the origin of the transit is within $2.47\arcsec$ of the location of the target. }

For our own transit detection analysis, we used the algorithms \textsc{DST} \citep[D\'etection Sp\'ecialis\'ee de Transits,][]{2012A&A...548A..44C} and TLS \citep[Transit Least Square,][]{2019A&A...623A..39H} to search for transit signals in the existing TESS data and found a signal with period of 
P=1.07d which were consistent with the detection reported by SPOC. We then masked the transits at 1.07 d and searched for further signals in the dataset but found no detection that indicates the presence of additional transiting planet candidate. This process was repeated later with a focus on signals with periods of $\approx$ 10 d and 27 to 30 d, corresponding to peaks in radial velocity periodograms reported in Sect.~\ref{sec:planmod} of this work, but again to no avail.

For all further analysis of lightcurves, we used the presearch data conditioned simple aperture photometry (PDCSAP) fluxes \citep{2012PASP..124..985S,2012PASP..124.1000S,2014PASP..126..100S,pdcsap} available at MAST. Flux points in which some\footnote{Cadences of expected low quality are identified by a bit-wise AND of the quality flag of a given data-point with the binary number 0101001010111111, as recommended in the TESS Archive Manual at \url{https://outerspace.stsci.edu/display/TESS/2.0+-+Data+Product+Overview}.} quality flags are raised were removed. Also, the fluxes were normalized to an average flux of 1 in each sector independently. This lightcurve was used for the fit using Gaussian Processes with \pyan\ described in Sect.~\ref{sec:joint}.

The field around \target\ is moderately crowded and the TIC indicates a contamination ratio\footnote{The contamination ratio $c_{TIC}$ is defined as the ratio of flux from nearby objects that falls in the aperture of the target star, divided by the target star flux in the aperture \citep{2018AJ....156..102S}. CROWDSAP is defined as the ratio of the flux from the target to the total flux in the aperture \citep{TESSdataprod}. A conversion is therefore given by {\tt CROWDSAP} $= 1/(1+c_{TIC})\,.$} of $c_{TIC}=0.193$. Very similar values for contamination are also indicated by the {\tt CROWDSAP} keyword$^4$ in the headers of the SPOC lightcurves from S16 and S23, whereas the S50 lightcurves indicates only very minor contamination. PDCSAP fluxes are in principle corrected against contamination \citep{2020ksci.rept....8S}. We evaluated however the impact that an error in $c_{TIC}$ (or in the corresponding {\tt CROWDSAP} values) might have onto the final system parameters reported in Tables~\ref{table:final_param_model} and \ref{table:final_param_derived}.  The impact of an error of $c_{TIC}$ was however found to be negligible as long as  $c_{TIC}$ is correct within $\approx 25\%$. Lacking any indications about the uncertainty of $c_{TIC}$ (or {\tt CROWDSAP}), 
we did not propagate this uncertainty into the finally given parameter errors.

\begin{figure}
\centering
	\includegraphics[width=1\linewidth]{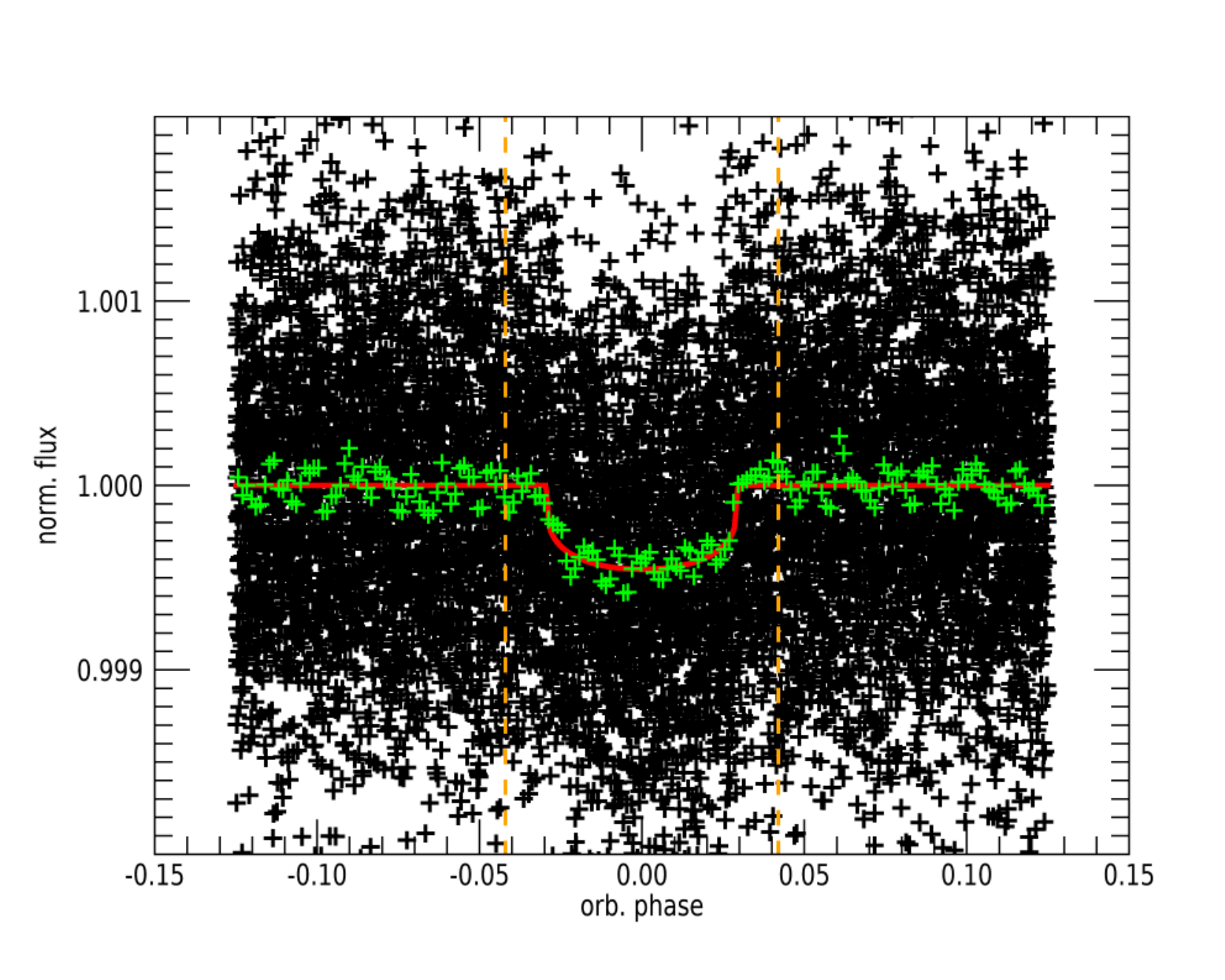}	
\caption{Black crosses: TESS lightcurve around the transits of planet $b$, after phasing by the planet's period against the adopted ephemeris and the correction against gradients in the off-eclipse sections (indicated by the orange vertical dashed lines) as described in Sect.~\ref{sec:tessphot}. Green crosses: The same curve, after a box-car smoothing over 100 phased data points and posterior binning over 50 points. We note that the average time-increment between the binned points is 126 seconds, which is very similar to the 120 s temporal resolution of TESS lightcurves. The red curve is the transit model generated with {\tt UTM/UFIT}, described in Appendix~\ref{app:utm}. 
}\label{fig:wrap_lc}\end{figure}

Individual transits of \pname\ have a S/N of $\approx$ 3.6 and they are too shallow to be individually detectable in the lightcurve. For the preparation of the lightcurve to be used in transit fits with {\tt UTM/UFIT}  (described in Appendix~\ref{app:utm}),   
we extracted short sections between orbital phases of $\pm$ 0.125 around the transit center of planet $b$ (initially using the ephemeris provided by SPOC, and then improved ones from our own transit fits), and performed a linear fit across both off-transit sections around each transit. The fluxes were then divided by that fit, which leads to an off-transit flux that is normalised to 1. Only transits that were fully covered by TESS have been included in the final lightcurve; see Table~\ref{table:tessobs} for the number of transits in each sector.  The phase-folded lightcurve {containing 48 transits} is shown in Fig.~\ref{fig:wrap_lc}. With the transit ephemeris that was finally adopted and  which is given in Table~\ref{table:final_param_model}, it shows a transit-shape that is much better defined --  with steeper in- and egress --  than one produced by a folding with the original period indicated by SPOC.  The standard deviation (or {\it rms} noise) of the unbinned off-eclipse data is 765 ppm, and the noise of a smoothed and binned version of the phased lightcurve, with a temporal resolution similar to TESS' 2-minute cadence (green crosses in Fig. ~\ref{fig:wrap_lc}) is 86 ppm, while the depth of the transits is $\approx$ 455 ppm {and the S/N of the phased transit is $\approx$ 25}.

\begin{figure}
\centering
	\includegraphics[width=1\linewidth]{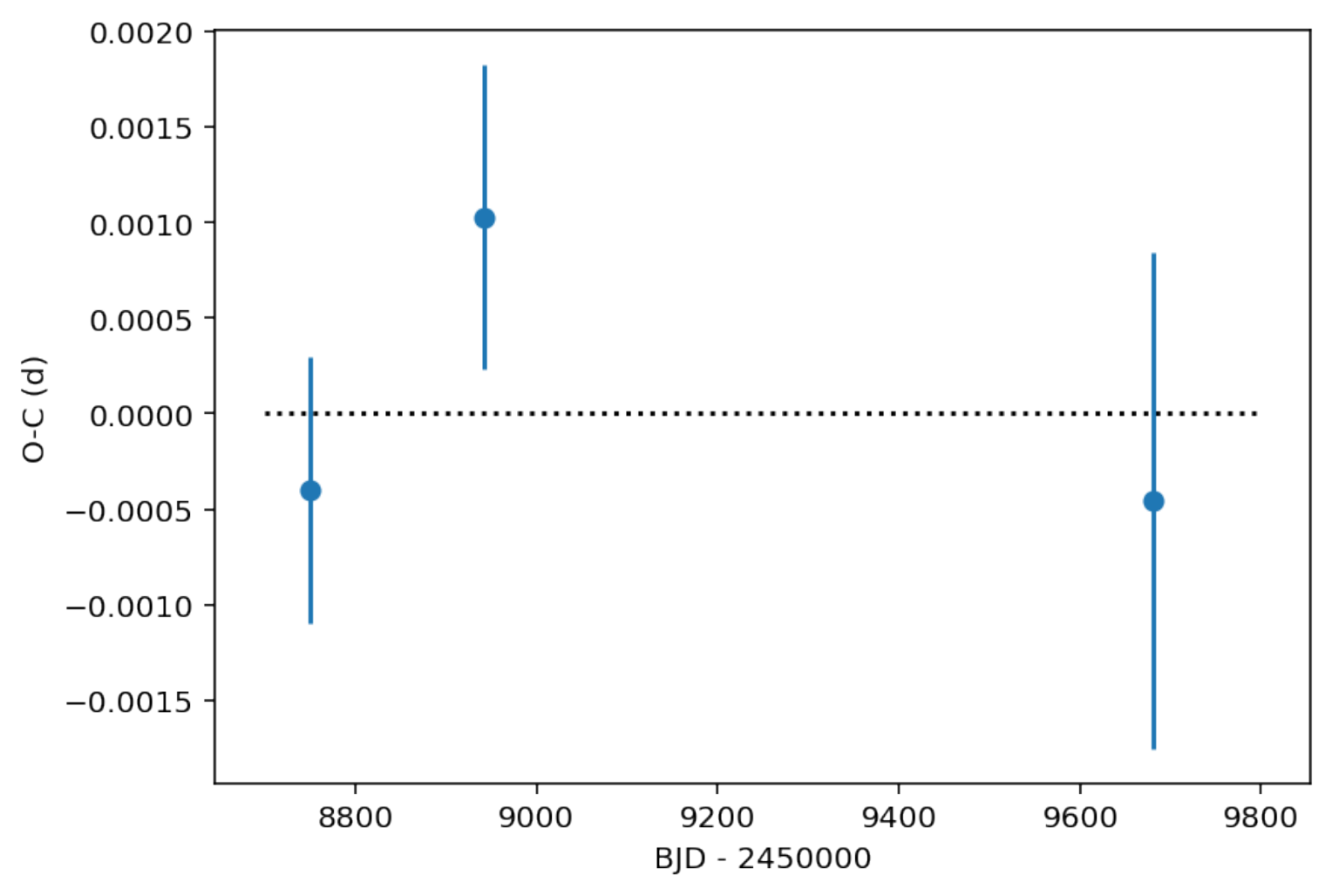}	
\caption{O-C diagram of the transit epochs of TESS Sectors 16, 23 and 50, against the adopted ephemeris (dotted black line).}
\label{fig:OC}\end{figure}

Table~\ref{table:tessobs} indicates also transit epochs for each sector  (corresponding to a transit near the middle of each sector's data), which had been derived using {\tt UTM/UFIT} with a set-up that was identical to the the transit-fit on the combined (S16 to S50) light curve described in Appendix~\ref{app:utm}. Against the adopted ephemeris from Table~\ref{table:final_param_model}, a diagram of observed minus calculated (O-C) times (Fig.~\ref{fig:OC}) shows no relevant deviation that might indicate the presence of transit timing variations.

.

.


\begin{table}
\centering
\small
\caption{{\em TESS} observations of TOI-1416} 
\label{table:tessobs}
\resizebox{\linewidth}{!}{%
\begin{tabular}{ccccccc}
\hline\hline
\noalign{\smallskip}
Sector   &  Camera   & CCD   & Start date & End date & N$_{tr}$&Epoch T$_{0,b}$\\
   &                      &               &        UT          &    UT      &        &BJD-2450000\\
\noalign{\smallskip}
\hline
\noalign{\smallskip}
16  &  4  &   4  &  2019-09-12 &  2019-10-06 &20& 8750.1592[7]\\
23  &  2  &   1  &  2020-03-21 & 2020-04-15 &17& 8942.7168[8]\\
50 &  2 &  2 &   2022-03-26& 2022-04-22 & 11& 9680.8473[13]\\
\noalign{\smallskip}         
\hline
\end{tabular}
}
\tablefoot {The start and end dates refer to the first and last points of the lightcurves after processing as described in Sect. 2. N$_{tr}$ is the number of complete transits of planet $b$. T$_{0,b}$ is the transit epoch of planet $b$  in the given TESS Sector}
\end{table}

\section{Ground-based follow-up}
\subsection{High-Resolution spectroscopy}
\label{sec:obs-hrs}

\begin{figure*}
\includegraphics[width=\linewidth]{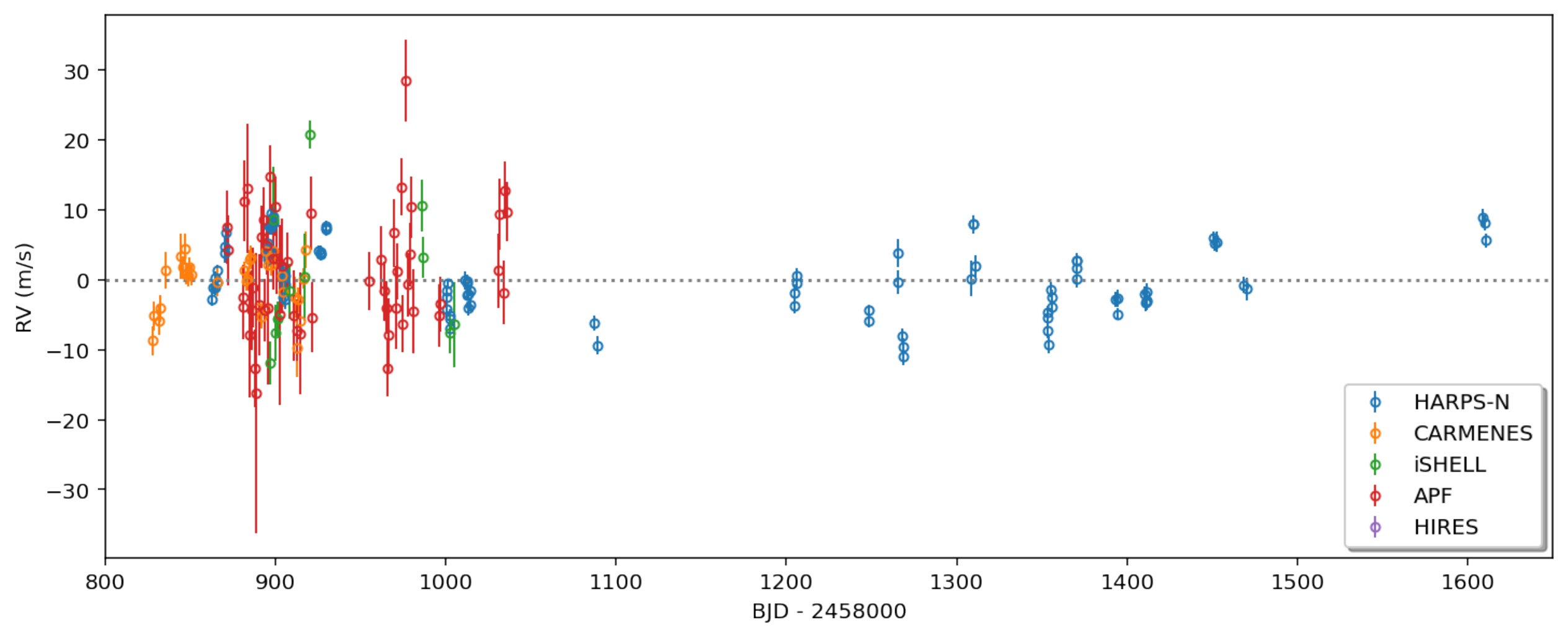}
\caption{Relative radial velocities of \target\ from all contributing instruments. Each instrument's set of RVs was offset separately to an average of zero.
\label{fig:allRVs}}
\end{figure*}

High-resolution spectroscopic observations of \target\ were obtained by several instruments, described in more detail in the following sections, with an overview on the observations given in Table~\ref{table:RVstats}.  Fig.~\ref{fig:allRVs} shows a time-series of all the RVs that have been collected. Corresponding tables  with the RVs from each instrument and -- if available -- spectral indices can be found at the CDS. Also provided at CDS is a joint table in which all acquired RVs are listed in temporal order.

\begin{table}
\centering
\small
\caption{RV observations of \target} \label{table:RVstats}
\resizebox{\linewidth}{!}{%
\begin{tabular}{ccccccc}
\hline\hline
Instrum. &spect. range& Start date & End date & t$_{cov}$ & N$_{RV}$ & $\sigma_{RV}$ \\
 &  $\mu$m&UT& UT & $\mathrm{d}$ &  & $\mathrm{m\,s^{-1}}$ \\
 \hline
CARMENES&0.52-0.96&2019-12-10 & 2020-03-09 & 90 & 34 & 1.96 \\
HARPS-N&0.38-0.69&2020-01-14 & 2022-01-31 & 748 & 96 & 1.06 \\
APF&0.37-0.90&2020-01-23 & 2020-07-05 & 164 & 52 & 5.67 \\
HIRES&0.41-1.02&2020-01-04 & 2020-08-05 & 214 & 12 & 0.85 \\
iSHELL&2.17-2.47&2020-02-17 & 2020-06-04 & 108 & 11 & 4.07 \\
\hline
\end{tabular}
}
\tablefoot{$t_{cov}$ is the time-span covered, N$_{RV}$ the number of RV values, and  $\sigma_{RV}$ is the mean of the formal uncertainties of individual RVs.}
\end{table}

\subsubsection{3.5\,m Calar Alto/CARMENES}
We started the RV follow-up of \target\ using the CARMENES instrument mounted on the 3.5\,m telescope at Calar Alto Observatory, Almer\'ia, Spain, under the observing programs F19-3.5-014 and F20-3.5-011 (PI Nowak), setting the exposure times to 1800 seconds. The CARMENES spectrograph has two arms \citep{2014SPIE.9147E..1FQ,2018SPIE10702E..0WQ}, the visible (VIS) arm covering the spectral range 0.52--0.96\,\microns\ and a near-infrared (NIR) arm covering the spectral range 0.96--1.71\,\microns. Due to the S/N that was obtained, only the VIS channel observations could be used to derive useful RV measurements. All observations were taken with exposure times of 1800\,s, resulting in a SNR per pixel ( at 4635,7 nm in  the VIS spectra) in the range of 42 to 113. CARMENES performance, data reduction and wavelength calibration are described in \citet{2018A&A...609A.117T} and \citet{2018A&A...618A.115K}. Relative radial velocity values, chromatic index (CRX), differential line width (dLW), and H$\alpha$ index values were obtained using {\tt serval}\footnote{\url{https://github.com/mzechmeister/serval}} \citep{2018A&A...609A..12Z}. For each spectrum, we also computed the cross-correlation function and its full width half maximum, contrast and bisector velocity span values, following \citet{2020A&A...636A..36L}. The RV measurements were corrected for barycentric motion, secular acceleration and nightly zero-points. 

\subsubsection{TNG/HARPS-N}
\label{sec:HN}
96 spectra in three observing seasons were collected with the \hn\ spectrograph with R$\approx$ 115\,000 \citep{2012SPIE.8446E..1VC}, mounted at the 3.58-m Telescopio Nazionale Galileo (TNG) of Roque de los Muchachos Observatory in La Palma, Spain. 
The exposure times were set to 636--2700 s, based on weather conditions and scheduling constraints, leading to a SNR per pixel (at 5500\,\AA) of 48--138. The spectra were extracted using the \hn\ {\tt DRS} pipeline version 3.7 \citep{2014SPIE.9147E..8CC}. Doppler measurements and spectral activity indicators (CCF\_FWHM, CCF\_CTR, BVS and the Mont-Wilson S-index) were measured using the {\tt DRS} and the {\tt YABI} tool\footnote{Available at \url{http://ia2-harps.oats.inaf.it:8000}}, by cross-correlating the extracted spectra with a K5 mask \citep{1996A&AS..119..373B}. Furthermore, we used {\tt serval} to measure relative RVs, chromatic RV index, differential line width, and the H$\alpha$ index, as defined in \citet{2018A&A...609A..12Z}. 
While both {\tt DRS} and {\tt serval} derive very similar RVs, we adopted those from {\tt serval} for further analysis, due to issues with the DRS in those exposures that were terminated prematurely; see also Fig.~\ref{fig:HNdrssrv} in Appendix~\ref{app:morefigs}. 
The table of \hn\ measurements available at CDS contains the 96 RVs from both pipelines, together with all activity indicators extracted by either pipeline, in the following columns (for most indicators, a column with the errors is also provided, not shown below):

\begin{tiny}
\begin{verbatim}
bjd_tdb       - BJD_TDB
rvs_srv       - SERVAL barycentric corrected relative RV
                 (against a zero average)
rvs_drs       - DRS  barycentric corrected absolute RV 
ccf_bis_drs   - DRS Bisector Inverse Slope (BIS) measured 
                 from Cross-Correlation Functions (CCFs) 
ccf_fwhm_drs  - DRS Full Width at Half Maximum of CCF 
ccf_ctr_drs   - DRS CCF contrast 
smw_drs       - DRS Mont-Wilson S-index
log_rhk_drs   - DRS log(R_{HK}) 
crx_srv       - SERVAL chromatic RV index (CRX) 
dlw_srv       - SERVAL differential line width (dLW) 
halpha_srv    - SERVAL H-alpha index 
nad1_srv      - SERVAL sodium Na~D1 index 
nad2_srv      - SERVAL sodium Na~D2 index 
snr_550nm_drs - DRS SNR at spectral order 46 (~550 nm)
expt            - exposure time from FITS header
\end{verbatim}
\end{tiny}


\subsubsection{IRTF/iSHELL}

A total of 11 observations of  \target\  was obtained in as many nights with the iSHELL instrument at NASA InfraRed Telescope Facility \citep[IRTF,][]{2022PASP..134a5002R} atop Mauna Kea, Hawaii, USA, using its KGAS mode covering the wavelengths of 2.17-2.47 $\mu$m. The exposure times were always set at 300 seconds, and exposures were repeated anywhere from 4-16 times consecutively per night, in order to obtain a signal-to-noise ratio (SNR) per spectral pixel of $\approx$ 120, though the actual results varied from 85-186 due to variable seeing and atmospheric transparency conditions. A methane isotopologue (13CH4) gas cell is used in the instrument \citep{2019AJ....158..170C} to constrain the line-spread function and to provide a common reference for the optical path wavelength. Along with each observation, a set of five 15-second flat-field images was also collected, with the gas cell removed for data reduction purposes, in order to mitigate flexure-dependent and time-variable fringing present in the spectra. The 11 RVs included in the electronic tables at CDS are nightly averaged values from the individual exposures.

\subsubsection{Keck/HIRES and Lick Observatory APF}
The High Resolution Echelle Spectrometer (HIRES) on the 10m Keck Observatory \citep{HIRES} was used to obtain 12 high-resolution spectra of \target, and the Automatic Planet Finder (APF) on the Lick Observatory \citep{APF} was used to obtain 52 
high-resolution spectra. Each exposure of \target\  was about 500 s on HIRES and 1200 s on APF. 
 We also obtained an iodine-free spectrum on HIRES as the template for the radial velocity extraction for both the HIRES and APF observations. The HIRES radial velocities collected using the telescope setup, the instrument setup, and the analysis pipeline described in \citet{Howard}. The APF radial velocities were collected using a 1$\arcsec$ decker and analyzed with the standard California Planet Search pipeline \citep{2015ApJ...805..175F}.

\subsection{Ground-based imaging and time-series photometry} 
The TESS pixel scale is $\sim21\arcsec$ per pixel and its photometric apertures typically extend out to roughly 1$\arcmin$, generally causing multiple stars to blend in the TESS aperture. To attempt to determine the true source of our detection in the TESS data, we conducted ground-based imaging and photometric time-series observations of the field around \target\ as part of the TESS Follow-up Observing Program\footnote{https://tess.mit.edu/followup} (TFOP) Sub Group 3 (High-resolution Imaging) and Sub Group 1 \citep[Seeing limited Photometry;][]{collins:2019}.

\subsubsection{High-resolution imaging at Palomar Observatory}\label{subsec:hri}

As part of our standard process for validating transiting exoplanets to assess the possible contamination of bound or unbound companions on the derived planetary radii \citep{ciardi2015}, we observed \target\ with infrared high-resolution adaptive optics (AO) imaging at Palomar Observatory.  The Palomar Observatory observations were made with the PHARO instrument \citep{hayward2001} behind the natural guide star AO system P3K \citep{dekany2013} on 2020-01-08 UT in a standard 5-point quincunx dither pattern with steps of 5\arcsec.  Each dither position was observed three times, offset in position from each other by 0.5\arcsec\ for a total of 15 frames.  The camera was in the narrow-angle mode with a full field of view of $\approx 25\arcsec$ and a pixel scale of approximately $0.025\arcsec$ per pixel. Observations were made in the narrow-band $Br-\gamma$ filter $(\lambda_o = 2.1686; \Delta\lambda = 0.0326\mu$m) with an integration time of 5.6 s per frame (118 seconds total). 

\begin{figure}
\centering
	\includegraphics[width=1\linewidth]{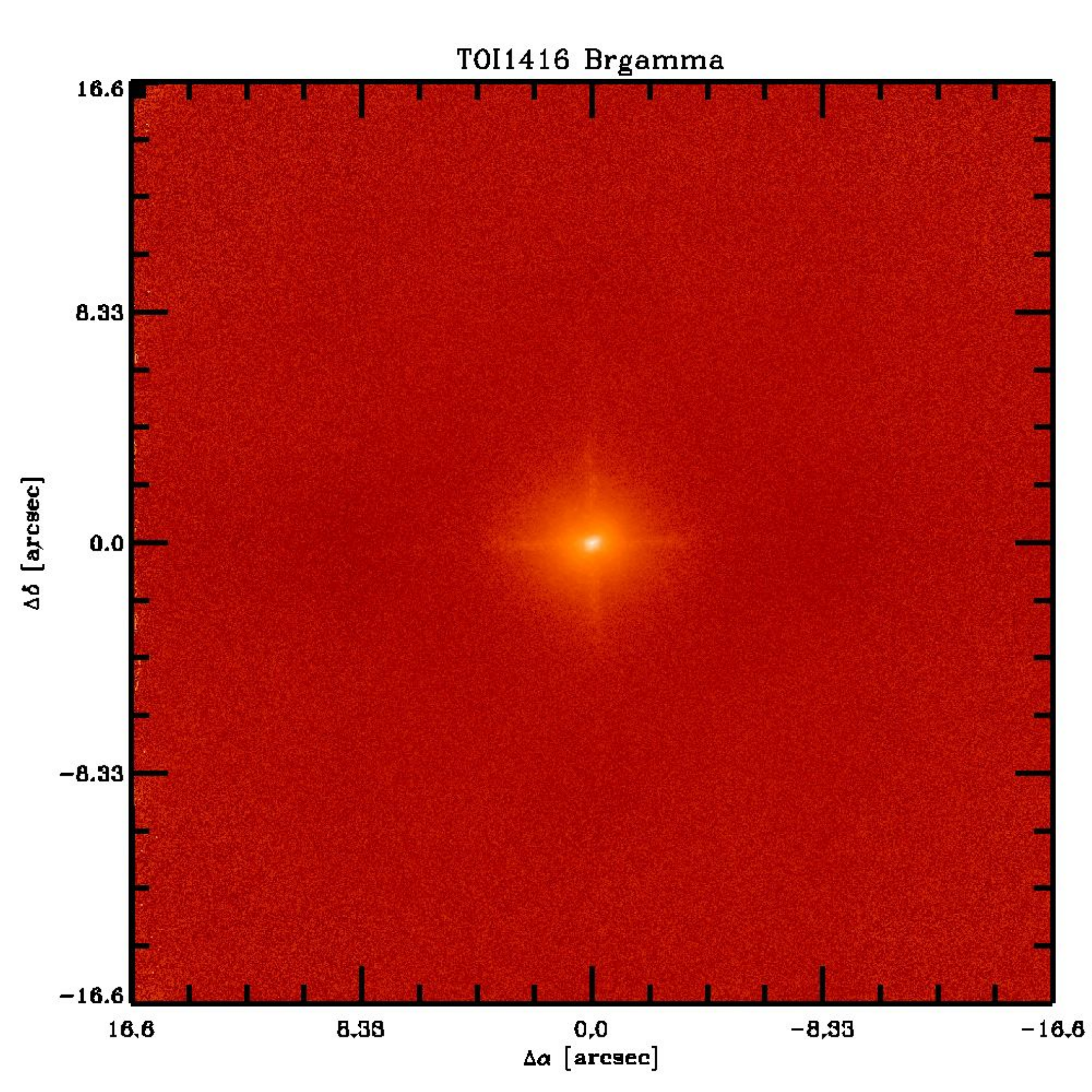}	
\caption{Full field of view image of the final combined dither pattern for the Palomar AO imaging.}\label{fig:ao_fullfov} 
\end{figure}

\begin{figure}
\centering
	\includegraphics[width=1\linewidth]{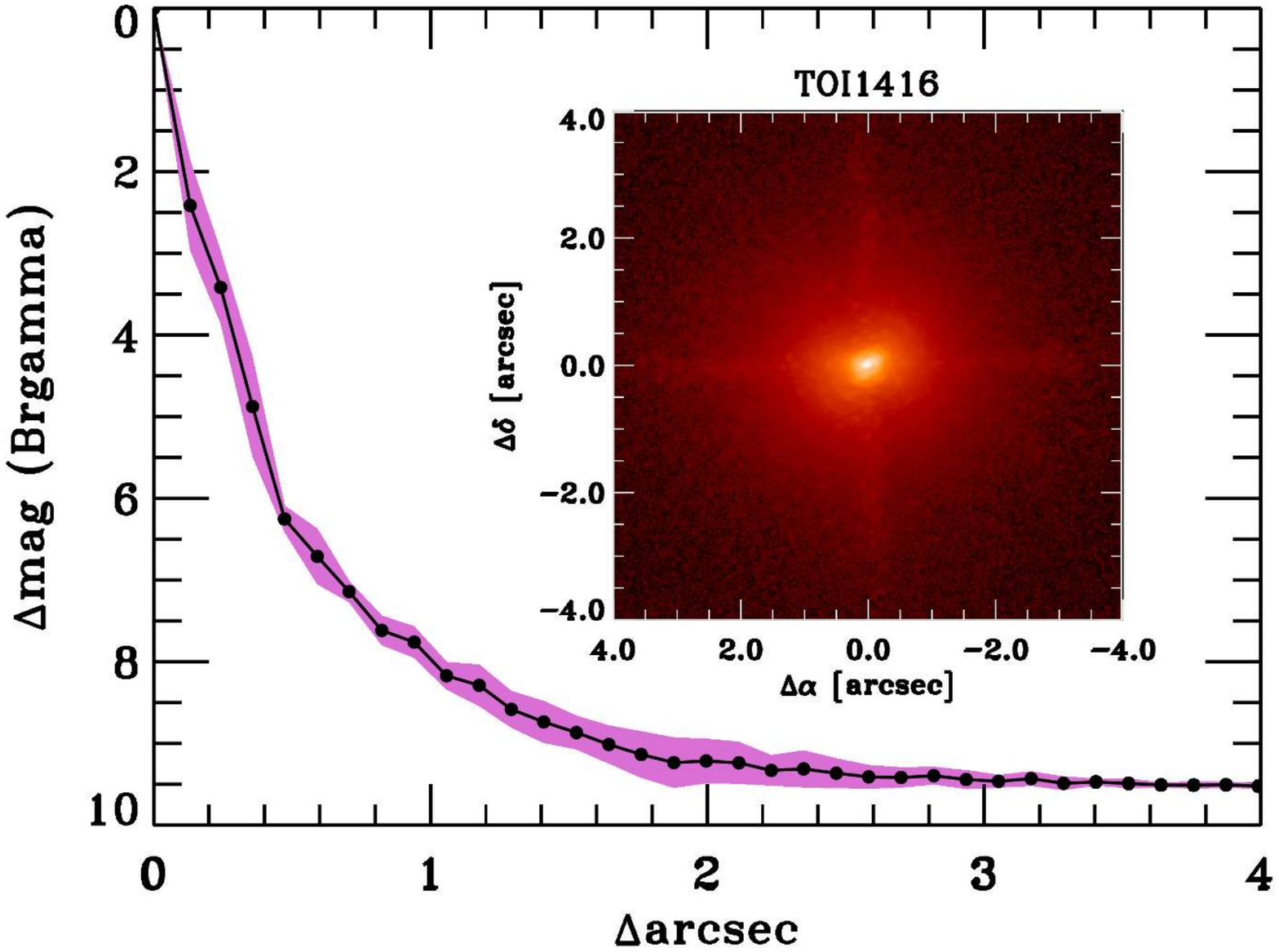}
\caption{Companion sensitivity for the Palomar AO imaging.  The black points represent the 5$\sigma$ limits and are separated in steps of 1 FWHM ($\approx 0.1$\arcsec); the purple zone represents the azimuthal dispersion (1$\sigma$) of the contrast determinations (see text). The inset image is of the primary target showing no additional companions within 3\arcsec\ of the target.}\label{fig:ao_contrast}  
\end{figure}

The AO data were processed and analyzed with a custom set of tools written in IDL.  The science frames were flat-fielded and sky-subtracted.  The flat fields were generated from a median average of dark subtracted flats taken on-sky.  The flats were normalized such that the median value of the flats is unity.  The sky frames were generated from the median average of the 15 dithered science frames; each science image was then sky-subtracted and flat-fielded.  The reduced science frames were combined into a single combined image using an intra-pixel interpolation that conserves flux, shifts the individual dithered frames by the appropriate fractional pixels, and median-coadds the frames (Fig.~\ref{fig:ao_fullfov}).  The final resolution of the combined dither was determined from the FWHM of the point spread function, of 0.11$\arcsec$ (Fig.~\ref{fig:ao_contrast}). 

No sources, other than the primary target, were detected. The sensitivities of the final combined AO image were determined by injecting simulated sources azimuthally around the primary target every $45^\circ $ at separations of integer multiples of the central source's FWHM \citep[][Lund, M.B. et al., in prep.] {furlan2017}. The brightness of each injected source was scaled until standard aperture photometry detected it with $5\sigma $ significance. The resulting brightness of the injected sources relative to the target sets the contrast limits at that injection location. The final $5\sigma $ limit at each separation was determined from the average of all of the determined limits at that separation and the uncertainty on the limit was set by the {\it rms} dispersion of the azimuthal slices at a given radial distance. The sensitivity curve is shown in Fig.~\ref{fig:ao_contrast} along with an image zoomed around the target, showing no other companion stars.
We also note that an interrogation of the GAIA EDR3 showed as the most nearby star one that is 23" W of the target and 11.4 mag fainter, whereas as the second closest one is 51" NE and 9.5 mag fainter; due to their faintness neither of these stars can be responsible for the transits on \target. 

\subsubsection{Time-series photometry with MUSCAT2}

\target\ was observed with the MUSCAT2 multi-colour imager \citep{MUSCAT2} mounted at the 1.5m Telescopio Carlos S\'anchez at Teide Observatory, Tenerife, Spain, on several dates: Between 2020-01-17 03:42 UT and 06:12 covering a full transit of planet $b$; 2021-05-03 22:19 and 2021-05-03 02:09 UT with a partial transit (ingress) and 2022-04-20 20:45 and 2022-04-20 00:07 UT for a full transit. The raw data were reduced by the MuSCAT2 pipeline \citep{2019A&A...630A..89P} 
which performs standard image calibration, aperture photometry, and is capable of modelling the instrumental systematics present in the data while simultaneously fitting a transit model to the light curve. Due to the target's brightness, only short exposure times could be used. Given the noise present,  no evidence for a transit could be found on the target. There are  77 sources listed in the GAIA DR3 in a radius of 2.5$\arcmin$ around the target, of which 7 have a  brightness large enough  that they could potentially be an eclipsing binary that mimics the transit observed by TESS. Of these, however, only the star TIC 158025007, which is the brightest nearby contaminant and about 1.5$\arcmin$ south of the target, could be excluded with certainty as a source for a false alarm.

\subsubsection{Time-series photometry with LCOGT}

We observed full predicted transit windows of \planet\ on 2020-05-21 UT and 2021-03-08 UT using the Las Cumbres Observatory Global Telescope \citep[LCOGT;][]{Brown:2013} 1.0\,m network node at McDonald Observatory. The 1\,m telescopes are equipped with $4096\times4096$ pixel SINISTRO cameras having an image scale of $0\farcs389$ per pixel, resulting in a $26\arcmin\times26\arcmin$ field of view. The images were calibrated by the standard LCOGT {\tt BANZAI} pipeline \citep{McCully:2018} and differential photometric data were extracted using {\tt AstroImageJ} \citep{Collins:2017}.

We extracted light curves from the 2020-05-21 UT data for all 6 known Gaia DR3 and TICv8 neighboring stars within $2\farcm5$ of TOI-1416 that are bright enough in the TESS band to produce detection by TESS. We thus checked all stars down to 8.4 magnitudes fainter than TOI-1416 (i.e. down to 17.5 mag in TESS band). We calculate the {\it rms} of each of the 6 nearby star light curves (binned in 5 minute bins) and find that the LCOGT light curve {\it rms} values are smaller by at least a factor of 5 compared to the expected NEB (Nearby Eclipsing Binary) depth in each respective star. We then visually inspected the neighboring star light curves to ensure no obvious deep eclipse-like signal. We therefore rule out NEBs as the cause of the \planet\ detection in the TESS data.

\begin{figure}
\centering
	\includegraphics[width=1\linewidth]{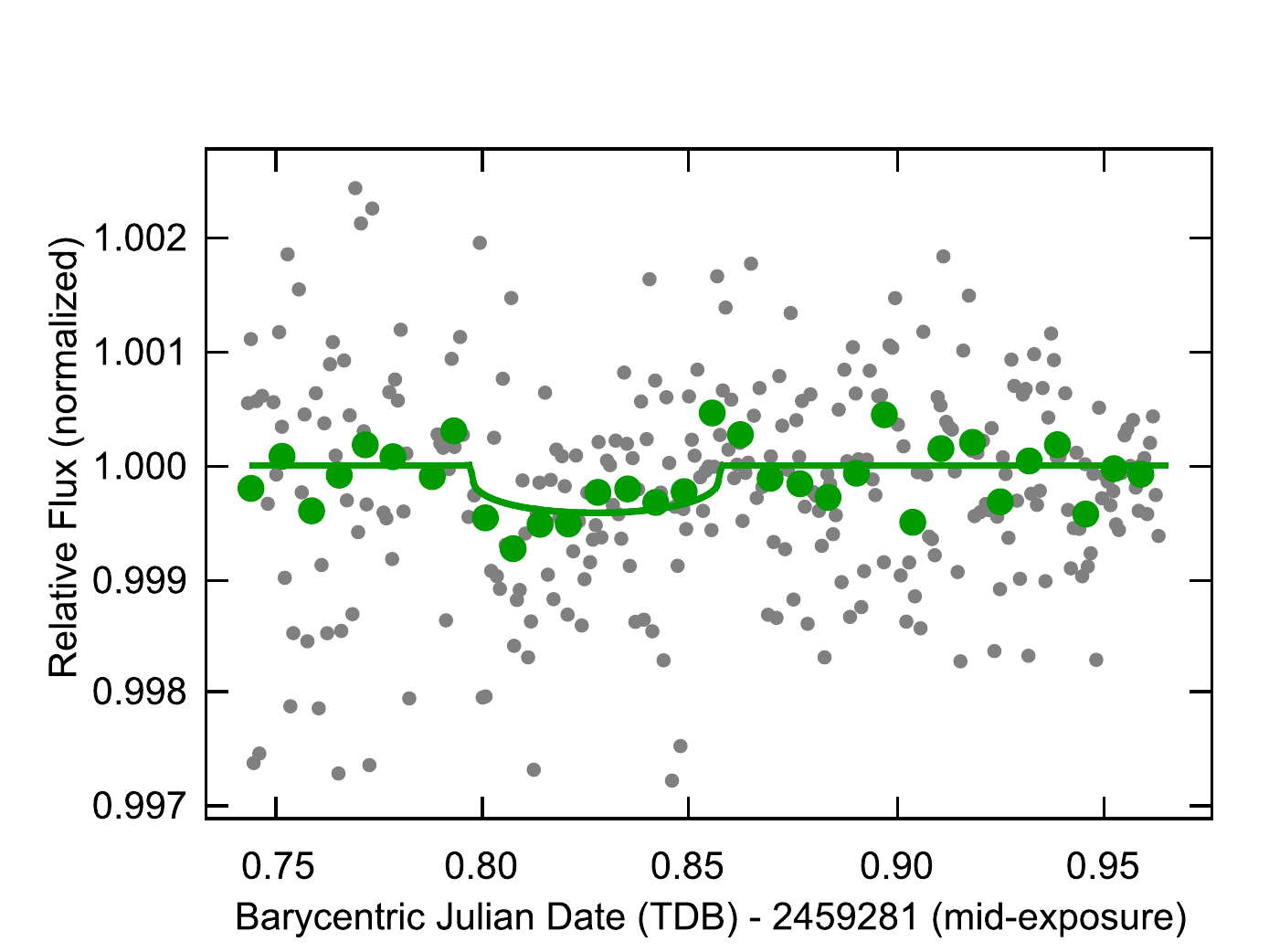}
\caption{Timeseries of a predicted transit of \planet\ on 2021-03-08 UT, observed by the LCOGT. The grey dots are the unbinned differential photometry (no detrending applied) and the green dots show the data in 10 minute bins. The green line is a transit-model fit to the data using priors from the Data Validation Report mentioned in Sect. 2, except the epoch and the size of the planet, which were unconstrained parameters. The differ
}\label{fig:lcogt}  
\end{figure}

For the second observation on 2021-03-08 UT, we defocused the telescope to improve photometric precision and attempt to detect the shallow \planet\ event on target. As shown in Fig.~\ref{fig:lcogt}, we find a likely transit detection centered at $2459281.827\pm0.005$\,$\rm BJD_{TDB}$ with a depth of $350\pm100$ ppm. {The difference between Bayesian Information Criterion (BIC) of the transit model shown and one without any transit was $\Delta$-BIC = - 43 in favor of the transit model.}

\section{Stellar modelling} \label{sec: Stellar modelling}
\subsection{Spectral analysis} \label{sec: spectral analysis}
We started our analysis of the host star by first deriving the stellar effective 
temperature, $T_\mathrm{eff}$, the stellar 
radius, $R_\star$, and the abundance  of 
the key species iron relative to hydrogen, [Fe/H], with the empirical {\tt {SpecMatch-Emp}} code \citep{2017ApJ...836...77Y}.
We modelled our co-added high resolution ($R=115000$) \hn\ spectra with a SNR of 346 at 6100~\AA. 
This software characterises stars from their optical spectra
and compares observations to a dense spectral library of  well-characterised FGKM stars observed with Keck/HIRES.

In addition to {\tt {SpecMatch-Emp}}, we analysed  the co-added \hn\ spectra with version 5.22 of the  spectral analysis  
package \href{http://www.stsci.edu/~valenti/sme.html}{{\tt{SME}}} \citep[Spectroscopy Made Easy;][]{vp96, pv2017}.
This software is fitting observed spectra 
to calculated  synthetic stellar spectra for a given set of parameters.  
We chose the Atlas12 \citep{Kurucz2013} atmosphere grids, and retrieved the 
atomic and molecular line data from \href{http://vald.astro.uu.se}{VALD} \citep{Ryabchikova2015}  
to synthesise the spectra. We modelled   $T_\mathrm{eff}$ from the line wings of the hydrogen 
$\lambda$6563 line, and 
the surface gravity, $\log g$, from the calcium triplet 
 at 6102, 6122, and 6162 \AA, and the 6439 \AA\ line. We fitted the iron and calcium 
abundances, the projected stellar rotational velocity, $V \sin i_\star$, and the macroturbulent velocity, $V_\mathrm{mac}$
from unblended lines between 6000 and 6600~\AA. 
The sodium abundance was fitted from spectral lines between 5600 and 6200~\AA. 
We found  similar abundances of iron, calcium and sodium, and determined $V \sin i_\star = 2.0\pm0.7$~km~s$^{-1}$ and $V_\mathrm{mac} = 1.5\pm1.0$~km~s$^{-1}$. 
To check and further refine our model, we used the  \ion{Na}{I}  doublet  at 5888 and 5895 \AA. 
The resulting model suggests that TOI-1416 is a an early K dwarf star.

Results from both models are listed in Table~\ref{table: spectroscopic parameters} and are in good agreement within
the uncertainties. They also agree well with the corresponding values from the Gaia DR2 and from the TESS Input Catalogue \citep[TIC, ][]{2018AJ....156..102S,2019AJ....158..138S}.

The metalicity and kinematics of \target point to a membership  in the galactic thin disk; following the precepts of \citet{2006MNRAS.367.1329R}, we obtain a thin-disk membership probability of 0.975$\pm$0.012.


\subsection{Stellar mass, radius and age} 
\label{subsect: Stellar mass and radius}
\begin{figure}
\centering
	\includegraphics[width=1\linewidth]{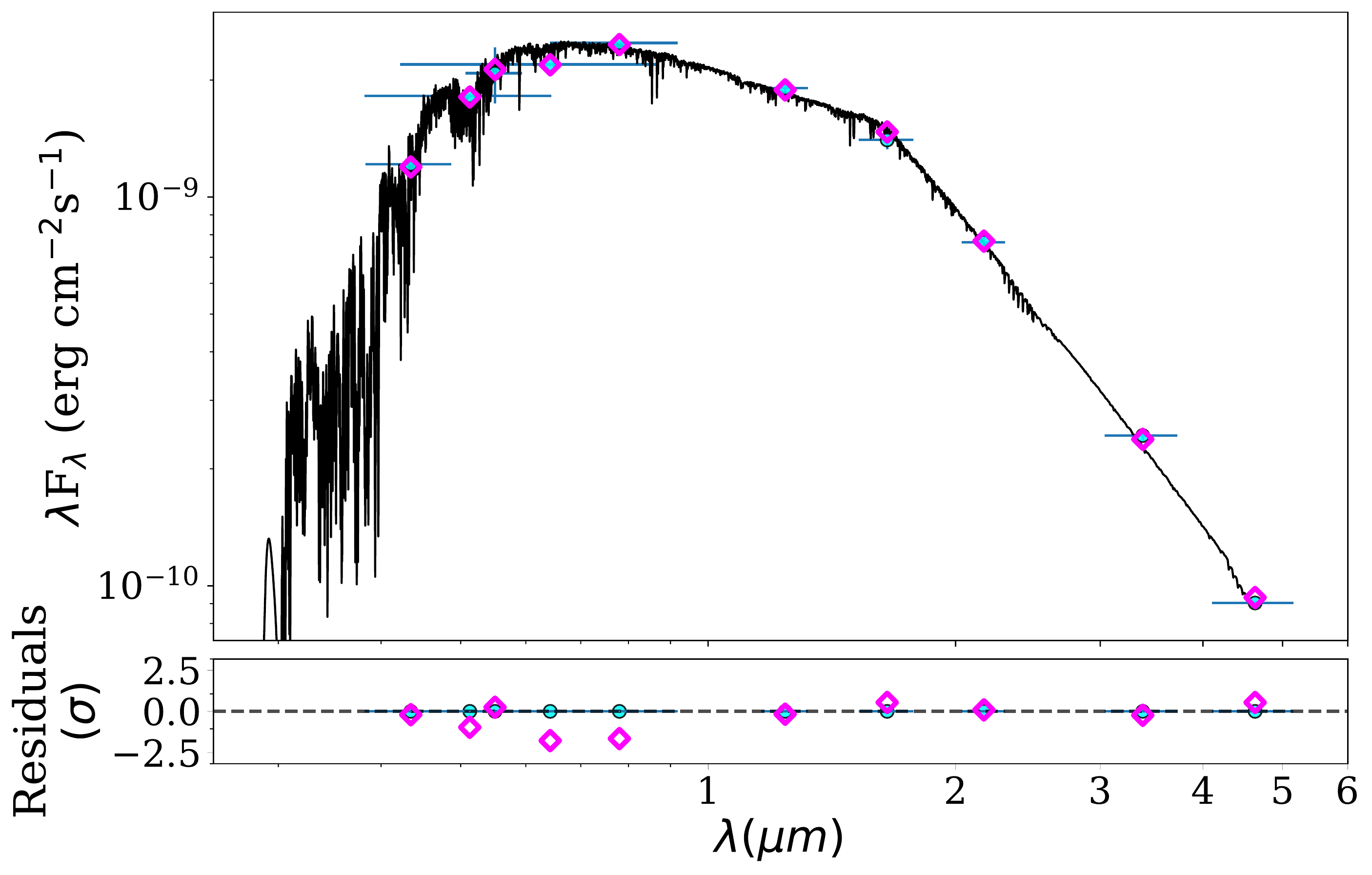}
    \caption{The  spectral energy distribution (SED) of TOI-1416. The best fitting model {\tt {Phoenix~v2}} is
    shown in black.  
    The observed photometry is marked with cyan circles, and the synthetic photometry with magenta diamonds. 
    The  horizontal bars of the observations indicate the effective widths of the passbands, 
    while the vertical bars mark the $1~\sigma$ uncertainties.
    The lower panel shows the residuals normalised to the errors of the photometry which implies that precise photometry shows the largest scatter.}
    \label{fig: sed}
\end{figure}

To obtain an independent estimate of the stellar radius, we analysed the  spectral energy distribution (SED) of
TOI-1416   
with the python code {\tt{ARIADNE}} \citep{2022MNRAS.513.2719V}. 
This software fits  broadband photometry   to the 
{\tt {Phoenix~v2}} \citep{2013A&A...553A...6H}, {\tt {BtSettl}} \citep{2012RSPTA.370.2765A}, 
\citet{Castelli2004}, and \citet{1993yCat.6039....0K} atmospheric model grids for stars with 
$T_\mathrm{eff} > 4000$~K convolved with various filter response functions.  
For TOI-1416, we utilised data in the bandpasses  $G G_{\rm BP} G_{\rm RP}$ from  Gaia eDR3,  
{\it WISE} W1-W2,     $JHK_S$ magnitudes from {\it 2MASS},
and the Johnson $B$ and $V$ magnitudes from APASS DR9 \citep[AAVSO Photometric All-Sky Survey;][]{2016yCat.2336....0H}. 
By interpolating the $T_\mathrm{eff}$, $\log g_\star$, and [Fe/H]  model grids,  
 SED models were produced  where 
 distance, extinction  ($A_V$), and stellar radius are treated as free parameters. 
The Gaia eDR3 parallax   was used to obtain the distance, and  
priors for  $T_\mathrm{eff}$, $\log g_\star$, and [Fe/H]   were taken from  {\tt{SME}}. 
We used   flat priors for $R_\star$  
between 0.05 and 20~$R_\odot$, 
and  for  $A_V$ 
between zero and   the maximum line-of-sight
value from the dust maps of \citet{1998ApJ...500..525S}. 
Each SED model was integrated to get the bolometric flux which together with $T_\mathrm{eff}$ and
the $Gaia$ eDR3 parallax gives the stellar radius for each fitted model. 
The weighted average of each parameter is computed based on the relative probabilities of the models, and 
the final value of the stellar radius is computed with Bayesian Model Averaging. 
The {\tt {Phoenix~v2}} model grid which has the highest probability was used to calculate the synthetic 
photometry. The model is  shown in Fig.~\ref{fig: sed} along with the fitted bands. 

In addition to the above modelling we used the python code 
\href{https://github.com/timothydmorton/isochrones}{\tt{isochrones}} \citep{2015ascl.soft03010M}
to obtain a homogeneous  model of TOI-1416. 
This code is fitting stellar parameters with an MCMC fitting tool and the 
MIST \citep{2016ApJ...823..102C}  stellar evolution tracks. We used the same bands and priors as in the 
{\tt{ARIADNE}} model. 
We find  $A_V$\,=\,\Avisochrones~mag, and a bolometric luminosity of \Lisochrones~$L_{\odot}$.
The resulting stellar properties are in very good agreement with the values found by the above models.

As a comparison, we used the {\tt {Param 1.5}}  on-line tool  \citep{daSilva2006, 2014MNRAS.445.2758R, 2017MNRAS.467.1433R} 
with the PARSEC isochrones
\citep{2012MNRAS.427..127B} and the same bands and priors as in the above models. 
And finally, we used 
 the empirical calibration equations of \citet{2010AJ....140.1158T} to compute
stellar mass and radius from $T_\mathrm{eff}$, $\log g$, and [Fe/] from {\tt{SME}}.

\begin{figure}
\centering
	\includegraphics[width=1\linewidth]{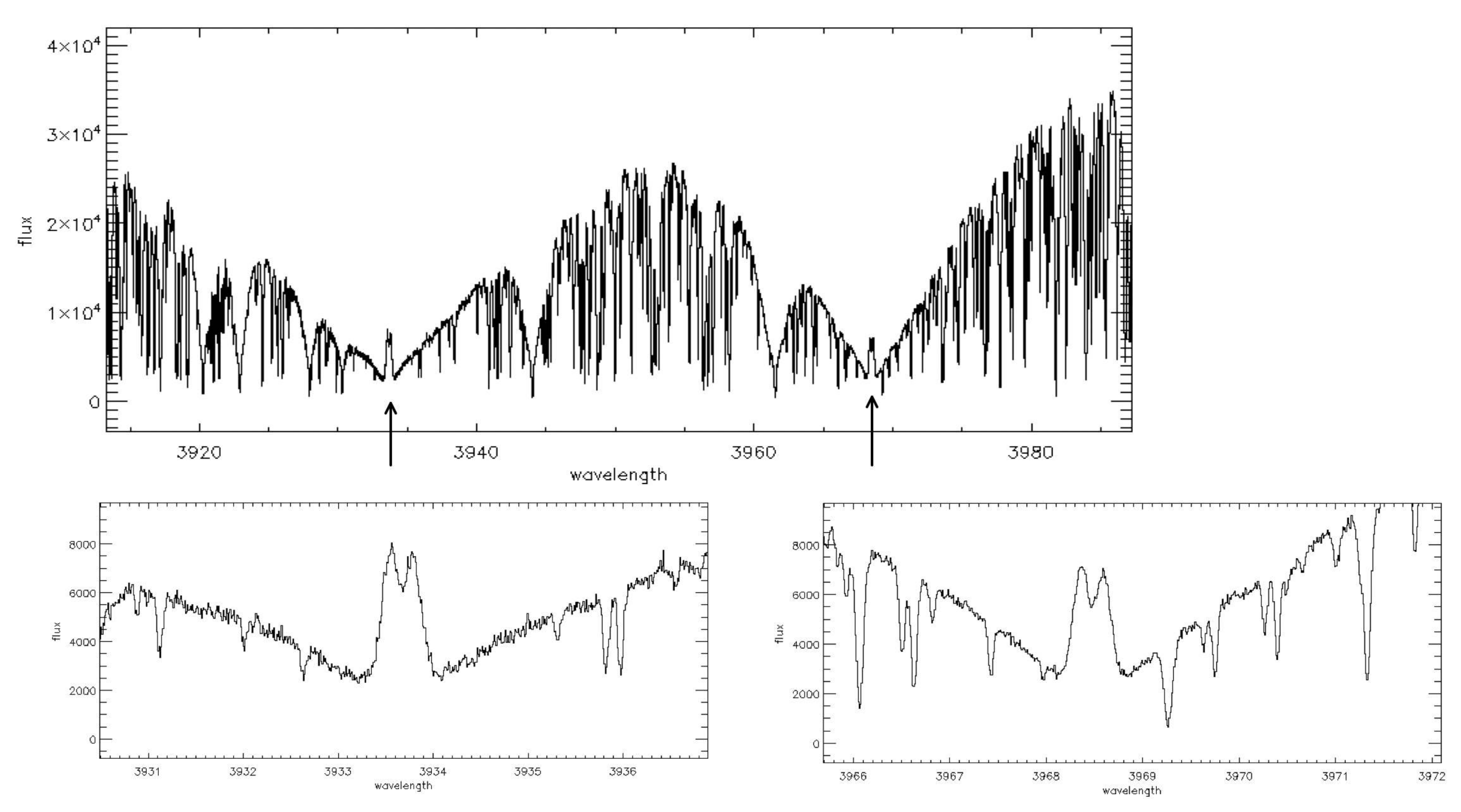}
    \caption{Upper panel:  Co-added \hn\ spectrum of \target \ analysed with {\tt{SME}} \citep{vp96, pv2017}, in the range of the Ca H \& K lines (arrows). Lower panels: Zooms around the Ca K (3933.66 \AA ) and Ca H (3968.47 \AA) lines.}
    \label{fig:CaHK}
\end{figure}

All results are in excellent agreement. The stellar masses, radii, and corresponding bulk densities, 
are listed in Table~\ref{table: comparison stellar parameters}  
together with the Gaia radius for comparison. 
The adopted values, which were also used in the joint modelling of the radial velocities and light curves in Sect.~\ref{sec:joint}, were derived by the 
  adding of simulated probability distributions that are associated to each of the values from the different methods (the values from the TIC were not used for this), using two-sided Gaussian distributions with 1 million elements. Hence, each of the methods has been taken with equal weight. In the resultant distribution, the percentiles at 15.9, 50, and 84.1 percent where then used to quote the median and the $\pm$ 1-sigma errors.
The derived values for the temperature place \target\ right at the border between spectral classes G and K, with a slight preference for spectral class G9V, due to the notable Ca H\& K lines (Fig.~\ref{fig:CaHK}), which are defining feature of the class G \citep[][p. 158]{1901AnHar..28..129C}. The mean $R'_{HK}$ index among the  96 \hn\ spectra of $log(R'_{HK} ) = -4.86 \pm 0.03$ indicates however only very moderate chromospheric activity.
This activity implies also an age in the range of 4 -- 7 Gyr, based on the activity-age relation by \citet{2008ApJ...687.1264M}. Ages from the aforementioned isochrone analyses are not very well constrained but indicate a similar evolutionary phase, with MIST isochrones indicating an age of $10.6^{+0.5}_{-3.2}$ Gyr and {\tt {Param 1.5}} one of $13.8^{+0.2}_{-3.9}$ Gyr, which in either case excludes that \target\ is a very young system. With \target\ being a likely thin-disk member and age estimates for the local thin disk being  6.8 -- 7.0 Gyr \citep{2017ApJ...837..162K}, the age of \target\ is most likely close to that value.

In Appendix ~\ref{sec:rotation} we also present an analysis of the stellar rotation based on the TESS lightcurves, leading to $P_\mathrm{rot}=17.6$ d, which is also compatible with a rotation period of $P_\mathrm{rot}/ \sin i = 20_{-5}^{11} $ d from the star's $V \sin i$, and which leads to a gyrochronological age of 1 -2 Gyr. This apparently young age might however be a consequence from a delay in the star's age-related spin-down due to the presence of the close planet $b$, given that its orbital period is shorter than the stellar rotation period, with a transfer of angular momentum from the planet to the host star \citep{1980A&A....92..167H}.

The work by \citet{2021A&A...650A.126A} indicates that a planet with the mass and orbital period of \pname\ might  have a moderate effect on the star's rotation through magnetic interactions \citep{2016ApJ...833..140S}, and hence invalidate its gyrochronological age. However, more detailed studies that include also mass-loss scenarios for the planet \citep[e.g.][]{2021A&A...647A..40A} would be needed for a better estimate of the planet's effects onto the stellar rotation throughout its evolution, which then might enable a correction of its gyrochronological age.

\begin{table}
\centering
  \caption{Spectroscopic  parameters for TOI-1416 derived with {\tt {SME}} and {\tt {SpecMatch-Emp}} and comparison values from Gaia and the TIC.   
 \label{table: spectroscopic parameters}}
\resizebox{\columnwidth}{!}{%
\begin{tabular}{lcccc}
 \hline\hline
     \noalign{\smallskip}
Method  & $T_\mathrm{eff}$   & [Fe/H]   &  $\log(g)$ & $V\sin (i)$  \\  
& (K)  &(dex) &(cgs) &(km~s$^{-1}$)  \\
    \noalign{\smallskip}
     \hline
\noalign{\smallskip} 
{\tt {SME}}$^a$  &\steffsme      &        \sfehsme&  \sloggsme &\svsinisme    \\
 {\tt {SpecMatch-Emp}}   &\steffspecm  &  \sfehspecm&      \ldots &   \ldots \\
 Gaia DR2 &$4909^{+97}_{-58}$ &\ldots &\ldots &\ldots  \\
 TIC&$4946\pm129$&\ldots&$4.54\pm0.09$&\ldots \\
\noalign{\smallskip} 
\hline
\noalign{\smallskip}
\multicolumn{5}{l}{$\footnotesize^a$ Adopted for the modelling of stellar mass and radius in Sect.~\ref{subsect: Stellar mass and radius}.} \\
 \end{tabular}%
 }
\end{table}

\begin{table}
\centering
  \caption{Stellar masses and radii with corresponding mean densities of TOI-1416 derived with different models with priors from {\tt {SME}}.  
\label{table: comparison stellar parameters}}
\resizebox{\columnwidth}{!}{%
\begin{tabular}{lccc}
 \hline\hline
     \noalign{\smallskip}
Method    & $M_\star$  &     $R_\star$   & $\rho_\star$     \\
  & ($M_{\odot}$)  & ($R_{\odot}$)  & (g~cm$^{-3}$)  \\
    \noalign{\smallskip}
     \hline
\noalign{\smallskip} 
{\tt {isochrones}}    & \smassisochrones & \sradiusisochrones  & \srhoisochrones  \\
{\tt {Param1.5}}$^{a}$    & \smassparam &\sradiusparam & \srhoparam    \\ \noalign{\smallskip}  
SED$^{b}$ &\smassSED & \sradiusSED& \srhoSED\\
 {\tt {SpecMatch-Emp}}  &\ldots & \sradiusspecm&\ldots \\
Torres$^{c}$ &\smasstorres &   \sradiustorres & \srhotorres \\
Gaia DR2 &\ldots & \sradiusgaia&\ldots \\
TIC$^d$&$0.81\pm0.10$&$0.80\pm0.05$&$2.21\pm0.60$\\
Adopted value&$0.798^{+0.035}_{-0.044}$&$0.793^{+0.036}_{-0.028}$& $2.21^{+0.32}_{-0.21}$\\
\noalign{\smallskip}  
\hline
\noalign{\smallskip}   
\multicolumn{3}{l}{$\footnotesize^a$\href{http://stev.oapd.inaf.it/cgi-bin/param}{Param1.5} with PARSEC isochrones.}  \\
\multicolumn{3}{l}{$\footnotesize^b$\href{https://github.com/jvines/astroARIADNE}{ARIADNE} SED fitting with Bayesian Model Averaging.} \\ 
\multicolumn{3}{l}{$\footnotesize^c$\citet{2010AJ....140.1158T} calibration equations.} \\
\multicolumn{3}{l}{$\footnotesize^d$Not used for adopted values.}\\
 \end{tabular}
 }
\end{table}

 
\section{Planet system modelling}
\label{sec:planmod}
{
In this section, we first provide an analysis of the periodicities and activity indicators in the RV data, with a detailed evaluation of a potential contamination of RV signals by lunar light given in Appendix~\ref{lunar_disco}. This is followed by a joint RV/transit RV fit  using Gaussian Processes, in which several models with and without a second planet were evaluated. A fit to the RVs using the Floating Chunk Offset (FCO) method \citep{2010A&A...520A..93H,2014A&A...568A..84H} provided a clear detection of the transiting plane $b$ and is described in Appendix  \ref{app:fco}. Also, a classical (non-Bayesian) fitting to the transit lightcurve was performed with the {\tt UTM/UFIT} package \citep{deegutm}.  Fits with {\tt UTM/UFIT}, which were also used in some other parts of this work, are described in Appendix \ref{app:utm}. The results from all methods are included in Table~\ref{table:final_param_model}.
}

\subsection{Periodicities in the RV data:  planetary signals or stellar activity?}
\label{sec:specanal}
Beyond the anticipated detection of RV signals from the P=1.06 d transit-candidate found by TESS, the RV data may contain further signals that need a revision about their nature, be they one or more further planet(s) in the system, or from other sources.
{The data acquired with \hn\ provide the most precise measurements (with the exception of data from HIRES, from which however only 12 RV points were acquired) and is it is the dataset with the most consistent coverage by far (see also Table~\ref{table:RVstats}); our analysis will hence concentrate on these data. Tests including other datasets showed in all cases a degradation in the detection of the 1.06 d signal. The data from the other instruments are however used in the evaluation of a potential contamination of the RV signals by the Moon mentioned later, which is described in more detail in Appendix~\ref{lunar_disco}.}

In Fig.~\ref{fig:bglsHNbase} we show Generalized Bayesian Lomb-Scargle periodograms (BGLS) of the HN RVs and of the more common activity indicators from the list in Sect.~\ref{sec:HN}. The BGLS \citep{BGLS2015}\footnote{The figures were generated with the latest version of the code for the BGLS and related plots, available from A. Mortier in \url{https://anneliesmortier.wordpress.com/sbgls/ }}  provides several improvements over the common LS periodograms: It weights the data-points by their errors, it is independent of the setting of the data's zero-point and lastly, it provides a quantifiable probability of the relevance of the periodogram peaks. Fig.~\ref{fig:bglsHNbase} shows also the spectral window function \citep{1987AJ.....93..968R,2010ApJ...722..937D}, whose peaks indicate the likely presence of artefacts due to the temporal distribution of the observations. In the periodogram of the RVs, of prominence is a double peak with the highest probability at 29.4 d, with a slightly lower peak (albeit by a factor of log $p \approx 10$) at 27.4 d (see also Fig.~\ref{fig:bglsHN29d} in the Appendix). Among the activity indicators, only the chromatic RV index (CRX) has peak near $\approx$ 30d, while the window function is rather flat in this region. We note that the period of the higher one of the double-peak is very close to the lunar synodic period of 29.53 days. The Appendix to this paper provides a more detailed evaluation of this signal as a candidate for a second planet $c$.

A further signal is notable at $\approx$ 10 d which corresponds to local maxima of most activity indicators. Hence it is likely due to stellar activity\footnote{Fits to the \hn\ RVs using Gaussian Processes, as described in Sect.~\ref{sec:joint}, were made for models including this 10 d signal as a Keplerian one arising from a further planet, but this led to fits that were significantly worse than those presented in Sect.~\ref{sec:joint}}, albeit at a shorter period than the stellar rotation period of $P_\mathrm{rot}/ \sin i = 20_{-5}^{11} $ d determined from $V \sin i_\star$ and $R_\star$ or the $17.6\pm 2$ d from the lightcurve analysis of Sect.~\ref{sec:rotation}.  
The same goes for an RV peak at 138 d, with several activity indicators showing maxima at a slight larger period of $\approx$ 160d, and which we will not consider further. The periodicity of the transits of 1.07 d does not appear well in the BGLS periodogram, which instead shows a series of peaks around P $\approx$ 1d, with the highest and second highest ones at  P = 1.035 d and P = 0.967d respectively. These are clearly aliases of the 29.4d signal due to a sample period of 1 day (Fig.~\ref{fig:bglsHN1d}), given by the aliasing equation $f_\mathrm{alias} = | f_\mathrm{real} + N f_\mathrm{sample} |$ with $N = \pm 1$, where the $f$ are the frequencies of the alias signal, the real signal and the sample frequency, respectively.  

In a further evaluation, we use the framework provided by the online-tool  {\tt Agatha}\footnote{\url{https://phillippro.shinyapps.io/Agatha/}} \citep{2017MNRAS.470.4794F}. With this tool, in a first step  a model comparison between different models describing the data is performed. In this process, {\tt agatha} evaluates 'MA' (Moving Average) models of varying complexity to describe the RV's red noise. These MA models are simplified Gaussian processes that only account for the correlation between previous data points and the current point, for which models with zero (corresponding to purely white noise), one or more 'MA' components are evaluated. We then used {\tt agatha} to evaluate models  with 0 to 2 MA componentes and also with one or several (or without any) noise-proxies among the activity indicators. 
For the different MA models (with or without the presence of proxies) that were evaluated, {\tt agatha} generated Bayes Factors which account for the varying complexity of the models. In the case of our \hn\ data, a one-component MA model without any noise proxies was indicated as the best model. This model was then also used by {\tt agatha} to generate a Bayes Factor Periodogram \citep[BFP, as defined by ][]{2017MNRAS.470.4794F}), shown in Fig.~\ref{fig:bfpHN}.  
In this periodogram, the highest peak by a wide margin corresponds to the 1.07 d period of the transits. Beyond the peaks around P=1 d and the aforementioned one near 135 days, next highest peak (albeit by a small margin) is again the signal near 29.4 d, identified previously with the BGLS periodograms. 

In a further evaluation, we generated correlations between the various activity indicators and the RVs, following the precepts of \citet{2018AJ....155..126D}, which was based on prior work by \citet{2014A&A...566A..35S}. 
Fig.~\ref{fig:corrHN} shows no strong correlation between any of these indicators and the RVs,  with a notable absence of any correlation between the RVs and the bisector inverse slope (BIS, in Fig.~\ref{fig:corrHN} labelled as  {\tt dsr\textunderscore ccf\textunderscore bis}). The only correlations of mention are the weak ones between the RVs and the differential line-width (dLW) and the H$\alpha$ index, with correlation coefficients of 0.39$\pm$ 0.08 and 0.36$\pm$ 0.08, respectively. 

Considering the significant differences between periodograms generated by different methods \citep[for further examples of strongly differing results among different periodograms, see][Figs. 1, 3 and 5] {2017MNRAS.470.4794F}, none of them should be taken to provide definite results. In any case, these periodograms suggest the presence of planet-like RV signals with periods of 1.07 and 29.4 days. A more detailed evaluation about the 29.4 d signal being caused by the Moon (see the Appendix) is not fully conclusive and a strong chance remains that it is a residual from contamination by Moon light; hence at most it may be treated as a tentative planet. Further modelling of the data concentrates therefore on the short-periodic transiting planet $b$.

\begin{figure}
\centering
	\includegraphics[width=1\linewidth]{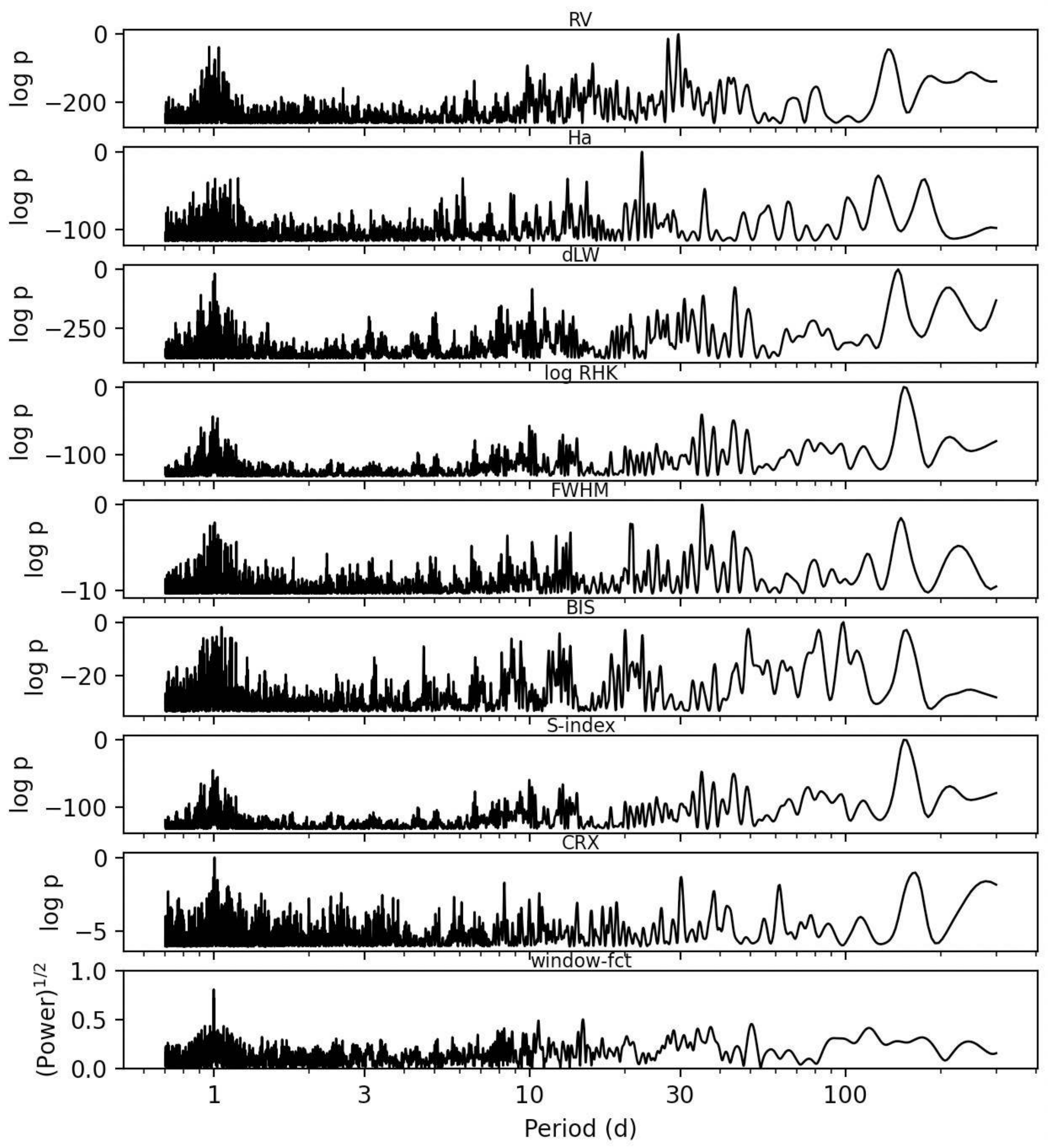}
    \caption{BGLS periodograms of the \hn\ observations, for the measured RVs and for several activity indicators. The vertical scale is given in units of the logarithm of the Bayesian probability of a signal with a given period, where the highest peak is normalised to log p = 0. The lowest panel shows the spectral window function of the sampled data. See also the Appendix, Figs.~\ref{fig:bglsHN1d} and \ref{fig:bglsHN29d} for zoomed-in views around the 1.07d and 29.4d periods of planet $b$ and the candidate $c$, respectively.}   
    \label{fig:bglsHNbase}
\end{figure}

 \begin{figure}
\centering
	\includegraphics[width=1\linewidth]{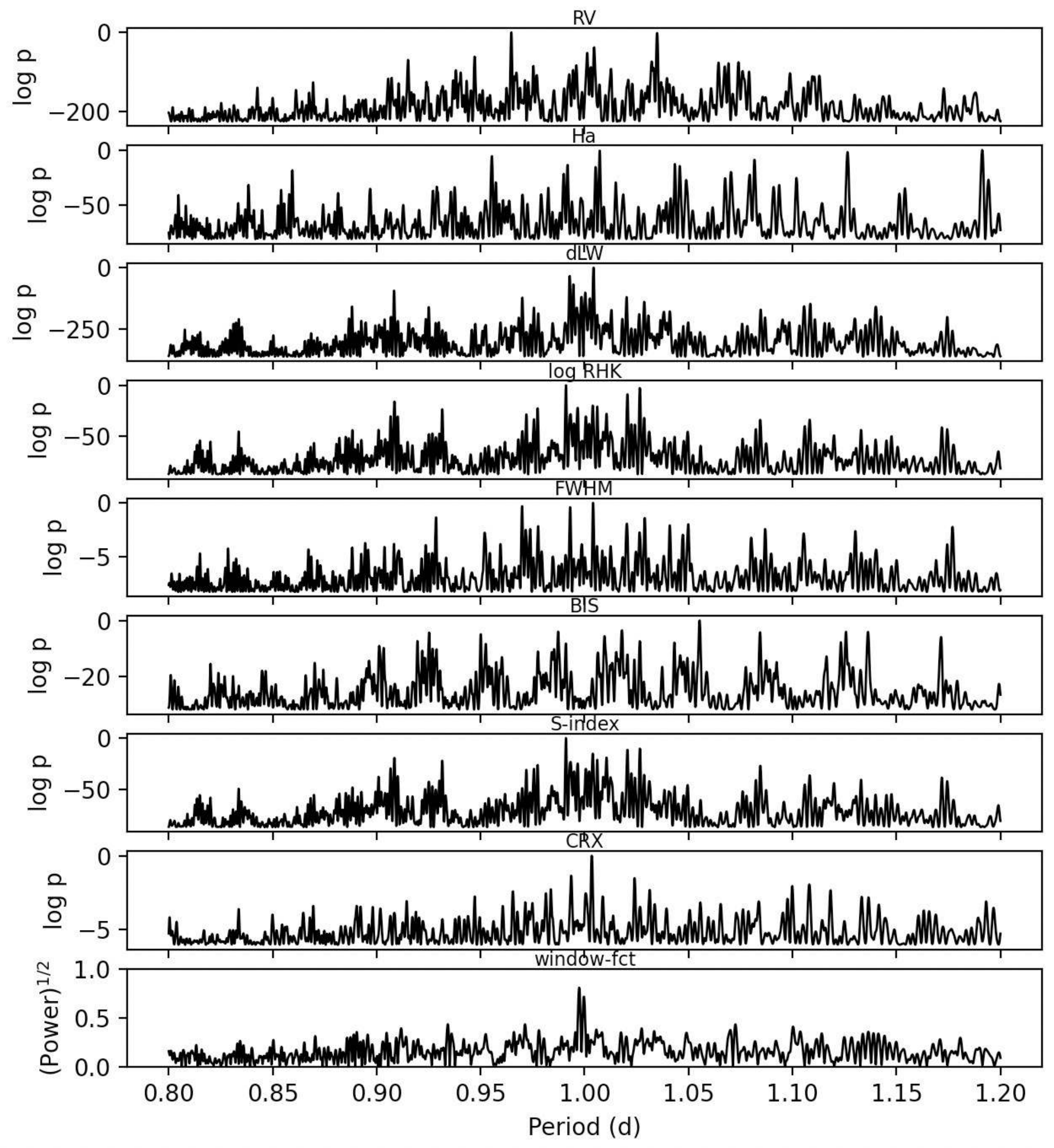}	
    \caption{Zoomed-in view of the BGLS periodogram of Fig.~\ref{fig:bglsHNbase}, around the 1.07 d period of the transiting planet $b$, where only a minor peak is discernible in the RVs (top panel).  The highest RV peaks at P=1.035 d and P=0.967d are aliases of the 29.4d signal over the sample period of the solar or the sidereal day. Their periods of 1.0 resp. 0.9973 days show up as the principal double peak in the window function (lowest panel).}   
    \label{fig:bglsHN1d}
\end{figure}

\begin{figure}
\centering
	\includegraphics[width=1\linewidth]{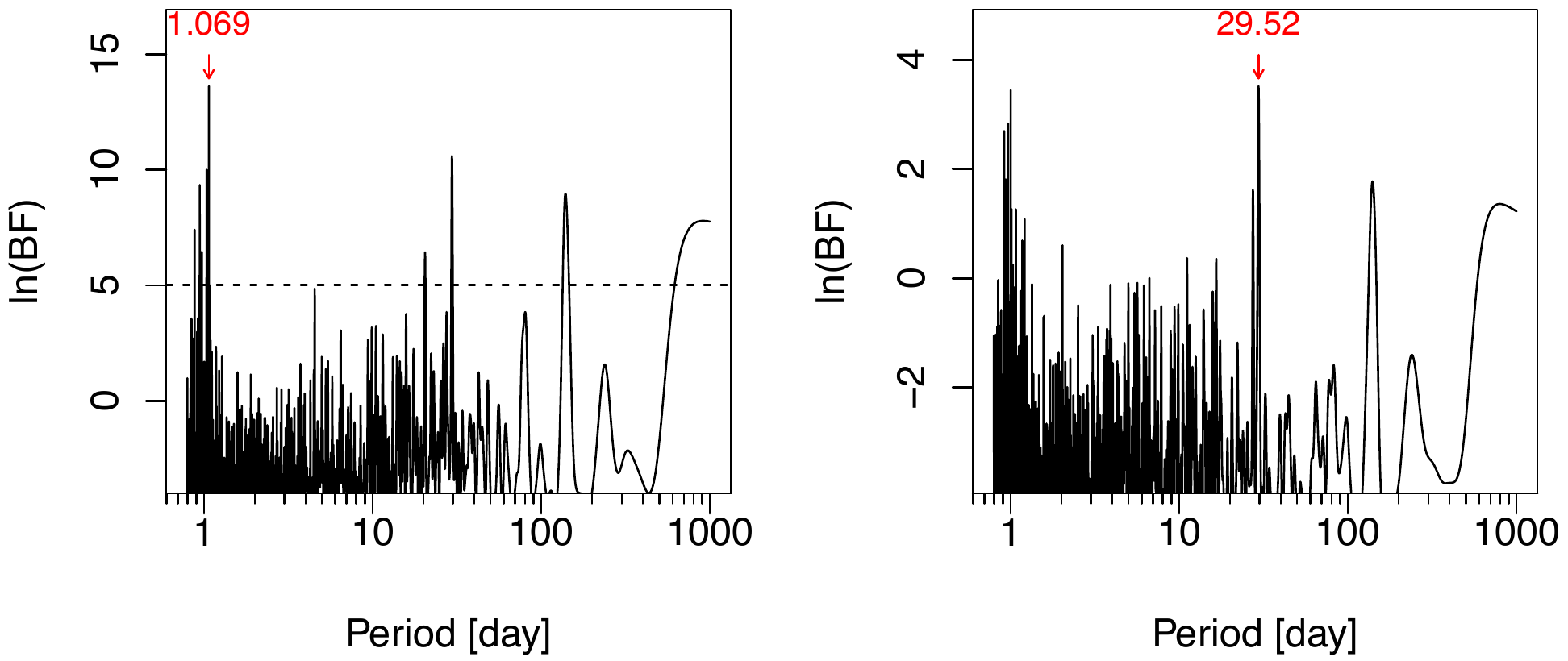}	
    \caption{Left panel: BFP periodogram of the \hn\ radial velocities generated by {\tt agatha} \citep{2017MNRAS.470.4794F}, using one MA component. The vertical axis provides the probability of peaks being real, in terms of the logarithm of their Bayes Factor (BF). The period of the highest peak is indicated, which corresponds to the period of the transits of \toib. Right panel: Like the left panel, after the removal of the 1.069d signal, showing now the signal near 29.5d as the highest one. }
    \label{fig:bfpHN}
\end{figure}

\begin{figure}
\centering
	\includegraphics[width=1\linewidth]{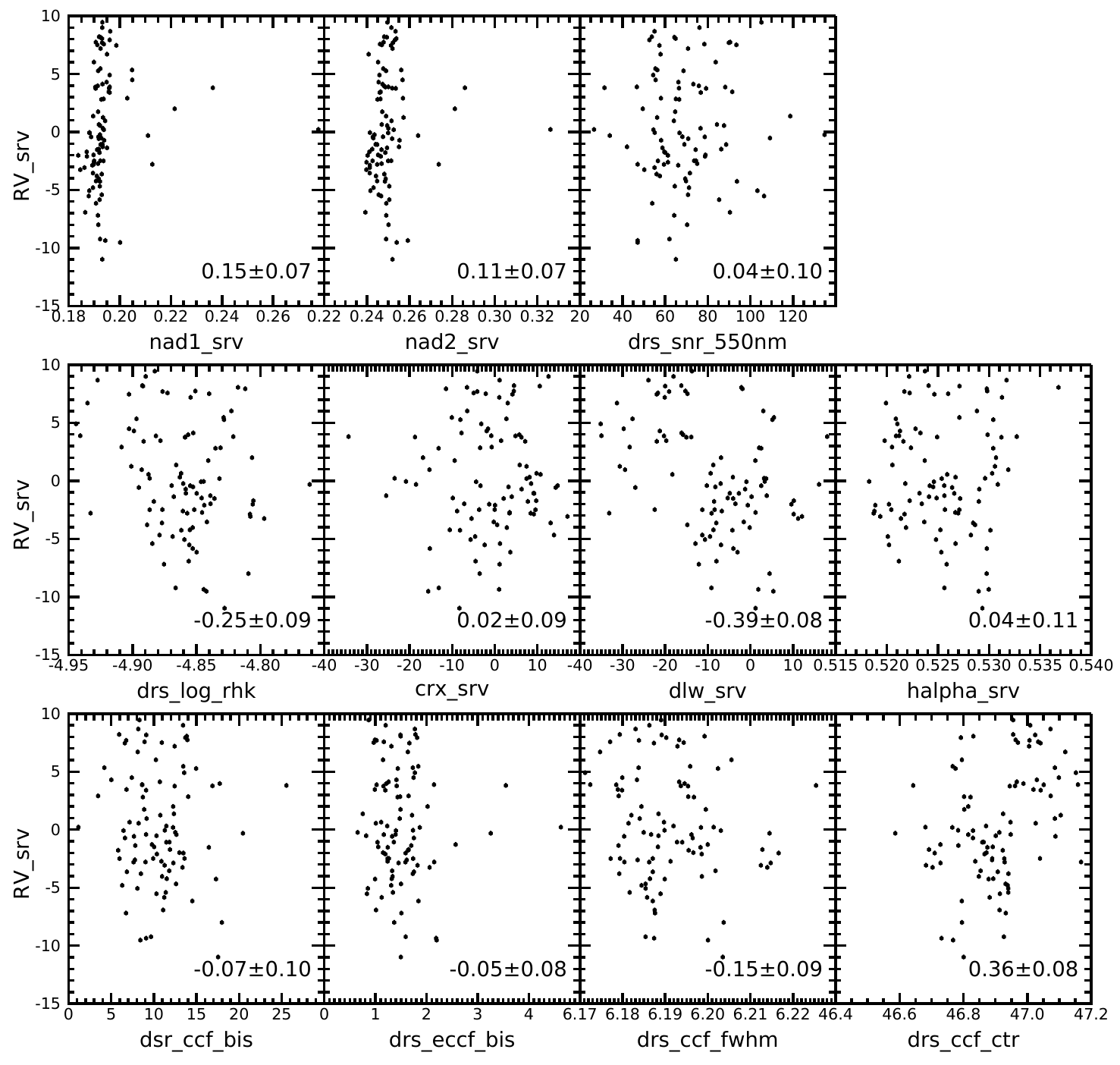}	
    \caption{Correlations in the \hn\ data between the RVs (labelled as {\tt RV\textunderscore srv} and the activity indicators listed in Sect.~\ref{sec:HN}. The Pearson correlation coefficient is indicated in each panel.} 
    \label{fig:corrHN}
\end{figure}


\subsection{Joint RV and lightcurve modelling}
\label{sec:joint}



'Classical' Keplerian RV fits that assume white noise in the jitter of the RV values performed well in fits to the \hn\ RVs from the first observing season, finding a distinct RV amplitude of $\approx$ 2 m\,s$^{-1}$ at the period and epoch of the transits. However, with the addition of RVs from subsequent observing sessions, the quality of these fits degraded substantially, implying the presence of activity and other longer-term variations in the data.

Hence, to allow for the presence of additional signals and especially those arising from stellar activity, we model the spectroscopic data from \hn\ (and jointly also the transit lightcurve) with \pyan, which uses the multi-dimensional Gaussian process (multi-GP) technique as described by \citet{Rajpaul2015}. This approach models the RVs alongside activity indicators, taking advantage of the fact that these indicators should only be coupled  to the RV components that arise from stellar variability. For the case of \target, we use the differential line width (dLW) -- a line shape indicator, and construct a two-dimensional GP model as follows:

\begin{equation}
    \begin{aligned}
    RV_{\rm ac} & = & A_{\rm RV} G(t) & + B_{\rm RV} \dot{G}(t), \\
    dLW & = & A_{\rm dLW} G(t),
    \label{eq:GP}
\end{aligned}
\end{equation}

\noindent where $RV_{\rm ac}$ is the RV component arising from stellar activity; $A_{\rm RV}$, $B_{\rm RV}$, and $A_{\rm dLW}$ are free parameters relating the individual timeseries to the GP-generated function $G(t)$ and its derivative $\dot{G}(t)$. $G(t)$, in turn, can be viewed as a function that describes the projected area of the visible stellar disc as covered by active regions at a given time.
The dLW indicator measures the width of the spectral lines and is mostly affected by the fraction of the visible stellar disc covered by active regions, and is thus represented by $G(t)$. The RVs, on the other hand, are affected by both the location of the active regions, and their temporal evolution. To account for this time dependence thus requires the addition of the first derivative term, $\dot{G}(t)$. 

The multi-GP regression was performed on the \hn\ RVs and dLW using a quasi-periodic (QP) covariance function,

\begin{equation}
    \gamma(t_i,t_j) = \exp 
    \left[
    - \frac{\sin^2[\pi(t_i - t_j)/P_{\rm GP}]}{2 \lambda_{\rm P}^2}
    - \frac{(t_i - t_j)^2}{2\lambda_{\rm e}^2}
    \right],
    \label{eq:qp}
\end{equation}

\noindent and its derivatives, as described in \citep{pyaneti2}.  $P_{\rm GP}$ is the period of the activity signal
, $\lambda_p$ the inverse of the harmonic complexity, i.e. the variability complexity inside each $P_{\rm GP}$, and $\lambda_e$ is the long term evolution timescale, or the lifetime of the active regions.

For the simultaneous transit analysis, we used the TESS lightcurve after being prepared as described in Sect.~\ref{sec:tessphot}. In \pyan, the transits are modelled using the \citet{Mandel2002} algorithm. The parameterisation of the transits is the same one as described in Appendix~\ref{app:utm} for the {\tt UTM/UFIT} fitter; most notably with a sampling of the limb darkening parameters using the  $q_1$ and $q_2$ parametrisation by \citet{Kipping2013} and the stellar density as a fundamental parameter to be fitted. 

Besides the generation of models for both the RVs and the lightcurves, \pyan\ employs a Markov Chain Monte Carlo (MCMC) sampling in a Bayesian framework to calculate posterior distributions of planetary system parameters. Using this setup, we sampled the parameter space with 500 independent Markov chains, out of which we built posterior distributions for each sampled parameter with a thinning factor of 20, using the last 10000 steps of the converged chains. Several planet-system models were then investigated; an overview of them is given in Table~\ref{table:pyanmodel}. In all of these models, parameters that are depending on the TESS light-curve turned up virtually identical and resulted in transit models that are visually indistinguishable from the one plotted in Fig.~\ref{fig:wrap_lc}, and only the parameters depending on the RVs had different outcomes among the models. 

\begin{figure*}
\centering
\begin{subfigure}{1\textwidth}
  \includegraphics[width=1\linewidth]{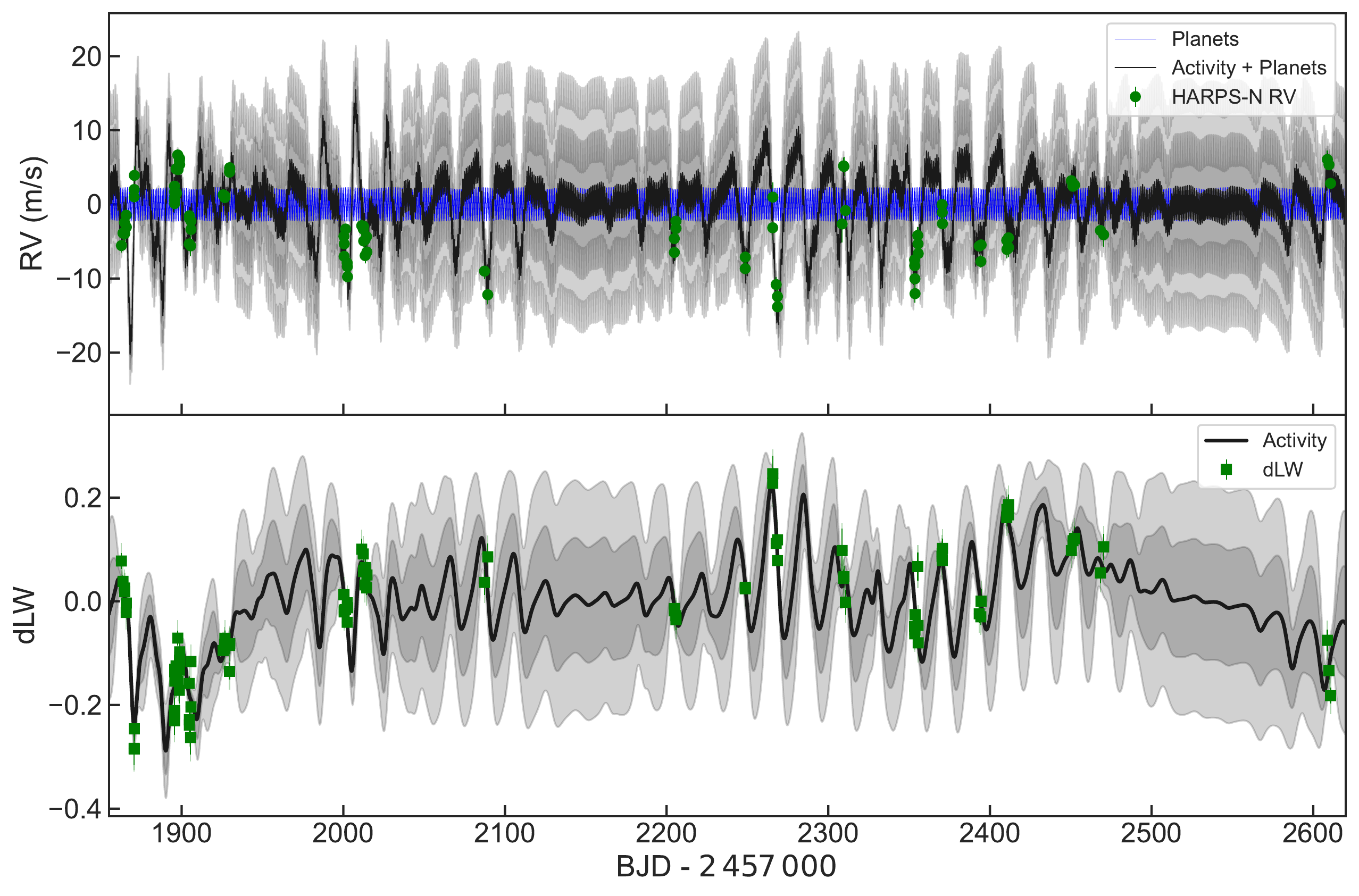} 
  \label{fig:sub-first}
\end{subfigure}
\begin{subfigure}{.6\textwidth}
  \includegraphics[width=\textwidth]{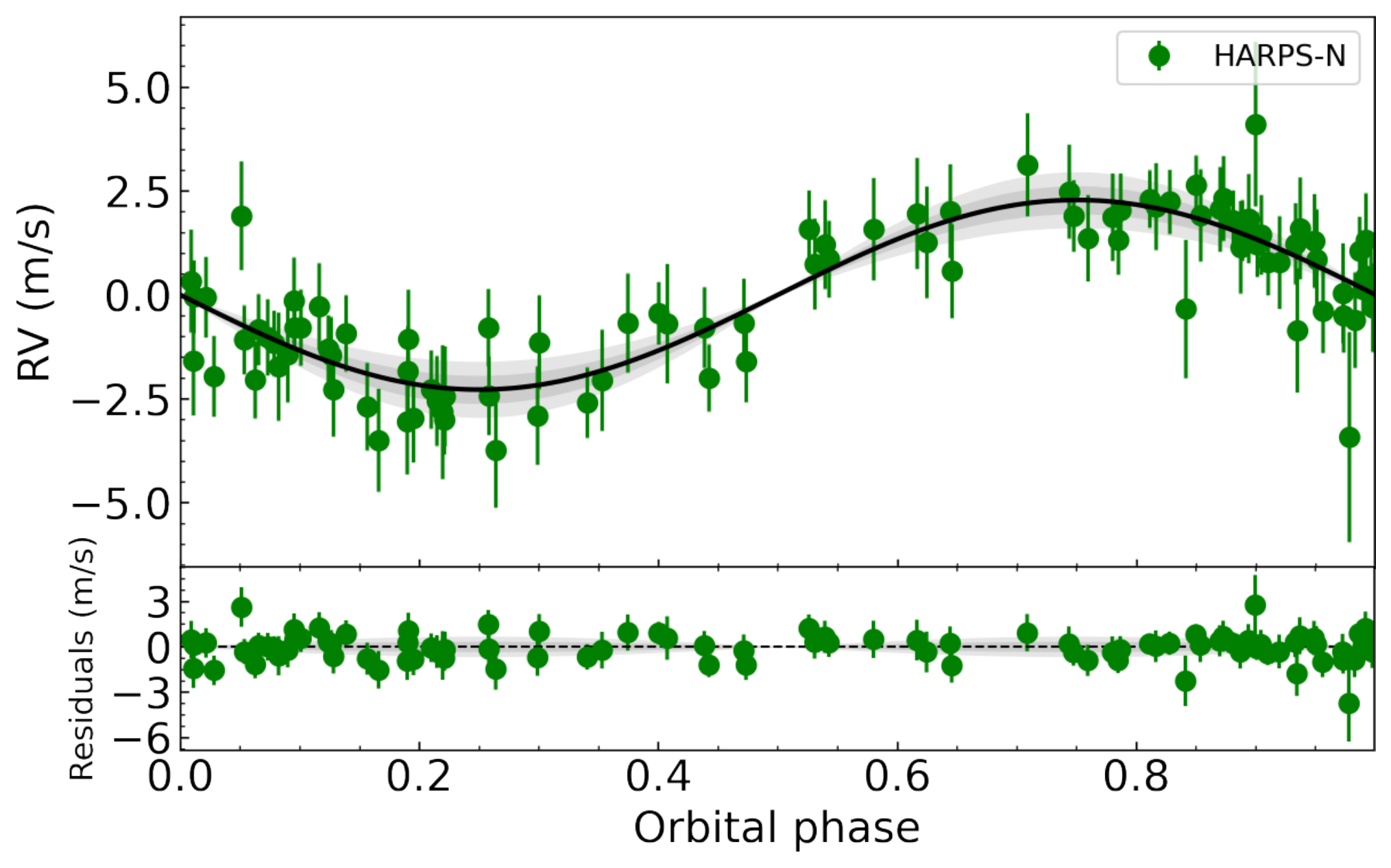}
  \label{fig:sub-second}
\end{subfigure}
\hn\
\caption{Upper panel: \hn\ RV  and differential line width (dLW) time-series for \pyan\ Model 1, assuming only the presence of a Keplerian signal with the 1.06d transit period. The green markers in each panel represent the RV and dLW measurements. The solid black curve shows the inferred multi-GP model, with dark and light shaded areas showing the one and two sigma credible intervals of the corresponding GP model. We note that the short period of the planet and the size of the plot make the RV sinusoids appear as a solid blue band. Lower panel: \hn\ RV data folded on the 1.07 day orbital period of planet $b$, after subtraction of the systemic velocity and the GP noise model. The inferred RV model is shown as a solid black curve with 1- and 2-sigma credible intervals (shaded areas).}
\label{fig:RVGP-1pl}
\end{figure*}

For Model 1, only the transits from TESS and an RV signal with an ephemeris based on the transits were modelled, which yields a clearly detected RV semi-amplitude $K_b$ of \IGkbonepl (see Fig.~\ref{fig:RVGP-1pl}), consistent within $1\sigma$ with an independent determination obtained by the FCO method (see Appendix~\ref{app:fco}). In this model and the following ones, the orbit of planet $b$ is consistent with a circular one ($e_b$ = \eb), which is unsurprising given its very short period. For further work in this paper we are therefore assuming a circular orbit of planet $b$.

For Model 2, we added a Keplerian signal (denoted as $c$) with a period of $\approx$ 29 days to our model, corresponding to the highest peak in the BGLS periodogram (Fig.~\ref{fig:bglsHNbase} and the discussion in Sect.~\ref{sec:specanal}). Using an uniform prior on this signal's period of [28.0~d, 30.0~d], 
the signal $c$ is well-detected, with a semi-amplitude of $\approx$ 5.2 m\,s$^{-1}$. 
Also, the amplitude of the 1.06 d signal increases slightly in Model 2, to  $K_b=$\kbPcuni, still well within the error bars of our previous estimates. Looking at the Bayesian Information Criterion (BIC), we further note that Model 2 has a significant advantage over Model 1, with its BIC being lower by 24, despite the increased complexity (see also Table ~\ref{table:pyanmodel}). While these results are encouraging for the confirmation of the longer period signal $c$ as a genuine planet, the derived period of \Pcuni\ is fully consistent with the lunar synodic period of 29.5306 days (see Appendix \ref{lunar_disco} for further discussion). 


We note that fitting for an eccentricity of signal $c$ in Model 2 yielded a value of $e_c=\ $\ec. However, {the revised Lucy-Sweeney test \citep{2013A&A...551A..47L} indicates this as compatible with the absence of eccentricity, with the value to be replaced by an upper (95\% confidence)  limit of  $e_c < 0.68$. }
Given also the lack of apparent improvement of an eccentric versus a circular model, and the suboptimally sampled phase-coverage (with RVs falling into two groups, see Fig.~\ref{fig:RVGP-2pl}, bottom right), we remain skeptical of the authenticity of a significant eccentricity and zero eccentricity is assumed.  Also, we point out that the GP period cannot be better constrained due to the fact that the lifetime of the active regions, $\lambda_e$, is comparable to the GP period.  

In Model 3, we repeat Model 2 but now the period of signal $c$ is fixed to the lunar synodic period. This leads to a BIC that is $\approx$ 11 lower against model 2, favouring this approach. Irrespective of the nature of the 29.5-day signal, the presence of this signal appears to be genuine, with a semi-amplitude similar to the one from Model 2. The fitting results for Model 3 have no relevant differences to those from Model 2; the corresponding RV and dLW timeseries plots, together with the inferred Keplerian RV models, are found in the Appendix in Figure~\ref{fig:RVGP-2pl}. The priors and fitting results of Model 3 are shown in Table~\ref{table:final_param_model}, and are taken as the adopted values in this work.

Model 4 is similar to Model 2, but assumes a signal with a period of $\approx$ 27.5 days, resulting however in a significantly higher BIC than models 2 or 3. Given however the fact that the 27.4 d signal displayed the strongest peak in the periodogram of RV data from all contributing instruments (Fig.~\ref{fig:BGLSallinstr}) and the potential aliasing between this signal and the 29.5d one (see discussion in Appendix \ref{lunar_disco}), we do not want to discard that an eventual planet $c$ might instead have this period.

Regarding the apparent contradiction in Table~\ref{table:pyanmodel} between Model 3 having the best (lowest) BIC and Model 1 the smallest  {\it rms} of the RV residuals, we note that the {\it rms} indicates only a goodness-of-fit of the model against the RV data, whereas the BIC derived by \pyan\ includes (besides the quality of the transit-fit to the lightcurve, which should be identical in Model 1 -- 4) also several more parameters related to the Gaussian processes, among them the assumed amount of RV jitter and the likelihood of the correlated noise; the {\it rms} and the BIC are therefore not directly comparable.  

In the light of this, we choose a conservative approach and for the further discussion we assume only a tentative planet $c$ with a period near 27.5 or 29.5 d and a mass of M $\sin i$ of  19 to 25 $M_{\oplus}$, whose confirmation as a second planet in \target\ remains pending.

As mentioned in Sect.~\ref{sec:specanal}, there is a significant signal at $\approx$ 10~days evident in the RV data, which is well pronounced in the activity indicators but unlikely to be caused by stellar rotation. We tried modelling it as a Keplerian to investigate the possibility that it may be an additional planet. Our fits, however, were convincingly inferior compared to all of the scenarios discussed thus far in this section. To further exclude it as a potential stellar rotation period, we tested placing a $P_{\rm GP}$ prior using that rotation period of 9.6$\pm$1.4 d. We find that this leads to significant changes in the GP hyperparameters, to the point that their interpretation becomes unphysical, while the detection significance of the $b$ and $c$ signals is practically unchanged. This scenario is also disfavoured with a $\Delta$BIC of $\approx$ 8 against the one it was derived from (Model 3). Lastly, we note that this 10-day signal would be approximately the first harmonic of our favoured $\approx$ 20-day rotation period. This is not surprising given that harmonics often dominate over the true signals. 
A likely explanation for this is the presence of two spotted regions on the stellar surface separated by $\approx$ 180~$\deg$, each thus manifesting at half the rotation period. 


\begin{table}
\centering
\small
\caption{Models evaluated with \pyan} 
\label{table:pyanmodel}
\begin{tabular}{lccc}
\hline\hline
\noalign{\smallskip}
Model&$\Delta$BIC&$\sigma_{RV}$\\
   &  & m\,s$^{-1}$\\
\noalign{\smallskip}
\hline
\noalign{\smallskip}
Model 1& 0 & 0.92\\
planet $b$ only &  &\\
\hline
Model 2& -23.8 & 1.11\\
planet $b$, signal $c$ of P$\approx$29.5 d&  &\\
with free ephemeris &  &\\
\hline
Model 3& -34.5 & 1.11\\
planet $b$,  signal $c$ fixed& &\\
 to P=29.5306 d&  &\\
\hline
Model 4& 5.5 & 1.04\\
planet $b$, signal $c$ of P$\approx$27.4 d & &\\
 with free ephemeris &  &\\

\noalign{\smallskip}         
\hline
\end{tabular}
\tablefoot {$\Delta$BIC indicates the BIC relative to model 1. $\sigma_{RV}$ is the \emph{rms} of the RV residuals  relative to the best-fit models.} 
\end{table}

\begin{table*}
\centering
  \caption{Priors and inferred parameters$^{(a)}$ from transit and RV modelling with \pyan\ (Model 3) and UTM/UFIT resp. FCO. \label{table:final_param_model}}  
  \begin{tabular}{lccc}
  \hline\hline
  Parameter & Prior$^{(\mathrm{b})}$ & \pyan&UFIT / FCO\\
  \hline
  \noalign{\smallskip}
  \multicolumn{4}{l}{\emph{\toib}} \\
    \noalign{\smallskip}
    Orbital period $P_{\mathrm{orb}}$ (days)  &  $\mathcal{U}[1.0690 , 1.0705]$ & \Pb[] & \UPb[]\\
    Transit epoch $T_0$ (BJD$_{\rm{TDB}}$ - 2,450,000)  & $\mathcal{U}[  8739.455 , 8739.466 ]$ & \Tzerob[]  & \UTzerob[]  \\
    Eccentricity $e$  &  $\mathcal{F}[0]$ & 0  &0 \\
    Scaled planetary radius $R_\mathrm{p}/R_{\star}$ &  $\mathcal{U}[0.01 , 0.10]$ & \rrb[] & \Urrb[]  \\
    Impact parameter, $b$ &  $\mathcal{U}[0,1]$  & \bb[] & \Ubb[] \\
    RV semi-amplitude $K$ (m s$^{-1}$) &  $\mathcal{U}[0, 25]$ & \kb[] & \Ukb[] \\
    \hline
\toic \\
    Orbital period $P_{\mathrm{orb}}$ (days)  &  $\mathcal{F}[29.5306]$ & \Pc[] & -- \\
   \noalign{\smallskip}
    Transit epoch $T_0$ (BJD$_{\rm{TDB}}$ - 2,450,000)  & $\mathcal{U}[  8868.00 , 8885.00 ]$ & \Tzeroc[]& --   \\
     Eccentricity $e$  &  $\mathcal{F}[0]$ & 0 & --  \\
    RV semi-amplitude $K$ (m s$^{-1}$) &  $\mathcal{U}[0, 25]$ & \kc[]& --  \\
    \noalign{\smallskip}
    \hline
    \noalign{\smallskip}
    GP Period $P_{\rm GP}$ (days) &  $\mathcal{U}[15,28]$ & \jPGP[]& --  \\
    $\lambda_{\rm P}$ &  $\mathcal{U}[0.1,5]$ &  \jlambdap[] & -- \\
    $\lambda_{\rm e}$ (days) &  $\mathcal{U}[1,200]$ &  \jlambdae[] & -- \\
    $A_{RV}$ (\mps)  &  $\mathcal{U}[0,100]$ & \jArvc & -- \\
    $B_{RV}$ (\mps) &  $\mathcal{U}[0,1000]$ & \jArvr & -- \\
    $A_{dLW}$ (${\rm 100\ m^2\,s^{-2}}$) &  $\mathcal{U}[0,1]$ & \jAdlw & -- \\
    Offset \hn$^{(c)}$ (\mps) & $\mathcal{U}[ -511 , 509 ]$ & \HN[] & \UHN[] \\
    Offset dLW (${\rm m^2\,s^{-2}}$) & $\mathcal{U}[-0.5351 , 0.5180 ]$ & \dLW[] & --  \\
    Jitter term $\sigma_{\rm HARPS-N}$ (\mps) & $\mathcal{J}[1,1000]$ & \jHN[] & -- \\
    Jitter term $\sigma_{\rm dLW}$ (${\rm 100\ m^2\,s^{-2}}$) & $\mathcal{J}[1,1000]$ & \jdLW[]& --  \\
    Limb darkening $q_1$ & $\mathcal{G}[0.413,0.091]$ & \qone& \Uqone \\
    Limb darkening $q_2$ & $\mathcal{G}[0.354,0.030]$ & \qtwo& \Uqtwo \\
    Jitter term $\sigma_{\tess}$ ($\times 10^{-6}$) & $\mathcal{U}[0,1 \times10^{3}]$ & \jtr & -- \\
    Stellar density $\rho_{\star}$ (${\rm g\,cm^{-3}}$) &   $\mathcal{G} [2.21, 0.27]$ & \denstrb[]& \Udenstrb[] \\
        \noalign{\smallskip}
  \end{tabular}
    \begin{tablenotes}\footnotesize
  \item \emph{Notes} --   $^{(\mathrm{a})}$  Inferred parameters and errors are defined as the median and 68.3\% credible interval of the posterior distribution. $^{(\mathrm{b})}$ $\mathcal{U}[a,b]$ refers to uniform priors between $a$ and $b$ (only for \pyan; for UFIT or the FCO method, no priors were set except on the impact parameter $b$); $\mathcal{G}[a,b]$ to a Gaussian prior centered on $a$ with a $1 \sigma$ width of $b$; $\mathcal{J}[a,b]$ to modified Jeffrey's priors calculated using eq. 16 in \citet[]{Gregory2005}; $\mathcal{F}[a]$ to parameters that are fixed to $a$. 
   $^{(c)}$ Offset against the zero-averaged \hn\ RVs from {\tt serval} (column {\tt rvs\_srv} in electronic data).
\end{tablenotes}
\end{table*}

    \begin{table*}
\centering
  \caption{Adopted derived parameters \label{table:final_param_derived}}  
  \begin{tabular}{lcc}
  \hline\hline

    \noalign{\smallskip}
    Parameter & \toib& \toic \\
    \hline
  \noalign{\smallskip}

    Planet mass ($M_{\oplus}$)  & \mpb[] & \mpc[] ($M\sin i$)\\
    Planet radius ($R_{\oplus}$)  & \rpb[] & -- \\
    Planet density (${\rm g\,cm^{-3}}$) & \denpb[] & -- \\
    Scaled semi-major axis $a/R_\star$ & 5.14$\pm$0.24 & 47.0$\pm$2.4 \\
    Semi-major axis $a$ (AU)  & 0.0190$\pm$0.0003 & 0.1734$\pm$0.0030 \\
    Orbital inclination $i$ (deg)  & \ib[] & < 88.7 \\
    Transit duration $t_{\rm tot}$ (hours) & \ttotb[] & -- \\
    Equilibrium temperature $^{(\mathrm{a})}$ $T_{\rm eq}$ (K) & \Teqb[] & 510 $\pm$ 20 \\
    Insolation $S/S_{\oplus}$   & \insolationb[] & 11.2 $\pm$ 1.3 \\
    Planet surface gravity (cm\,s$^{-2}$)  & \grapparsb[] & -- \\
    \noalign{\smallskip}
    \hline
   \noalign{\smallskip}
  \end{tabular}
    \begin{tablenotes}\footnotesize
  \item \emph{Note} -- Adopted stellar parameters from Tables~\ref{table: spectroscopic parameters} and \ref{table: comparison stellar parameters} were used for values that are dependent on them. {  $^{(\mathrm{a})}$ Assuming an albedo of 0 and uniform heat redistribution over the entire surface. See also Sect~\ref{sec:secondary}.}
\end{tablenotes}
\end{table*}


\subsection{Limits to secondary eclipses}
\label{sec:secondary}
Here, we first estimate the maximum secondary eclipse depth of planet $b$ that can be expected, and then revise their presence in the data. The depth of a planet's eclipse behind its host-star is given by the brightness of the planet relative to the star, with the planet's brightness being the sum of its emitted thermal emission and the amount of stellar light that is reflected from the planet. Regarding thermal emission, Table~\ref{table:final_param_derived} indicates an equilibrium temperature of 1517K for planet $b$, which was calculated for a zero Bond albedo and assuming a uniform heat redistribution over its entire sphere (corresponding to a heat recirculation efficiency of $f=1/4$, e.g. \citet{2011ApJ...729...54C}). For the estimation of the maximum secondary eclipse depth from thermal emission, we assume however a realistic maximum temperature of 1900K, which is based on the assumption that with none of the absorbed radiation gets circulated to the planet's night-side (corresponding to a value of $f=2/3$).  Based on that temperature, and using again the adopted parameters from Table~\ref{table:final_param_derived}, we find that thermal emission from planet $b$ may generate eclipses with a depth of only 1.2 ppm in the wavelengths of the TESS bandpass. For a maximum value of secondary eclipse depth from reflected light, a geometric albedo of 1 is assumed, which leads to an eclipse depth of 14 ppm.

Combining thermal and reflected light, we conclude that secondary eclipses of \pname\ may not exceed a depth of 15 ppm. This value might barely be detectable in the lightcurve. For its detection, we assume that the secondary eclipse is well centred on an orbital phase of 0.5, and generated a phase-folded lightcurve similar to the one that was prepared for the transit-fits with {\tt UTM/UFIT} in Sect.~\ref{sec:tessphot} and shown in Fig.~\ref{fig:wrap_lc}, with off-eclipse fluxes that are normalised to 1, but now centred at phase 0.5.  The fluxes within the expected phase-range of total eclipse (phases from 0.48 to 0.52) were then obtained, which resulted in a flux that is 30$\pm$25 ppm higher than the off-eclipse flux. Hence, a secondary eclipse was not detected, and we may estimate that secondary eclipses deeper than $\approx$ 20 ppm can be excluded with a high (2-sigma) confidence from the observed data.

\section{Results and their interpretation}
\label{sec:results}
\emph{Final system parameters: }
As the two sets of analysis performed with \pyan\  and {\tt UTM/UFIT} showed, no relevant differences arose in those parameters that arose the TESS lightcurves, with \pyan\ employing Gaussian Processes and {\tt UTM/UFIT} a white-noise model on a lightcurve that had undergone a prior filtering against signals that were significantly longer than the transit-duration. The same goes for the RV fit to \pname, where the FCO method -- which is essentially a pass-through filter at the planet's period -- and \pyan\ obtained a very similar result. This outcome is similar to one on TOI-1235 b, where \citet{2020A&A...639A.132B} adopted a white-noise-only fit to the TESS lightcurves, after finding no relevant difference to results obtained from fits based on Gaussian processes. 
For the finally adopted values in Table~\ref{table:final_param_derived}, we quote however those from \pyan, as only this procedure produced an integral analysis of the combined set of lightcurves and RVs that was also suitable to evaluate the various models involving a signal from a potential further planet $c$. This planet remains however tentative due to strong doubts that its signal might arise for contamination from the Moon. Furthermore, with the current data we are not able to ascertain if the tentative planet's period would be 29.5 or 27.4 days. Such a second planet with a large period ratio of $\approx 26$ against the inner planet would however not be unexpected; the preference for USPs for companions with relatively large period-ratios has been known since the first description of USPs \citep{2014ApJ...787...47S,2018NewAR..83...37W}. From the absence of transits of $c$, a maximum orbital inclination of 88.7$^{\circ}$ can be determined. \citet{2018ApJ...864L..38D} find that in USPs with a further transiting planet, the systems with the largest period-ratio also tend to have larger mutual inclinations of $\gtrsim7^{\circ}$. However, the \target\ system is inconclusive in that respect: With \pname's inclination of 85.7$^{\circ}$, even a fully coplanar planet $c$ would not have caused any transits and no conclusions about the system's mutual inclination, or about limits to it, can be made. The RV fits for an eventual planet $c$ were compatible with eccentricities up to 0.5, which upon the availability of more reliable RV results might lead to the establishment of a formation pathway for \pname.

\begin{figure}
\centering
	\includegraphics[width=1\linewidth]{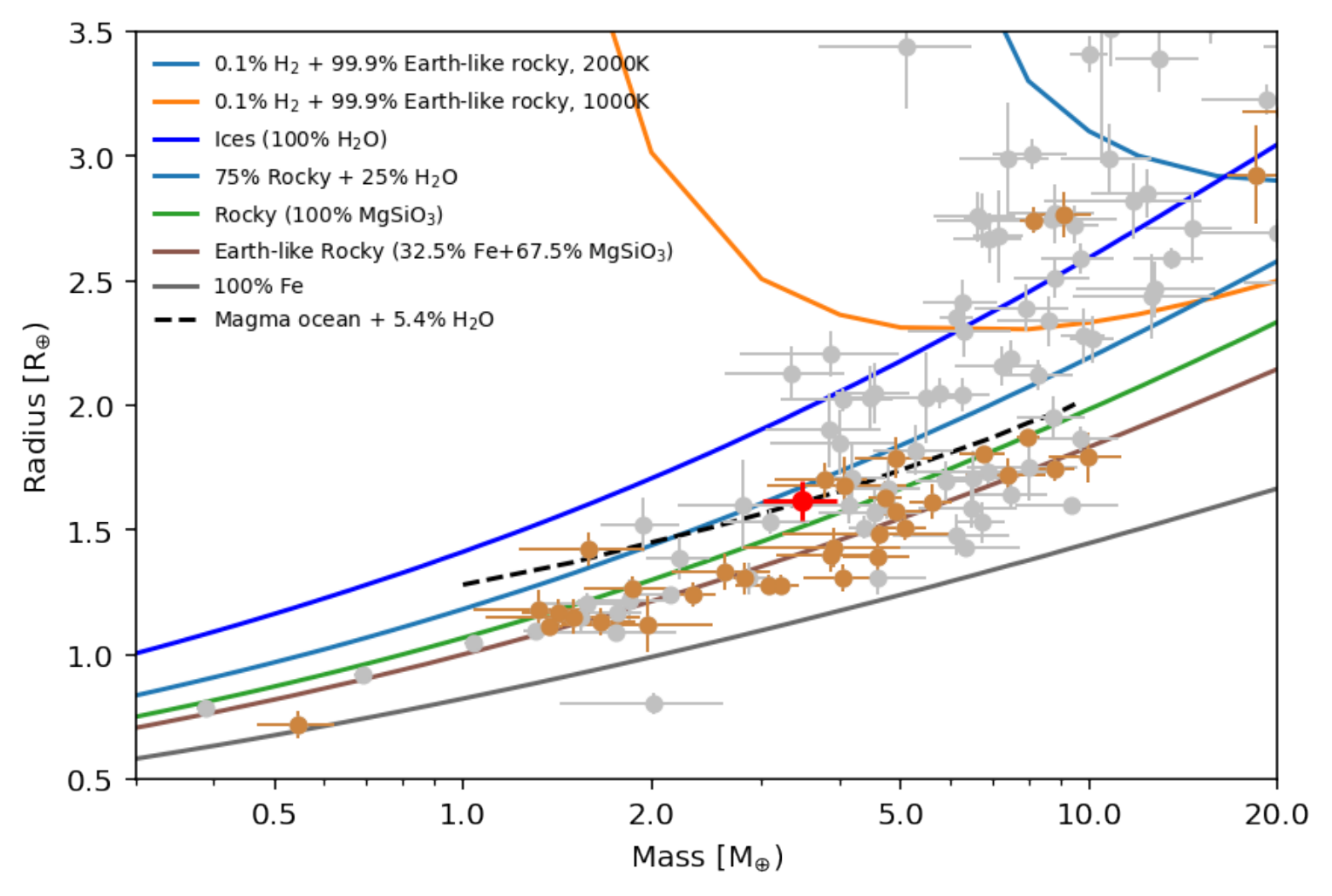}	
   \caption{Planet masses and radii, versus composition models: Grey markers: planets with well-determined masses (errors smaller than 30\%, from adopted values in the NASA Exoplanet Archive). Planets with periods smaller than 2d are shown with brown markers. Composition models indicated by solid lines are from \citet{2016ApJ...819..127Z,Zeng2019}, whereas the dashed line is a model from \citet{2021ApJ...922L...4D} for an Earth-like rocky composition (66\% Mg-Si oxides and silicates and 33\% iron), where the molten rock contains a water mass fraction of 5.4\%. \pname\ is indicated by the red dot.}    \label{fig:MRmod}
\end{figure}

\begin{figure}
\centering
	\includegraphics[width=1\linewidth]{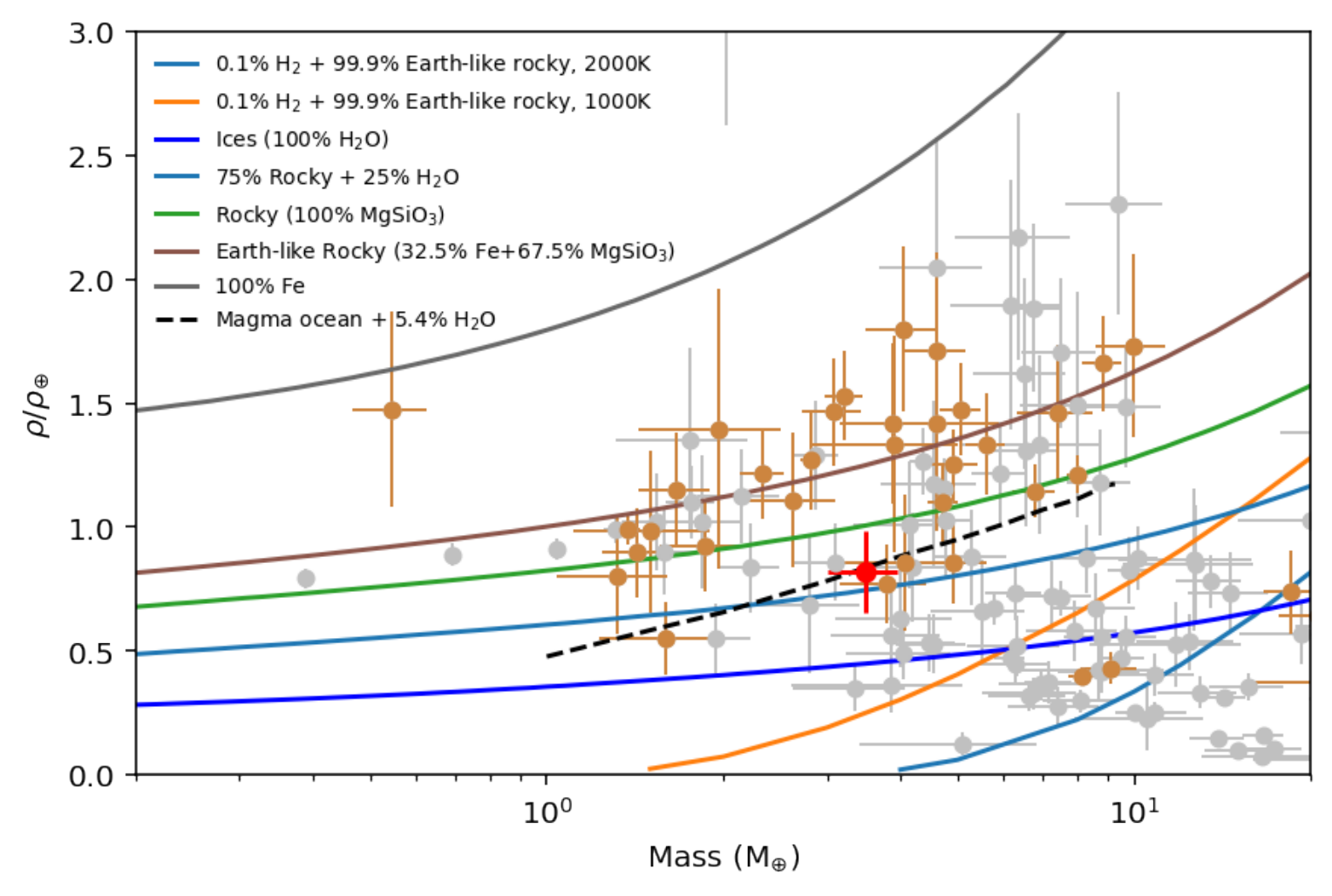}	
   \caption{Like Fig.~\ref{fig:MRmod}, but in mass - density space.}    \label{fig:Mdens}
\end{figure}

\emph{Composition of \pname: } 
For the transiting planet \pname, its radius of  \rpb[] $R_{\oplus}$ and mass of \mpb[] $M_{\oplus}$ indicate that it is a short-periodic Super-Earth like planet, with a density of  \denpb[] ${\rm g\,cm^{-3}}$. Fig.~\ref{fig:MRmod} shows a mass-radius (MR) diagram with several composition models from \citet{2016ApJ...819..127Z,Zeng2019}
, while Fig.~\ref{fig:Mdens} shows the same in mass-density space; We note that \pname\ is above the line for a purely rocky (100\% Mg Si O$_3$) composition, with a density relative to an Earth-like composition (scaled to the mass of \pname) of $\rho\rho_{\oplus,S} \approx 0.67 $. This separates \pname\  from most other short-period planets; \citet{2019ApJ...883...79D} found for a sample of comparable Hot Earths (11 planets with insolations > 650 times that of the Earth  and periods of $\lesssim$2 days)  that most of these are consistent with an Earth Like composition of 30\% Fe - 70\% Mg Si O$_3$.   
We also use the HARDCORE tool \citep{2018MNRAS.476.2613S} which is exploiting boundary conditions to bracket a planet's minimum and maximum core radius fraction (CRF), assuming a fully differentiated planet and iron to be the core material. For \pname\ we obtain a  marginal (most likely) CRF of $0.35\pm0.20$. Similar to the planet's density, this is slightly less but within the limits of the Earth's CRF of 0.55, whereas the potential minimum and maximum values of the CRF are zero and 0.71, respectively. Following \citet{2017ApJ...837..164Z}, we may also derive the core mass fraction (CMF) from the approximation  CMF $\approx$ CRF$^2$, leading to a value of CMF = $0.12^{+0.18}_{-0.10}$.  This value is again relatively small in comparison to the sample of Hot Earths by  \citeauthor{2019ApJ...883...79D}, who determined for them a mean CMF of 26\% with a standard deviation of 23\%.

 We also determine the planet's restricted Jeans escape parameter, given by $\Lambda = \frac{G M_p m_H }{k_B T_{eq} R_p}$, where  $T_{eq}$ is the planets' equilibrium temperature, $m_H$ the mass of the hydrogen atom, $G$ the gravitational constant, and $k_B$ the Boltzmann constant \citep{2017A&A...598A..90F}. The parameter $\Lambda$ is a global one for a given planet, without dependence on altitude within the atmosphere, for which Fossati et al. find from empirical study a critical value of $\Lambda_T = 15-35$, below which a planet's atmosphere is  unstable against evaporation, by lying in a boil-off regime that would shrink its radius within a few hundreds of My. For \pname, $\Lambda = 10.7$; it is hence unlikely to have retained a hydrogen-dominated atmosphere that could contribute significantly to its mass or radius. For highly irradiated planets, the evaporation of hydrogen might however lead to an enrichment of other light elements, be it Helium, or Oxygen from the thermolysis of H$_2$O. For these elements, the hydrogen mass $m_H$ in the equation above can be replaced with the element's atomic mass . For \pname, we then obtain values of $\Lambda \approx$ 40 and 160 for Helium and Oxygen , respectively, meaning that these elements are not affected by evaporation.
 
With \pname\ having at most a small core and a density that is less than a composition exclusively of silicates would require, but orbiting also too close to the central star to enable the retention of  a significant H - He atmosphere, the most likely outcome is the presence of a significant mass-fraction of H$_2$O or other volatiles.
Under this assumption, several types of planet compositions have been brought forward: For one, the original and widely discussed models of rocky cores of various fractions of iron and silicates, with mantles of condensed water, \citep[e.g.][]{2007ApJ...669.1279S,2012A&A...547A.112M,2013PASP..125..227Z,2016ApJ...819..127Z}. 
 For planets that are more irradiated than the runaway greenhouse irradiation limit of $\approx 1.1 S_{\oplus}$,  \citet{2020A&A...638A..41T} provide mass-radius models of silicate cores with a mantle of various fractions of H$_2$O  the form of steam, which lead to larger planet sizes for a given mass-fraction of H$_2$O than in the condensed-water models. The work by \citeauthor{2020A&A...638A..41T} 
 provides a procedure to generate MR relations of steam planets for insolations from $\approx 1$ to $30\,S_{\oplus}$. An extension of this work to highly irradiated planets, like \pname\ with $880\, S/S_{\oplus}$ is still pending, and the feasibility of a steam atmospheres at the insolation resp. temperature  of \pname\ would have to be evaluated. 
 
 With its equilibrium temperature of 1517$\pm$39 K, \pname\ is likely to consist of molten rock (magma) at or closely below the surface. We also note that tidal heating might have contributed a significant further source of internal heating that is potentially capable of melting a USPs entire interior \citep{2021A&A...653A.112L}. In any case, magma has recently been shown \citep{2021ApJ...922L...4D} to be able to absorb significant quantities of H$_2$O, which may lead to radius-increments of up to 16\% over the common interior compositions that do not take dissolved water into account. In Figs.~\ref{fig:MRmod} and Fig.~\ref{fig:Mdens} we include the MR relation from \citeauthor{2021ApJ...922L...4D} for their favoured 'wet-melt'  interior (their 'model C'), which assumes the dissolution of water in an Earth-like magma, with various water mass-fractions. This model provides a close agreement with the mass and radius of \pname, and hence provides the interpretation of the composition of \pname\ that we favour in this work: A planet of partially solid and molten interior of Earth-like composition, with water being distributed between the mantle belt and a a surface steam layer, with a total water mass-fraction of 1-15\% of water\footnote{Water mass-fractions derived from interpolation within Fig. 4 of \citet{2021ApJ...922L...4D}, considering the mass and radius uncertainties of \pname} in the melt. A more detailed modelling of \pname's composition is beyond the scope of our present work and would have to take into account the potential range in values of the CMF, and hence in the fraction between iron and silicates. Potential outcomes could be a relatively small core, with the average density of \pname\ dominated by silicates, or a larger core, that requires a then a larger contribution of H$_2$O to offset the high density of iron.




\emph{Suitability for atmospheric characterisation:} The suitability of a target for its atmospheric characterisation by transmission spectroscopy during a transit has been parametrised by \citet{2018PASP..130k4401K} with the  transmission spectroscopy metric (TSM).  The TSM of \planet\ is 83, so it could be a suitable target for such observations with the JWST\footnote{\citet{2018PASP..130k4401K} give a suggested cutoff of 92 in their Table 1, but we note that \planet's radius of $1.6 R_\oplus$ is near the lower limit of their  $1.5 < R_p < 2.75 R_\oplus$ radius-bin.}. We also note that its emission spectroscopic metric  (ESM) is 13.8, which is well above the threshold of 7.5 that  \citeauthor{2018PASP..130k4401K} recommend for the top atmospheric characterisation targets for JWST follow-up, albeit for a sample of slightly smaller planets with $R_p < 1.5 R_\oplus$. Neither the TSM nor the ESM consider the orbital period, with the TSM relating to the S/N from observing a single transit. Hence USP planets have the further advantage that more transits or orbital revolutions can be acquired in a given time-span. In conclusion, \planet\ might be a very suitable target for JWST follow-up.

 \begin{figure}
\centering
	\includegraphics[width=1\linewidth]{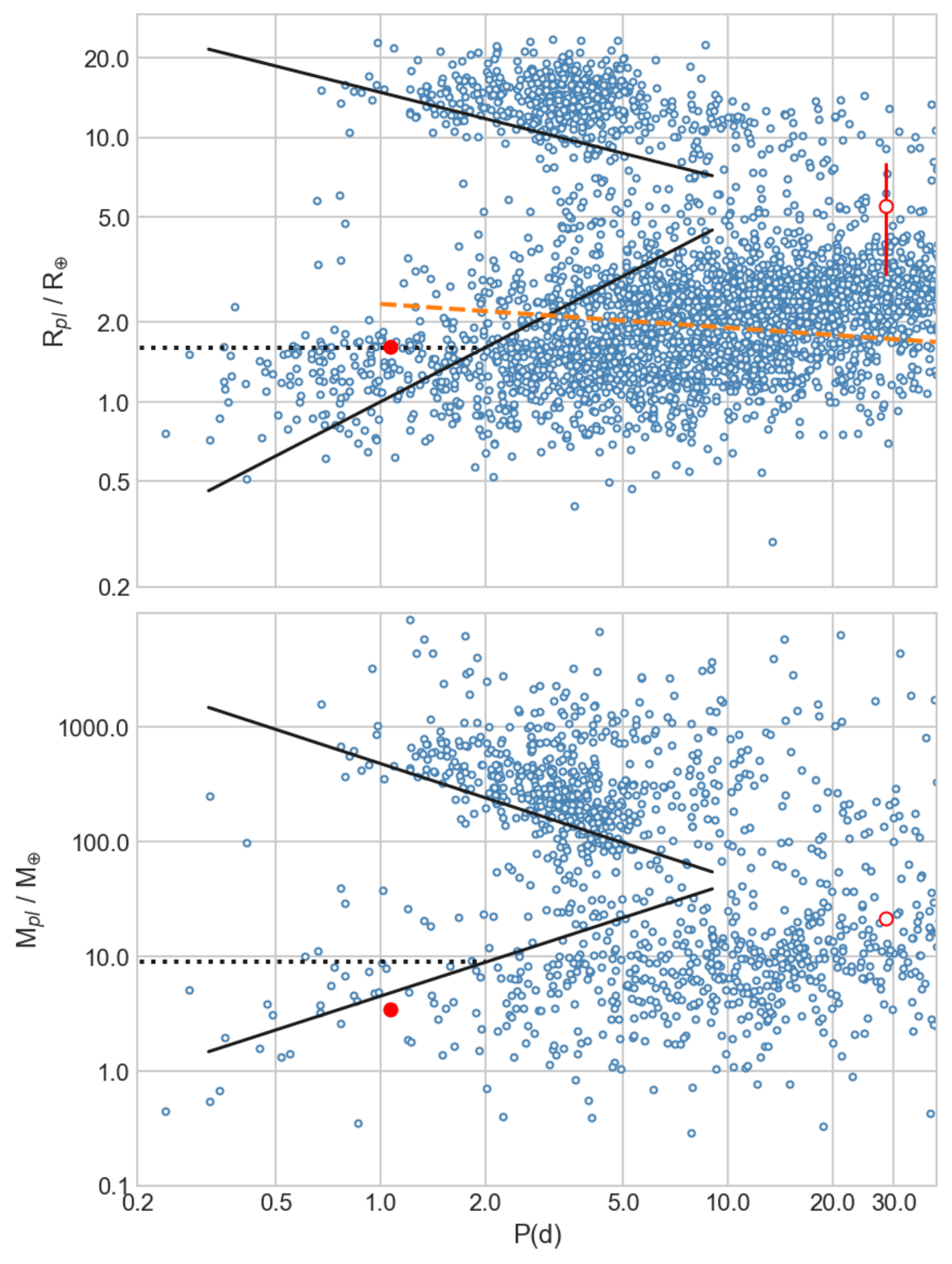}	
   \caption{Diagram of the radii (top panel) and. masses (bottom panel) versus period of the known planets, from the NASA Exoplanet Archive. The solid black lines show the delineation of the 'Neptune Desert' from \citet{2016A&A...589A..75M}, whereas the horizontal dotted black lines show the lower limits to the Neptune Desert for periods $\leq 2$ days that are proposed in this work. The dashed orange line in the upper panel indicates the period-radius valley from \citet{2018MNRAS.479.4786V}. \pname\ is indicated by the filled red circle and the tentative planet $c$ by the unfilled one.}   \label{fig:radiusperiod}
\end{figure}

 \begin{figure}
\centering
	\includegraphics[width=1\linewidth]{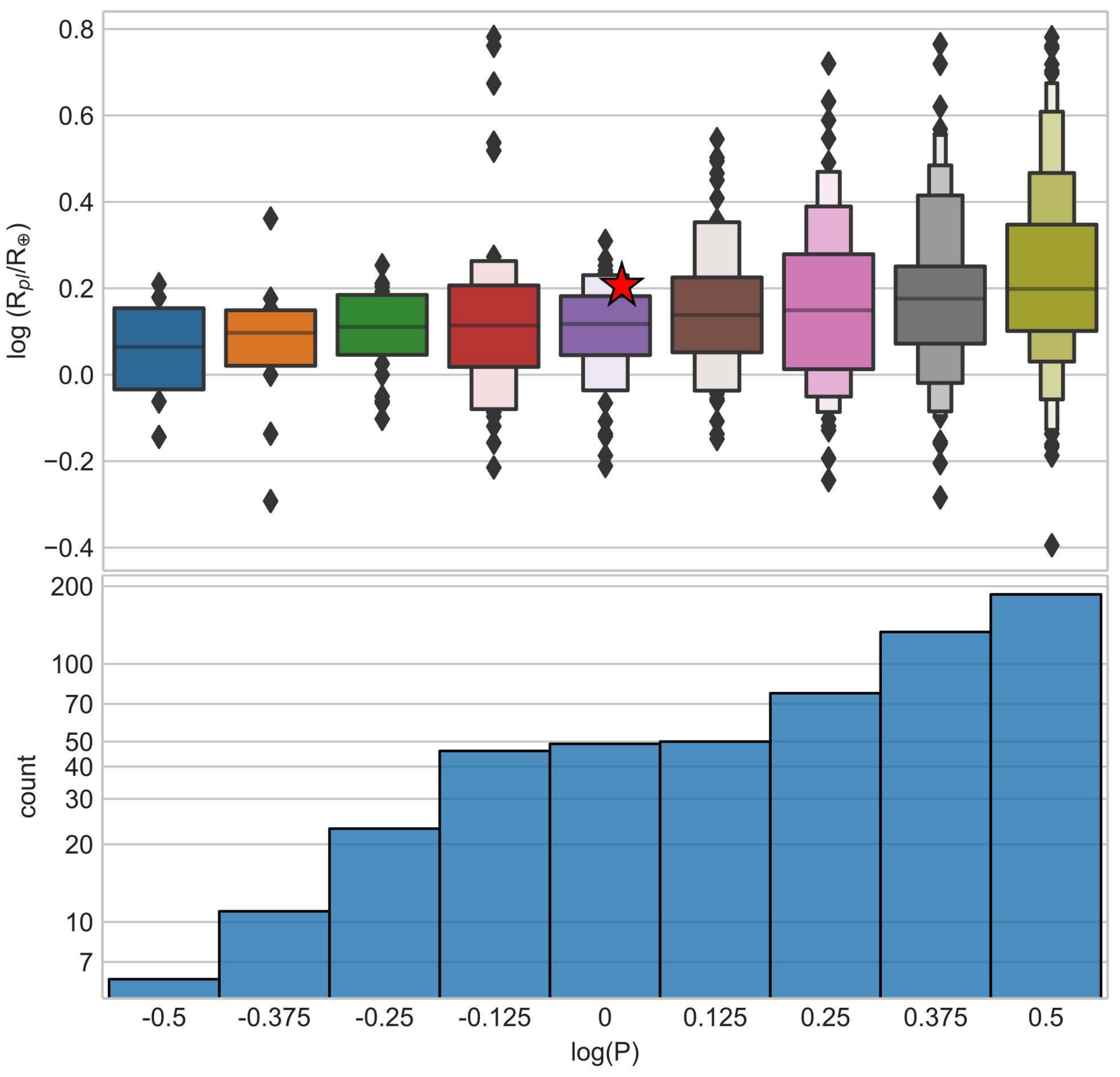}	
      \caption{Top panel: Distribution of planet-radii for small planets of $\log R/R_\oplus  < 0.8$ (resp. $R/R_{\oplus} < 6.3$), versus the orbital period $\log P $(day), after categorising the planet population into bins with a width of $\log P= 0.125$ and for periods shorter than 3.6 d. The distributions are shown as 'Boxenplots' or 'Letter Value Plots' \citep[][]{letter-value-plot}. \pname\ is indicated by the red star. {Bottom panel: Counts of the small planets versus the same bins in orbital period.} } 
   
      \label{fig:radiusdist}
\end{figure}

\emph{Position of \pname\ and $c$ relative to the radius valley: }
Planet $b$ is located slightly below (Fig.~\ref{fig:radiusperiod}, top panel) the mass-radius valley (also known as radius gap or Fulton gap) near radii of 2 $R_{\oplus}$\citep{2017AJ....154..109F, 2018MNRAS.479.4786V, 2022AJ....163..179P} that separates the population of Super Earth planets from the larger Sub-Neptune-like planets. At the short orbital period of \pname\ however, the valley is only poorly defined and only a population of smaller planets remains; see also Fig.~\ref{fig:radiusdist}. On the other hand, for the tentative planet $c$, with its mass of $M \sin i \approx 22 M_{\oplus}$, we estimate a radius of $5.5\pm2.5 R_{\oplus}$ from the radius-mass relation by \citet{2017ApJ...834...17C}. This indicates a Neptune-like planet which would lie well above the mass-radius valley and would convert \target\ into a system with a USP planet below the radius valley and a second planet that is above it. Of course, we do not know the size of the tentative planet $c$, but a radius that would place it below the radius valley would have to be smaller than $\approx 1.7 R_{\oplus}$. Such a small radius is unrealistic from both the observed radius-mass relation and from the required densities in excess of 20 $\rm g\,cm^{-3}$; hence this outcome can be excluded with near-certainty.

\emph{The Neptune Desert and its borders:}
In the planet radius and planet mass versus period diagrams (Fig.~\ref{fig:radiusperiod}), we note the well-known 'Neptune Desert' as defined by \citet{2016A&A...589A..75M}, with the lower boundary for the planet radius given by $\log R_{lo}/R_{\oplus} = 0.68 \log P$, with the period $P$ in days, and the lower boundary in the mass-period planet given by $\log M_{lo}/M_\mathrm{jup} = 0.98 * (\log P) - 1.85$.  
However, given the mass and period distributions in Fig.~\ref{fig:radiusperiod}, which contain many recently discovered planets with periods $\lesssim$ 1 day, we doubt the validity of the lower boundaries for periods shorter than $\approx$ 2 days, because most of the known USP's, including \pname, would be within the 'desert'. Indeed, only a few years ago the period regime below 1 to 2 days was only sparsely populated, with relatively small planets of $< 1.6 R_{\oplus}$. This also gave rise to statistical evaluations claiming that P$\approx$ 1 days separates the shortest period planets regarding their size and numbers against the slightly longer-periodic planets \citep{2019MNRAS.488.3568P, 2018NewAR..83...37W,2017ApJ...842...40L}. One of the principal impacts of the TESS mission has however been the discovery of over 20 planets with $P \lesssim 1$ d, with nearly all of them happening since the year 2020 and also counting on mass measurements from ground-based follow-up.
In the period regime of P $\leq$ 2\,d, we hence propose to replace the desert's lower boundary with a constant corresponding to the desert's lower boundary at P = 2 days for both radius and mass, leading for P < 2\,d to a boundary at a radius of 1.60 $R_{\oplus}$ ($\log R_{lo}/R_{\oplus} = 0.2$) and a mass of 0.028 $M_\mathrm{jup}$ ($\log M_{lo}/M_\mathrm{jup}$ = -1.55) resp. 8.9 $M_{\oplus}$ (dotted lines in Fig.~\ref{fig:radiusperiod}). In support of these lower limits to the desert, Fig.~\ref{fig:radiusdist} (top panel) shows the radius distribution of the short-period small planet population  $\log R/R_{\oplus} < 0.8$ resp. $R/R_{\oplus} < 6.3$ ), where we note that the radius distribution has little dependence on the orbital period, with the planet's median size following the relation 
\[R/R_{\oplus} = 1.4\ P^{0.11}\ ;\ \  0.3 \lesssim  P(\mathrm{day}) \lesssim 3\, .\]

In Appendix C we show in Fig.~\ref{fig:insolation} a plot similar to that of Fig.~\ref{fig:radiusperiod}, but against the planets' insolation {and effective temperature}, where the upper boundary of the Neptune desert has become notably better defined, and propose corresponding limits of the Neptune desert against these parameters.

From these distributions, it appears that \pname\ belongs to a continuous distribution of super-Earths with periods ranging from the shortest known ones up to $\approx$ 30 days, with neither the period-radius nor the period-mass distributions showing any signs for a discontinuity near the common limit of P = 1 days for USPs. The maximum radii of Super Earths are delimitated at the shortest periods by the Neptune desert (for which we propose a lower limit of $\approx 1.6\,R_{\oplus}$ for periods shorter than 2 days, albeit planets with radii up to $\approx 2\,R_{\oplus}$ would belong to the same population\footnote{We note that the limits for the Neptune Desert given by \citet{2016A&A...589A..75M} do not attempt to delineate an area that is empty of planets, but rather they were placed to produce the best contrast between the lower-density 'desert' and its more densely populated surroundings.}), while for longer periods, Super-Earth radii are delimited by the period-radius valley that separates them against Sub-Neptune type planets.

\emph{Distribution of small planets against period and USP formation pathways:}
{Regarding the abundance of small planets against period (Fig.~\ref{fig:radiusdist}, bottom panel), we note the emergence of a plateau between $\log P$ of -0.1.25 and +0.125 (P $\approx$ 0.6 to 1.4 d). This plateau might correspond to a bump that was previously noted as an excess of 50 \% more planets just below P = 1 than above it (\citealt{2019MNRAS.488.3568P}, based on the work by \citealt{2017ApJ...842...40L})\footnote{We note that the 1-d bump in \citet{2017ApJ...842...40L} might be a result from the integration of two different studies, one for period of less than one day, and one for periods larger than 1 day, with different stellar host types.}, with abundance slopes that are steeper below 1 day than above 1 day. The newer planet discoveries imply however that this bump has smoothed out into the observed plateau, but the abundance slope remains somewhat steeper to the left than to the right of the plateau.  Alternatively, there might be uniform slope in abundances against period, with an additional accumulation of planets at periods between 0.6 and 1 days. 

 \citet{2019MNRAS.488.3568P} proposed the formation of USPs within multi-planetary systems from low-eccentricity migration due to secular interactions among the planets. This pathway involves the initial birth of an innermost planet with a period of several days and a moderate eccentricity of 0.05 to 0.15. Through tidal interactions with further outer planets, the eccentricity of the innermost one is gradually damped to nearly zero, while its semi-major axis undergoes a quasi-equilibrium shrinkage. As a result of this process, the innermost planet transforms into a USP, while the outer planet stabilises at an orbital period that is larger by $\gtrsim$ 15 times. \citeauthor{2019MNRAS.488.3568P} also provide synthetic planet distributions that have undergone this formation pathway, with a variety of initial parameters (varying the mass and eccentricity of the innermost planets and also the tidal quality factor Q of both stars and inner planets). It is of note that their simulation with the highest initial orbital eccentricity, of 0.15$\pm$0.025 (their Fig. 15), agrees very well with the observed abundances from Fig.~\ref{fig:radiusdist}, with the reproduction of the abundance plateau around $P \approx 1$ d and the steeper slope to the left than to the right of it. Notably, the initial eccentricity was identified by \citealt{2019MNRAS.488.3568P} as the parameter that most clearly affected the final results of their simulations. This leads to the suggestion that USP formation from inwards migration of inner planets with an initial eccentricity of  $\approx$ 0.15 might be a common one. Several further formation pathways have been proposed in the literature, with a notable contrast being the high-eccentricity pathway by \citet{2019AJ....157..180P} that requires an initial eccentricity of $e \gtrsim 0.8$. However, without simulated planet distributions against basic parameters such as period, radius and mass being available, the presence of these pathways needs to be evaluated from other diagnostics, such as ratios of orbital periods or  mutual inclinations between inner and outer planets, or measurements of spin-orbit angles, which are beyond the scope of the present work.  
 }

\section{Conclusions}

We report the discovery of the Super-Earth planet \pname\ orbiting with a period of 1.07 days around a middle-aged G9V star of likely membership in the galactic thin disk, with a tentative second planet $c$ of Neptune-like mass and a period of 27.4 or 29.5 days. The highest peaks in RV periodograms and keplerian fits for $c$ indicate a best-fit period that coincides very closely with the lunar synodic period. Consequently, the true nature of $c$ had to remain tentative despite an intense campaign of RV observations, because contamination of the RV data by a signal arising from Moon-reflected solar light cannot be ruled out. If planet $c$ is real, its radius of 3 - 8 $R_{\oplus}$ would position it above the period-radius valley, while planet $b$ is below the valley, albeit in a zone in the period-radius plane where the valley is only poorly defined.

Several composition models are discussed for \pname. Given the expected high temperature of both planet surface and interior, we consider a model describing a molten interior in which a significant fraction of water is dissolved in magma as the most promising one to explain the planet's density, which is significantly below the expected one from a pure silicate composition.  An eventual atmosphere is unlikely to contribute significantly to the planet's mass but could be suitable to observation by transmission spectroscopy with the JWST, while the planet's surface might also be within the reach of emission spectroscopy. 

The lower limit of the Neptune desert, initially identified by \citet{2016A&A...589A..75M}, is revised for planets with periods of less than 2 days. For these, the original definition of the lower boundary is clearly inconsistent with recent discoveries of significant numbers of short-periodic planets. For periods of P < 2 days, a lower boundary to the desert at radii of 1.60 $R_{\oplus}$ and masses of 8.9 $M_{\oplus}$ is therefore proposed.. We also delimit the desert using the planets' insolation instead of the period as a basic parameter. In both radius vs. insolation and mass vs. insolation distributions, the upper limit of the desert is more pronounced and corresponding relations limiting the desert are given.


The borderline position of \pname\ just outside the conventional definition of USPs, as planets with periods of less than 1 day motivated an evaluation of its position within the planet population, in period-radius and period-mass diagrams. From these, we deduce that planets with periods of less than one day do not constitute a special group of planets. Rather, USPs appear to be the extreme end of a continuous distribution of super-Earths, with periods extending from the shortest known ones up to around 30 days, with upper radii limited by the Neptune desert for periods shorter than $\approx$ 2 days, and by the period-radius valley for longer periods. Within the super-Earths, sub-groups with specific properties may however become increasingly better characterised, depending e.g. on the insolation, type or age of central star, and/or the presence of further planets. {One such hint is the plateau that has emerged in the small-planet abundance against period, in a range from 0.6 and 1.4 days, and which is compatible with the low-eccentricity formation pathway proposed by \citet{2019MNRAS.488.3568P}. The recent discoveries of numerous short-period planets, such as \pname, should inspire comprehensive investigations to assess the suitability of the various proposed formation mechanisms in explaining the present distribution of these planets across the broadest spectrum of parameters feasible.}

\begin{acknowledgements}

This work was supported by the KESPRINT collaboration, an international consortium devoted to the characterization and research of exoplanets discovered with space-based missions (\url{http://www.kesprint.science}).
This paper includes data collected by the \tess{} mission. Funding for the \tess{} mission is provided by the NASA Explorer Program. We acknowledge the use of public TOI Release data from pipelines at the \tess{} Science Office and at the \tess{} Science Processing Operations Center. Resources supporting this work were provided by the NASA High-End Computing (HEC) Program through the NASA Advanced Supercomputing (NAS) Division at Ames Research Center for the production of the SPOC data products. 
This research has made use of the Exoplanet Follow-up Observation Program (ExoFOP; DOI: 10.26134/ExoFOP5) website, which is operated by the California Institute of Technology, under contract with the National Aeronautics and Space Administration under the Exoplanet Exploration Program.

This work has made use of data from the European Space Agency (ESA) mission {\it Gaia} (\url{https://www.cosmos.esa.int/gaia}), processed by the {\it Gaia} Data Processing and Analysis Consortium (DPAC, \url{https://www.cosmos.esa.int/web/gaia/dpac/consortium}). Funding for the DPAC has been provided by national institutions, in particular the institutions participating in the {\it Gaia} Multilateral Agreement.

Based on observations made with the Italian Telescopio Nazionale Galileo (TNG) operated on the island of La Palma by the Fundaci\'on Galileo Galilei of the INAF (Instituto Nazionale di Astrofisica) at the Spanish Observatorio del Roque de los Muchachos of the Instituto de Astrof\'{\i}sica de Canarias under programmes CAT19A\_162, CAT21A\_119, CAT22A\_111 and ITP19\_1.

CARMENES is an instrument for the Centro Astron\'{o}mico Hispano-Alem\'{a}n de Calar Alto (CAHA, Almer\'{\i}a, Spain). CARMENES is funded by the German Max-Planck-Gesellschaft (MPG), the Spanish Consejo Superior de Investigaciones Cient\'{\i}ficas (CSIC), the European Union through FEDER/ERF FICTS-2011-02 funds, and the members of the CARMENES Consortium (Max-Planck-Institut f\"{u}r Astronomie, Instituto de Astrof\'{\i}sica de Andaluc\'{\i}a, Landessternwarte K\"{o}nigstuhl, Institut de Ci\`{e}ncies de l'Espai, Institut f\"{u}r Astrophysik G\"{o}ttingen, Universidad Complutense de Madrid, Th\"{u}ringer Landessternwarte Tautenburg, Instituto de Astrof\'{\i}sica de Canarias, Hamburger Sternwarte, Centro de Astrobiolog\'{\i}a and Centro Astron\'{o}mico Hispano-Alem\'{a}n), with additional contributions by the Spanish Ministry of Economy, the German Science Foundation through the Major Research Instrumentation Programme and DFG Research Unit FOR2544 "Blue Planets around Red Stars", the Klaus Tschira Stiftung, the states of Baden-W\"{u}rttemberg and Niedersachsen, and by the Junta de Andaluc\'{\i}a.

This article is partly based on observations made with the MuSCAT2 instrument, developed by ABC, at the Telescopio Carlos S\'{a}nchez operated on the island of Tenerife by the IAC in the Spanish Observatorio del Teide.
This work makes use of observations from the LCOGT network. Part of the LCOGT telescope time was granted by NOIRLab through the Mid-Scale Innovations Program (MSIP). MSIP is funded by NSF.
We thank the following iSHELL observers: Kevin I Collins, Michael Reefe, Farzaneh Zohrabi, Eric Gaidos, Angelle Tanner and Claire Geneser.

We thank Annelies Mortier for the provision of the latest versions of her code for the BGLS and related plots.
HJD and SM acknowledge support from the Spanish Research Agency of the Ministry of Science and Innovation (AEI-MICINN) under grant 'Contribution of the IAC to the PLATO Space Mission' with references ESP2017-87676-C5-4-R and PID2019-107061GB-C66, DOI: 10.13039/501100011033. S.M. and D.G.R acknowledge support form the same source with the grant no.~PID2019-107187GB-I00. SM acknowledgtes from the same source support through the Severo Ochoa Centres of Excellence Programme 2020--2023 (CEX2019-000920-S).  PGB acknowledges from the same source support with the \textit{Ram{\'o}n\,y\,Cajal} fellowship number RYC-2021-033137-I.
G.N. thanks for the research funding from the Polish Ministry of Education and
Science programme the "Excellence Initiative - Research University"
conducted at the Centre of Excellence in Astrophysics and Astrochemistry
of the Nicolaus Copernicus University in Toru\'n, Poland. This work is partly supported by JSPS KAKENHI Grant Number P17H04574, JP18H05439 and JP20K14521, and JST CREST Grant Number JPMJCR1761. J.M.A.M. is supported by the National Science Foundation Graduate Research Fellowship Program under Grant No. DGE-1842400. J.M.A.M. acknowledges the LSSTC Data Science Fellowship Program, which is funded by LSSTC, NSF Cybertraining Grant No. 1829740, the Brinson Foundation, and the Moore Foundation; his participation in the program has benefited this work. KAC and DWL acknowledge support from the TESS mission via sub-award s3449 from MIT. K.W.F.L. was supported by Deutsche Forschungsgemeinschaft grant RA714/14-1, within the DFG Schwerpunkt SPP 1992, 'Exploring the Diversity of Extrasolar Planets'.

\end{acknowledgements}


\begin{appendix}

\section{Stellar rotation period}
\label{sec:rotation}

\begin{figure}
\centering
	\includegraphics[width=1\linewidth]{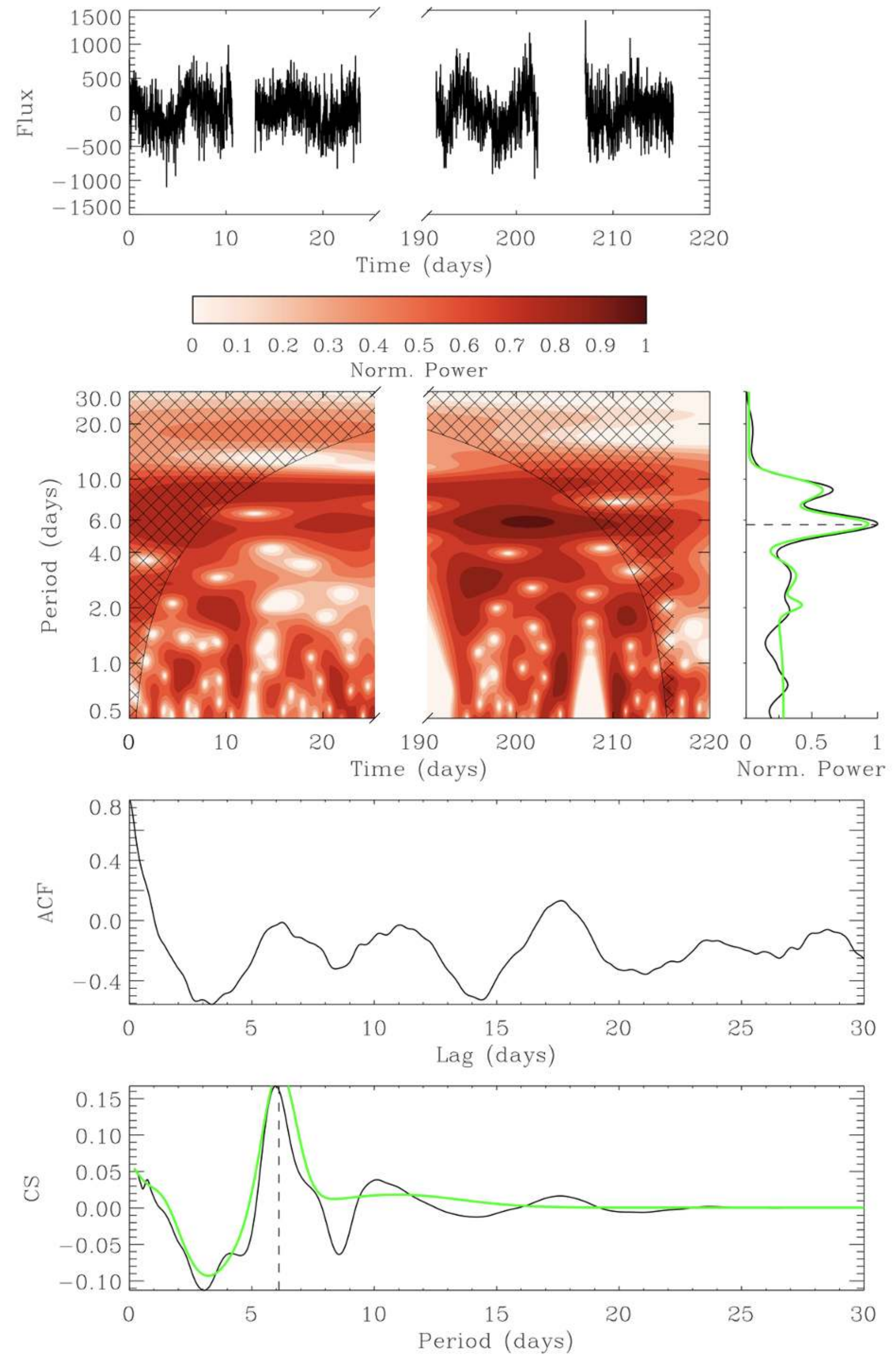}	
    \caption{Analysis of TESS lightcurve for stellar rotation of \target. The top panel shows the lightcurve from Sector 16 and 23 that was used for the analysis. The following panels show the three methods used for the period determination (see text): wavelet power spectrum (GWPS) and its projection onto the period axis; the autocorrelation function (ACF); and the composite spectrum (CS). The hatched region in the panel for the wavelet power spectrum indicates the zone where the method is not valid.} 
    \label{fig:rotation}
\end{figure}

We determine the rotation period from the lightcurve of \target\ by following the procedure described in \citet{2019ApJS..244...21S} and \citet{2021ApJS..255...17S} 
 ; see also \citet{2014A&A...562A.124M} 
 and \citet{2014A&A...572A..34G}. 
 The analysis was based on TESS light curves that have undergone after the same processing as described in Sect.~\ref{sec:tessphot} for the analysis with \pyan, from which we removed the exoplanet transits to avoid spurious signals (using the best-fit model obtained with {\tt UTM/UFIT} as described in Appendix~\ref{app:utm}). Due to the small number of data points that remained in sector 50 after removal of the bad quality data, only sectors 16 and 23 were used for the rotational analysis. Also, gaps in the light curve longer than 81 days (three consecutive TESS sectors) were removed, and inpainting techniques were used to fill in gaps shorter than 5 days \citep{2014A&A...568A..10G}, leading to the light curve shown in the top-panel of Fig.~\ref{fig:rotation}. From this curve, we derive three estimates of the rotation period: The first estimate is obtained from the global wavelet power spectrum \citep[GWPS;][]{1998BAMS...79...61T, 2010A&A...511A..46M}, which examines the correlation between the data and the mother wavelet (taken to be a Morlet wavelet), and its projection onto the period axis. The second estimate is obtained via the autocorrelation function \citep[ACF;][]{2013MNRAS.432.1203M, 2014ApJS..211...24M}, which computes the correlation between the light curve and itself for a range of time shifts. The third estimate is obtained from the composite spectrum \citep[][]{2016MNRAS.456..119C}, which is calculated as the product between the GWPS and the normalized ACF and which helps to enhance the periods that are present in both methods.

 
 Fig.~\ref{fig:rotation} shows the results from all three methods. From the ACF analysis, we can see three peaks with prominent absolute amplitudes. However, as shown in \citet{2017A&A...605A.111C}, one of the criteria to select reliable rotation periods is based on the relative amplitudes of the peaks, called H\_ACF, with significant periods having values of H\_ACF > 0.3. Computing the H\_ACF for these three peaks, the largest value is found for the period corresponding to 17.6 $\pm$ 2 days (with a value of H\_ACF = 0.5). That is the period we adopt, which approximately corresponds to the third harmonic of the $\approx$ 5 day signal seen in both the ACF and the global wavelet power spectrum (GWPS). Moreover, \citet{2021tsc2.confE.180G} applied the same method to over 2-million "Kepler-seen-as-TESS" light curves, for stars for which rotation periods had been measured by \citet{2019ApJS..244...21S,2021ApJS..255...17S}. They divided the full Kepler light curves into 27-day chunks to mimic the TESS observations, and their results showed that periods of up to $\approx$ 20 days can be retrieved with one sector of data. For instance, for peaks with H\_ACF > 0.3, they recovered periods in the 10 to 15 day window with a reliability of $\approx$ 70\%.

 We note that our adopted 17.6 $\pm$ 2 days ACF period is also compatible with the rotation period of  $P_\mathrm{rot}/ \sin i = 20_{-5}^{11} $ d determined from $V \sin i_\star$ and $R_\star$ of Tables~\ref{table: spectroscopic parameters} and \ref{table: comparison stellar parameters}. While the 17.6 day period does not show up in the GWPS, this is unsurprising as it would have been filtered out due to falling outside the cone of validity (hatched regions in  Fig.~\ref{fig:rotation}; see also \citealt{2021tsc2.confE.180G}).   Regarding a $\approx$ 10 day stellar rotation that would correspond to the second harmonic of the $\approx$ 5 day signal and for which activity indicators from the RV data indicate at notable peak in spectrograms (Sect.~\ref{sec:specanal}), it is argued at the end of Sect.~\ref{sec:joint} that this period is unlikely to be associated with stellar rotation.
 
 
 From the adopted period of 17.6 d we furthermore derive ages from several rotation-age relations reported in the literature, resulting in ages of:  $0.84 \pm 0.18$ Gyr \citep{2007ApJ...669.1167B}; $1.26 \pm 0.29$ Gyr \citep{2008ApJ...687.1264M}; $1.58 \pm 0.7$ Gyr \citep{2015MNRAS.450.1787A}; $1.49 \pm 0.23$ Gyr \citep{2019AJ....158..173A} and $1.75 \pm 0.25$ Gyr \citep{2020A&A...636A..76S}. Ignoring the value from \citet{2007ApJ...669.1167B} as the most discrepant one, gyrochronology indicates hence an age of 1 - 2 Gyr. We note however that the lightcurve analysis does not exclude a longer rotation period that is not perceived due to the limited coverage of the TESS lightcurves and which would also indicate older ages for \target.
 

\section{The RV double peak at periods of 27.4 and 29.5 days: Planet candidate or influence from the Moon?}
\label{lunar_disco}

 \begin{figure}
\centering
	\includegraphics[width=1\linewidth]{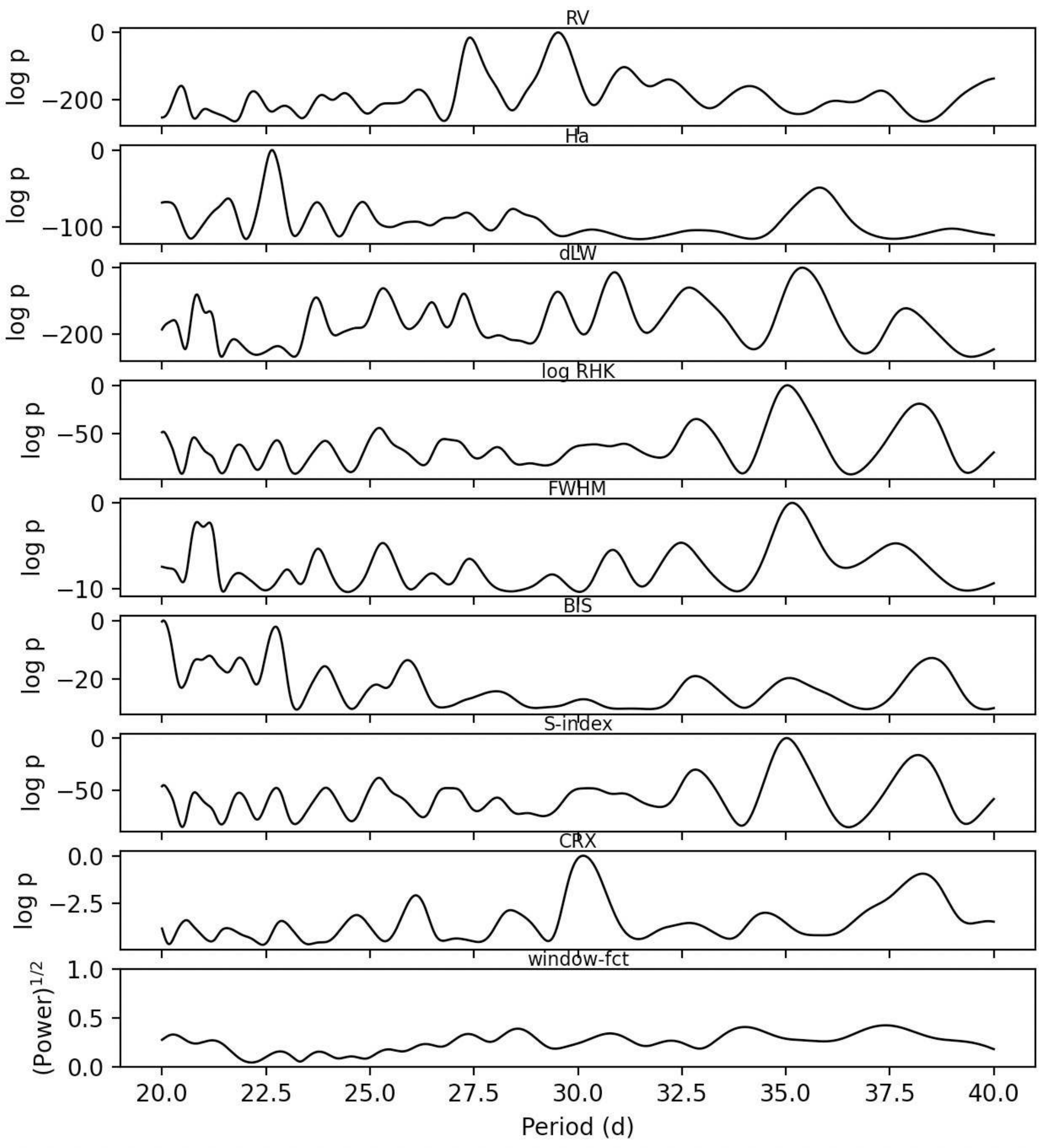}	
   \caption{Zoomed-in view of the BGLS periodogram of Fig.~\ref{fig:bglsHNbase}, around the 29.4 d period of planet $c$.}   
    \label{fig:bglsHN29d}
\end{figure}


The spectral signatures presented in Sect.~\ref{sec:specanal}, from both the BGLS and the Bayes Factor periodogram from {\tt agatha} indicate an RV signal in the \hn\ data with a period of $\approx$ 29.4 d as the most promising one for an additional planet in \target. Fig.~\ref{fig:bglsHN29d} shows a zoom of the BGLS periodogram near that period, which also shows the somewhat lower neighbouring RV peak with P= 27.4 days. Potentially, one of these peaks (more likely the lower 27.4 d one) is an alias of the other one,  related to each other by a seasonal sampling with a frequency of 1/365 d$^{-1}$. 

Of principal concern regarding the interpretation of the 29.4 d peak is its close match with the length of the lunar synodic month of 29.53 d, which in the case of the 29.52 d signal found by {\tt agatha} (see Fig.~\ref{fig:bfpHN}) is matched to the fourth digit. We also note a relative strong peak of the chromatic index (CRX) activity indicator near that period. Considering also \target's small systemic RV of $\approx$ 1.1 km s$^{-1}$, this leads to a strong suspicion that the observed RV peak might be due to a contamination by the Moon, or more precisely, be due to the influence of solar light that is reflected by the Moon. In Fig.~\ref{fig:absrv_vs_moonrv} we show a plot of the uncorrected \emph{absolute} RVs of \target\ against the RV of the Moon-reflected solar spectrum at the moment of observation. For differences between these two RVs of $\lessapprox$ 10 - 15 km/s, spectral lines in the reflected solar spectrum might overlap with similar lines in the target's spectrum\footnote{Assuming a spectral line broadening of \target\ of $7.0\pm1.7$ km/s \citep[GAIA DR3, see also][]{2022arXiv220610986F} and of the Sun of $\approx$ 5.6 km/s \citep[][ sum of rotational broadening and macro-turbulence]{2018ApJ...857..139G}.} and hence might affect the measured RVs. We note that the 'above horizon' RVs in Fig.~\ref{fig:absrv_vs_moonrv} appear to be on a down-wards slope; this effect is however due to observing when \target\ had a positive absolute RV preferentially during waning lunar phases (when the Moon moves towards the Earth and the Moon-reflected solar spectrum has a positive RV); whereas observations when  \target\ had a negative RV happened  mainly at waxing lunar phases. This is a consequence of observing a target preferentially in the morning at the begin of an observing season (when a waning moon is seen), whereas towards the end of a season, a target is observed in the evening, when only a waxing Moon can be seen.

 \begin{figure}
\centering
	\includegraphics[width=1\linewidth]{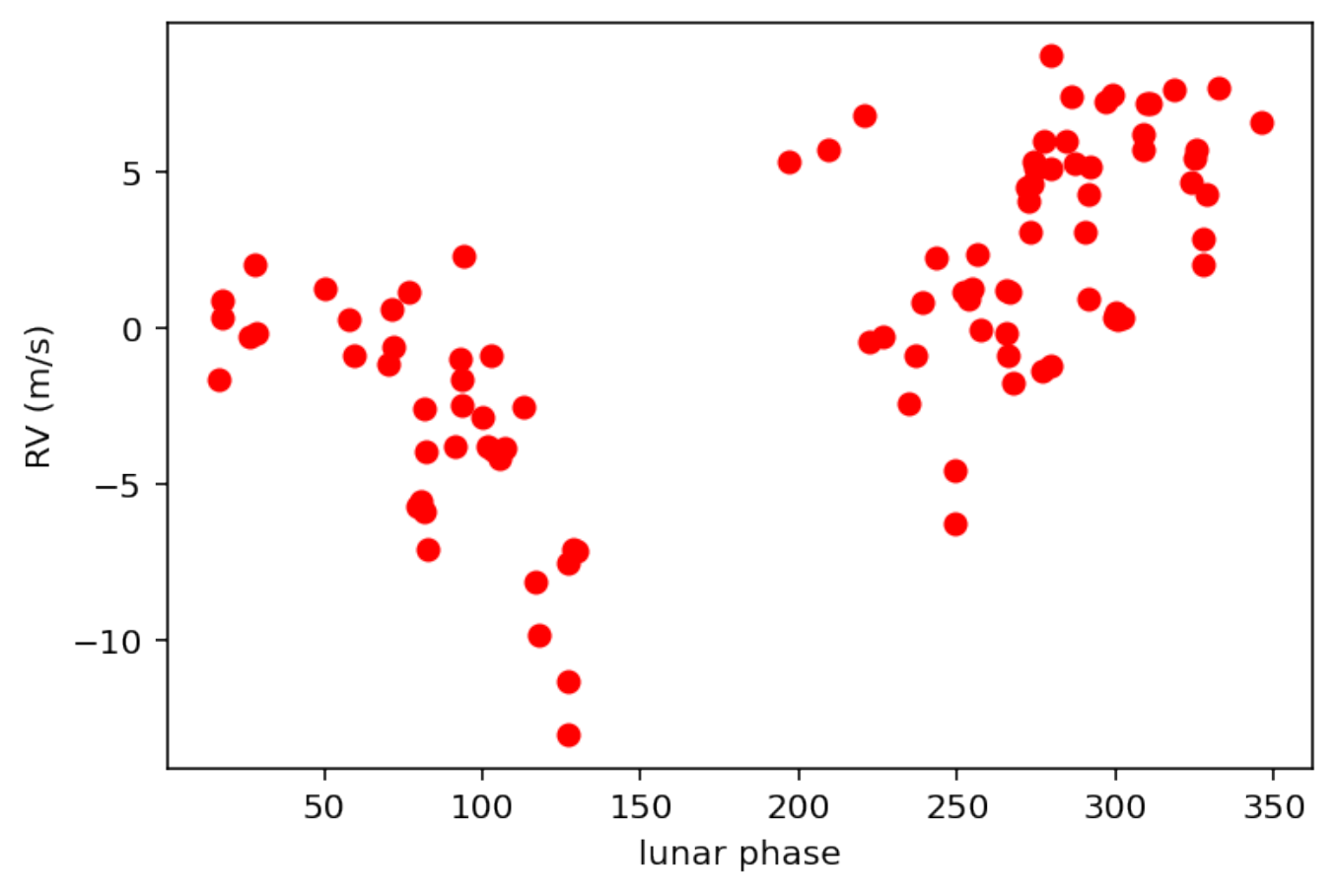}	
   \caption{\hn\ RV's folded against the lunar phase, where 0$^{\circ}$ or 360$^{\circ}$ corresponds to New Moon and 180$^{\circ}$ to Full Moon. The clumping of the RV data in two regions of lunar phases, with an avoidance of  Full Moon and lesser coverage near New Moon, is a consequence of the scheduling of the \hn\ observations, which were mostly executed in lunar grey time.} 
  
    \label{fig:RVmoonphase}
\end{figure}

 \begin{figure}
\centering
	\includegraphics[width=1\linewidth]{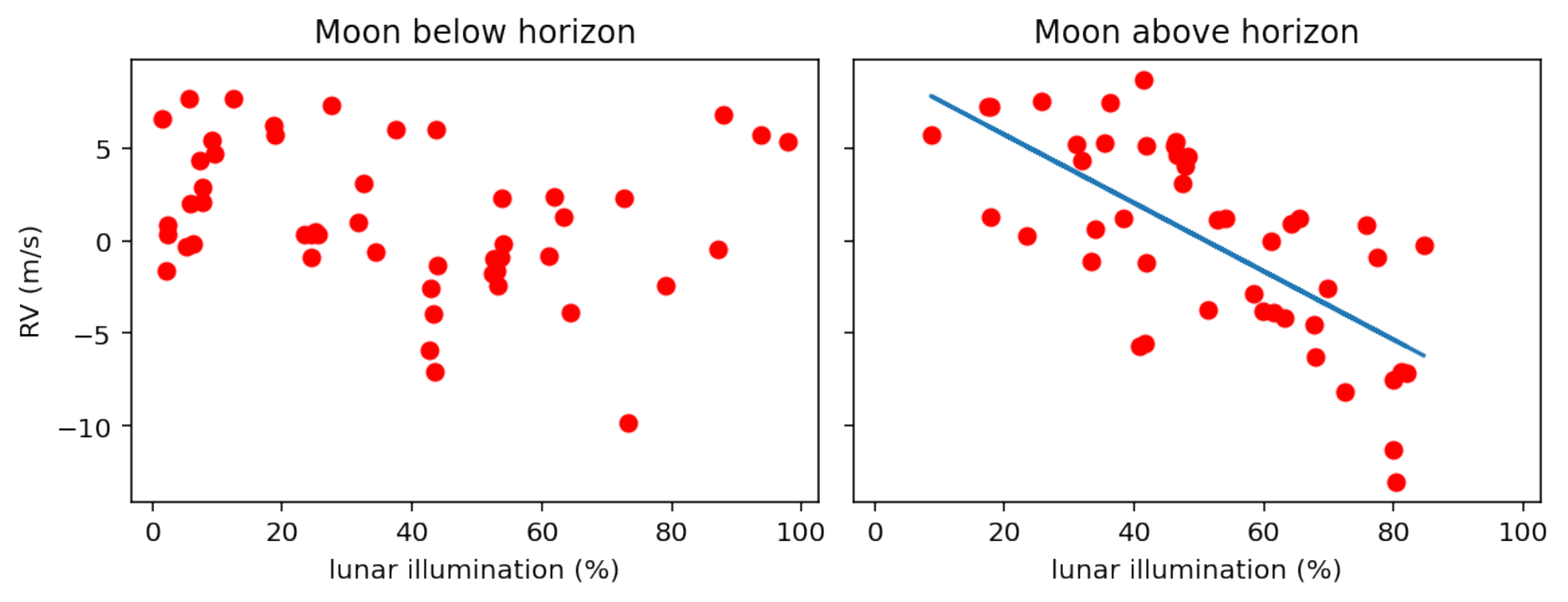}	
   \caption{Similar to Fig.~\ref{fig:RVmoonphase}, but the \hn\ RVs are plotted against the lunar illumination at the time of observation, and the data are separated into panels containing only RVs that were taken when the Moon was below resp. above the horizon. The blue line in the right panel shows a linear fit to the RV versus illumination dependency, which has a correlation coefficient of -0.69.} 

    \label{fig:RVmoonillum}
\end{figure}

\begin{figure}
\centering
	\includegraphics[width=1\linewidth]{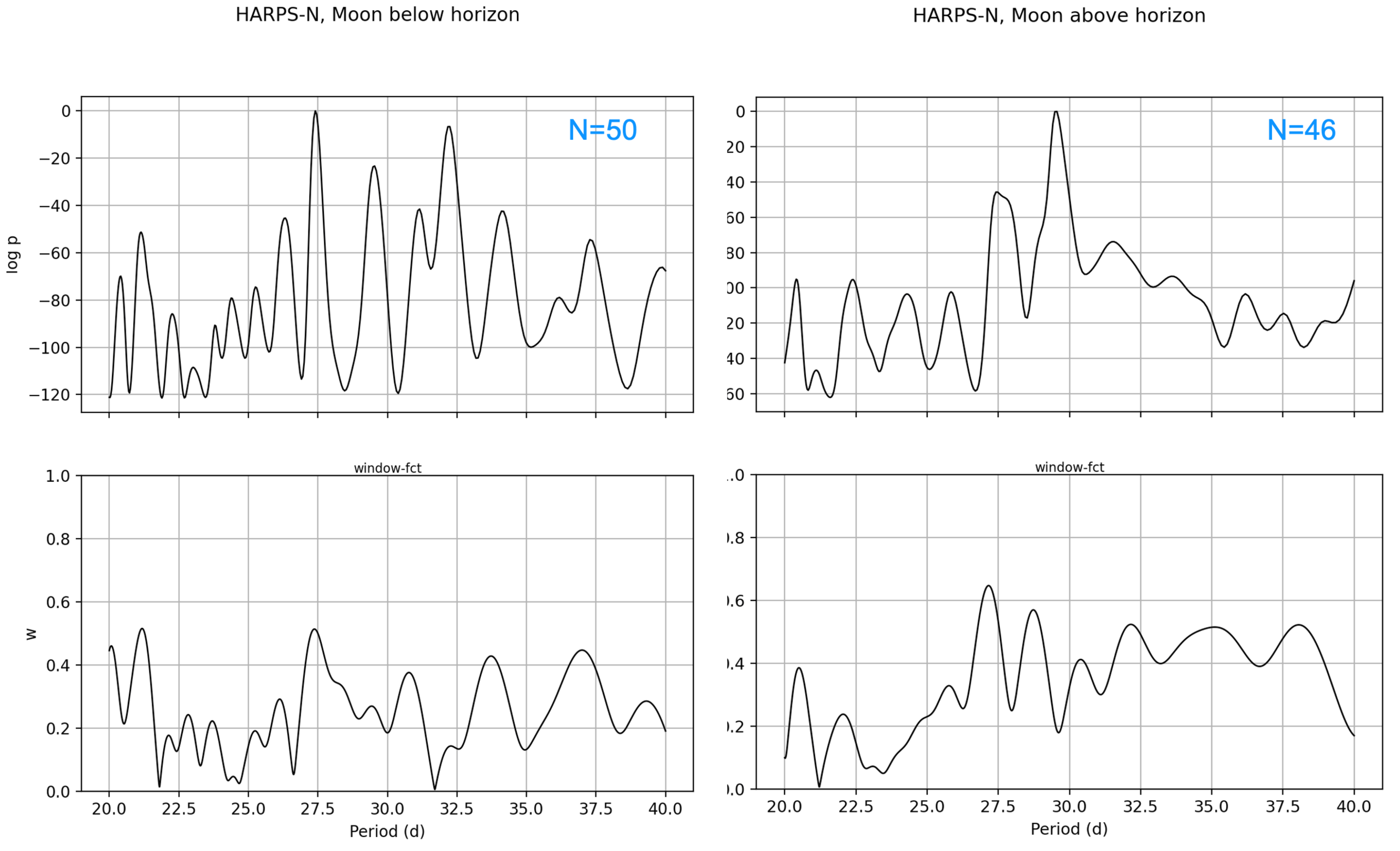}	
       \caption{BGLS periodograms of the \hn\ RVs  and window functions, with the RV data being separated into those taken when the Moon was below resp. above the horizon. The blue numbers indicate the number of RV points.}   
    \label{fig:hnspec_moon}
\end{figure}

\begin{figure}
\centering
	\includegraphics[width=1\linewidth]{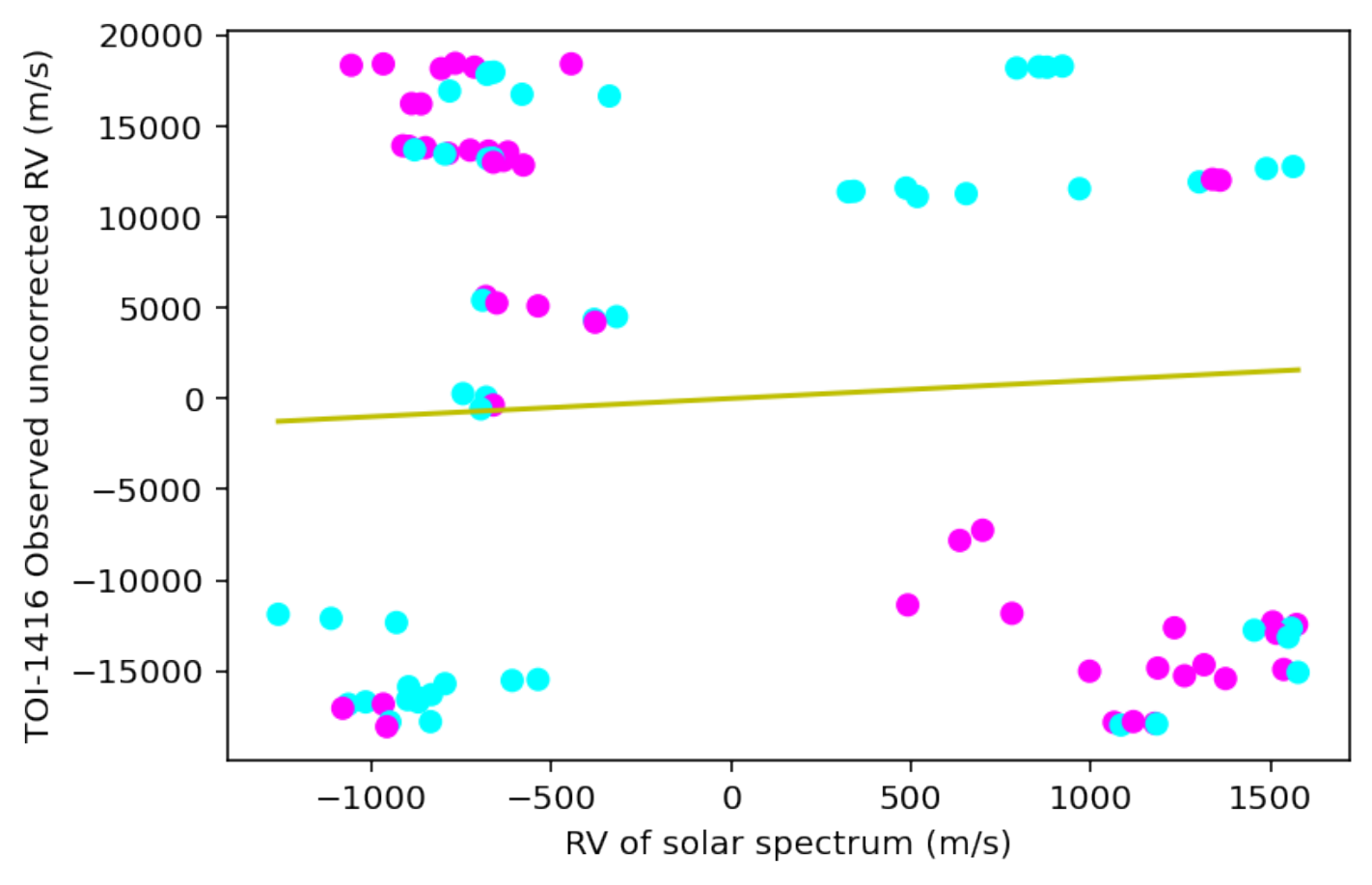}	
   \caption{Absolute uncorrected RVs of \target\ versus the RV of the Moon-reflected solar spectrum. The symbol colors indicate if the Moon was above (pink) or below the horizon  (blue) at the moment of observation. The green line corresponds to identical RV values on both axes.}   
    \label{fig:absrv_vs_moonrv}
\end{figure}

Using the hypothesis of a contamination by the Moon, the barycentric-corrected \hn\ RV values\footnote{The Keplerian signal corresponding to planet $b$ was subtracted from these RVs. However, the presence or absence of the planet $b$ signal does not alter the shown plots and the conclusions in any relevant way.} were folded against the lunar synodic period, with their time-stamps converted to corresponding values of lunar phases. The result (Fig.~\ref{fig:RVmoonphase}) shows a clear dependency between lunar phase and RV, with a symmetry against the full or the new Moon. However, this does not disprove that by coincidence, a planet in \target\ might have a period that is very close to the lunar one. 
In a further step, we divided the RVs into those which are taken with the Moon being above horizon (46 RV points), and those were the Moon was below horizon\footnote{The {\tt skyfield} python package \citep{2019ascl.soft07024R} was used to calculate all values related to the Moon's position or velocity at the time of the observations} (50 points). Also, instead of the lunar phase, we plot the RVs against an approximation of the lunar illumination, given by the relation 
\begin{equation}
\mathrm{illum(\%)} =(1-\cos \phi) *50 ,
\end{equation}
where $\phi$ is the lunar phase in radians, with $\phi=0$ at New Moon. The result, shown in Fig~\ref{fig:RVmoonillum}, shows no relevant correlation (with a correlation coefficient  of -0.23) for the RVs against illumination (or phase) when the Moon was below the horizon\footnote{We also note that the three outliers near the lunar phase of 200$^{\circ}$ in Fig.~\ref{fig:RVmoonphase} agree now well with the other RVs; these were taken in twilight when a nearly full Moon was just below horizon}. However, a relevant correlation  (with a coefficient of  -0.69) is present when the Moon was above the horizon. Corresponding BGLS spectra for the RVs with/without Moon (Fig.~\ref{fig:hnspec_moon}) show the 29.5 d peak very prominently in the 'above horizon' spectrum, whereas in the 'below horizon' spectrum, the 29.5 day peak is insignificant while the peak at 27.4 d has become more prominent and a second one at 32.2 d has appeared. The 32.2 d peak might be another alias of the 29.4 d peak against a yearly sampling frequency, but we also note the strongly disparate window-function between the 27.4 and the 32.2d peaks, which weakens any conclusions regarding the relations between these peaks. In any case, the prominence of the 29.5 d signal in the 'above horizon' spectrum and its disappearance in the 'below horizon' one, together with the  correct phasing of this signal against the Moon's illumination is a strong indicator that the Moon is indeed responsible for this signal.

Dependencies of the RVs against the Moon altitude at the time of observations or against the angular separation of the Moon and \target\ were evaluated as well, but these do not show any relevant correlation. 
Attempts were made to correct the above-horizon RVs against the illumination-dependency, using the linear fit shown in Fig.~\ref{fig:RVmoonillum} (and also higher-order fits, not shown)  and to perform a modelling with {\tt pyaneti} on the corrected RVs. The results were however unsatisfactory, showing only degraded fits for a Keplerian signal at either the 29.5 or at 275d period.


In order to identify potential RV shifts due to contamination by the Moon, we evaluated the effect of the Moon on the \hn\ high resolution spectra's cross correlation function (CCF). Only in 65 of the 96 HARPS spectra, a second fiber (B) was placed on the sky, and only in a minority of these 65 fiber-B spectra, a signal from the Moon-reflected spectrum could be identified and the RVs be corrected against it. The difference from that correction was almost always below 1 m/s, which is small against the $\approx$ 5 m/s amplitude of the 29.5 d signal. Consequently, periodograms with or without this correction in the 65 \hn\ RVs for which this could be done do not show relevant differences. We also investigated if there might a relation between the the RVs and the SNR in the spectra (e.g. due to sky-brightness from the Moon) but there is no correlation apparent (for the RVs taken with the most frequent exposure time of 1200 sec, a correlation coefficient of -0.05 was found). Hence, an \emph{identifiable} effect of the Moon-reflected solar spectrum onto the measured RVs is hence very minor. 

\begin{figure*}[t]
\includegraphics[height= 90mm]{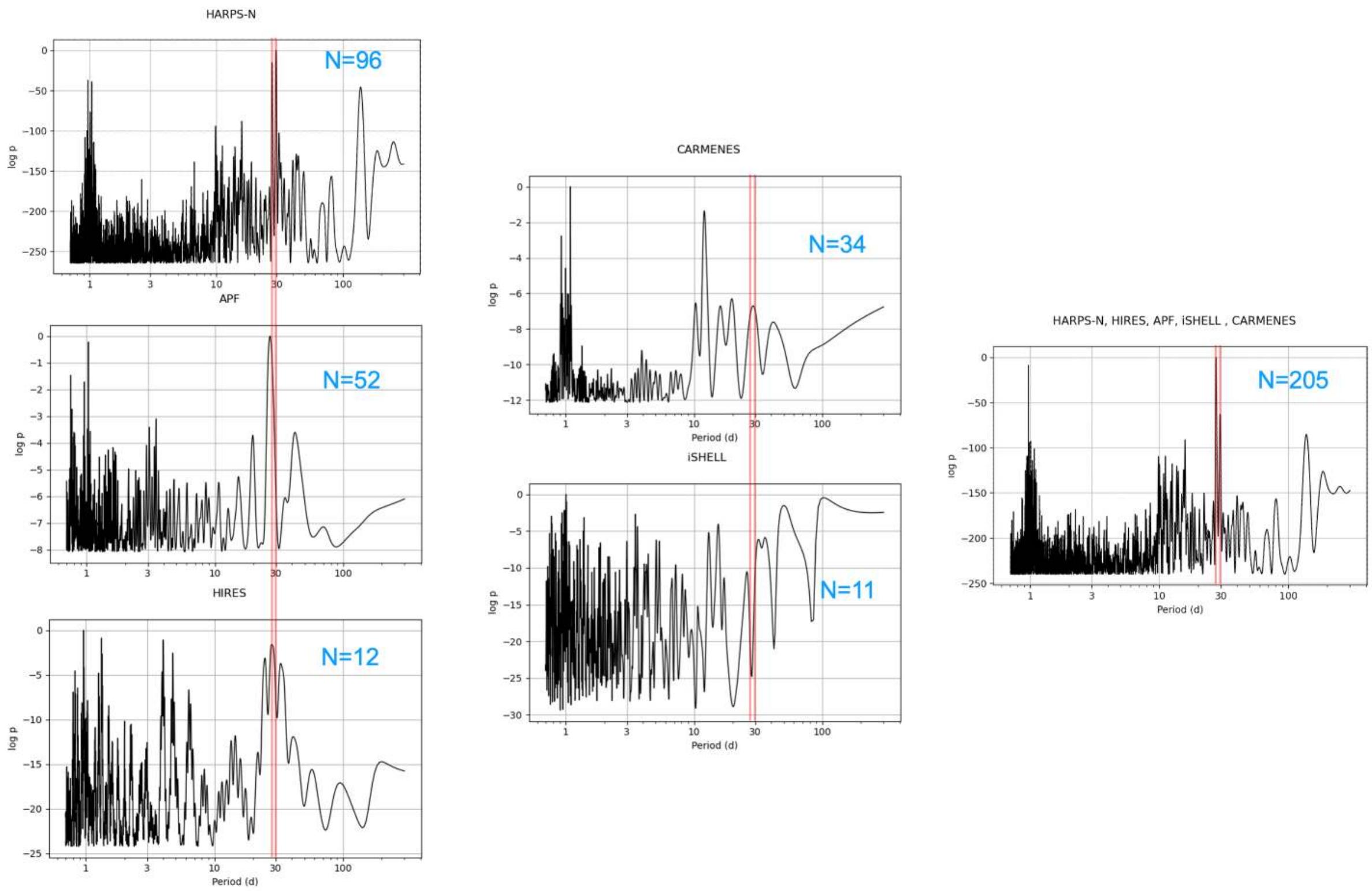}
\caption{BGLS periodograms of the RVs of all contributing instruments. In the left column are those that show a peak near 27 or 29 days (vertical red lines); in the center are those that don't, and the right panel shows a periodogram with the RVs from all instruments. The blue numbers indicate the number of RVs used.}
\label{fig:BGLSallinstr}
\end{figure*}

However, we consider that the 29.5 d peak in the \hn\ spectrograms remains of questionable nature, and now turn our attention to the neighbouring peak at $\approx$ 27.4 days. Fig.~\ref{fig:BGLSallinstr} shows BGLS spectrograms of all contributing instruments, and it is of note that data from the APF -- which contributed with the the second largest set of RVs  -- have their strongest peak at 26.8 d. Also, the HIRES RVs show a peak in the same period-range, which is very broad due to the small sample of only 12 RVs. However, peaks in that range are absent in periodograms from CARMENES and iSHELL data. We note that these are also the instruments whose spectral coverage is the most red-wards (see Table~\ref{table:RVstats}), and a Moon-reflected reflected solar spectrum would generate a weaker signal in them. Lastly, in a combination of all available RVs, the peak at 27.4 d is the highest overall, and is significantly stronger than the one at 29.5 d.   

\begin{figure}[H]
\centering
	\includegraphics[width=1\linewidth]{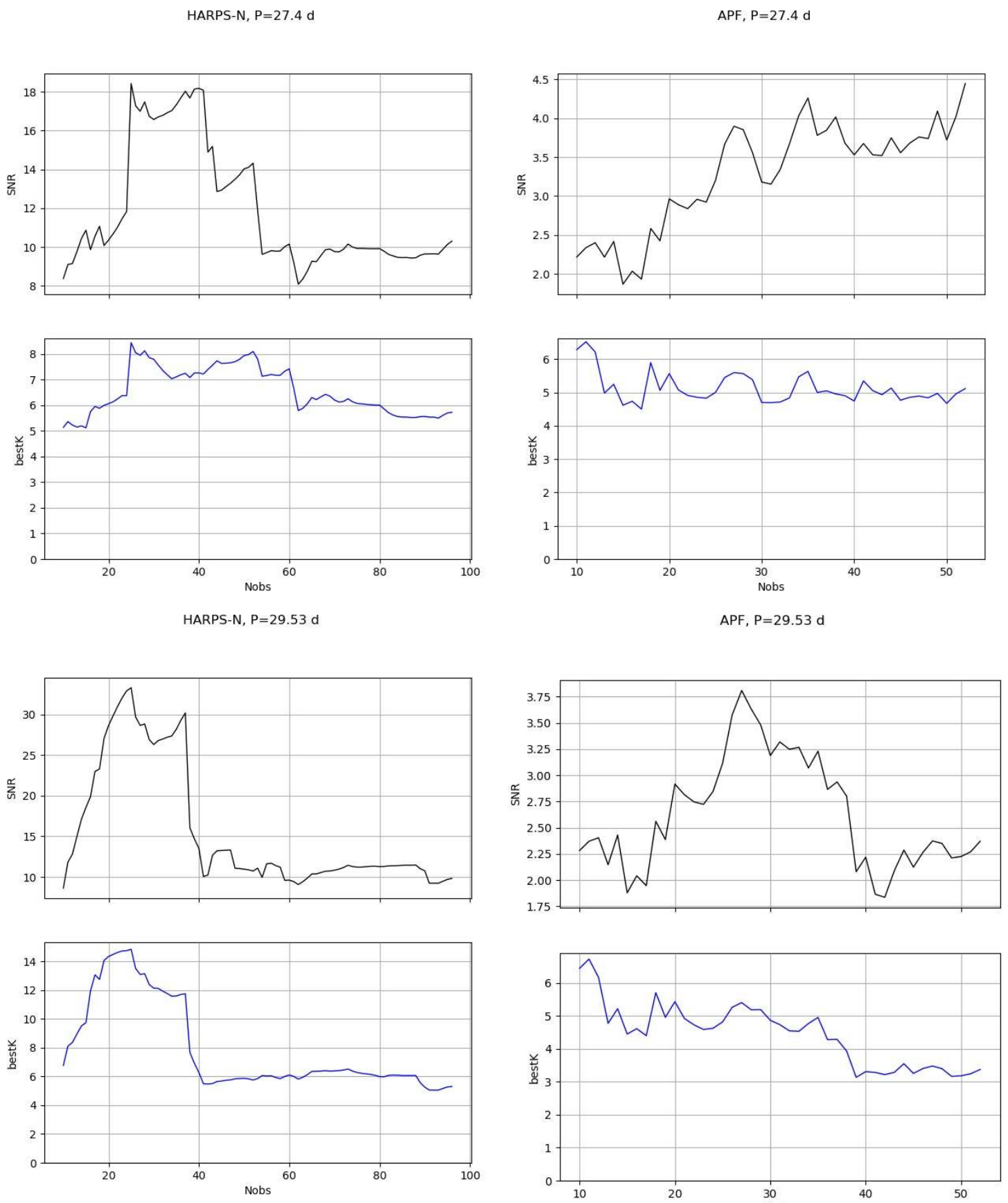}	
   \caption{Development of the SNR (signal to noise ratio) and the best-fitting amplitude 'bestK' of RV signals of the 27.4d (top panels) and the 29.5 d period (bottom) , versus the number of RV points since the first measurement.  The left panels are based on RVs data from \hn\ and the right ones on data from the APF.
      }   
    \label{fig:SNRplot_HN_APF.pdf}
\end{figure}

In Fig.~\ref{fig:SNRplot_HN_APF.pdf} we provide plots of the development of the SNR (signal to noise ratio) and the best-fitting amplitude $K$ of RV signals at the 27.4 d and 29.53 d periods, versus the number of RV points (counting from the first measurement), following the precepts of \citet{sBGLS2017}. These plots are shown for the two largest sets of RVs, those from \hn\ and from the APF. In the plots for \hn, the SNR degrades at either period near the 40th point, which is likely due to a lesser consistency of these signals across RV coverages spanning more than one observing season. On the other hand, in the plots for the APF (which cover only one observing season), the 27.4 d signal shows a steady increase in SNR and a rather constant amplitude $K$, whereas the 29.5 d signal shows a less consistent picture, more similar to the one from \hn. As is stated in Sect.~\ref{sec:joint}, fits of Keplerian orbits to the 27.4 d signal where however significantly worse than those to the 29.5 d one. 

In summary, we cannot decide on a clear preference that either of these signals represent a true signal from \target, nor about their actual nature, and conclude that the RV signals with a 27.4 or 29.5 day period are at most tentative of a further planet at either of these periods. 

\section{Modelling of the lightcurve with {\tt UTM/UFIT}} \label{app:utm}

The normalised and gradient-corrected lightcurve around transits of planet $b$, whose  preparation is described in Sect.~\ref{sec:tessphot}, was used for transit fits using the Universal Transit Modeller / Fitter \citep[{\tt UTM/UFIT},][]{deegutm}\footnote{Available at \url{https://github.com/hdeeg/utm_ufit/}}.  In brief, {\tt UTM} is a lightcurve modeller for all kinds of eclipsing or transiting configurations between any number and kind of objects, such as stars, planets, moons and rings. {\tt UFIT} was developed as a wrapper to UTM to perform fits, albeit it has been extended to accept several further modelling modules (such as the one used for the FCO fit described in Appendix~\ref{app:fco}). As the core modelling engine, \utm\ may use either pixelised object representations suitable for arbitrary configurations of multiple occulters, or an analytical 'fast mode' suitable for basic transit configurations, which employs the {\tt exofast\_occultquad.pro} routine from the EXOFAST library \citep{2013PASP..125...83E,2019arXiv190709480E}; the latter one was used in this work. \ufit\ (Universal Fitter) permits the fitting of any of \utm's input parameters, either with the Amoeba algorithm or through the generation of MCMC chains using the  Differential Evolution Markov Chain method of \citet{2006S&C....16..239T}. Its implementation is based on the {\tt EXOFAST\_DEMC} routine from the same library, but with an extension that permits the constraining of free parameters by several types of symmetric and asymmetric priors.  

\utm\ permits the modelling of transit curves from any set of input parameters that is fully able to describe an orbiting system. For this work, we modelled the light-curves curves against the following set of parameters (these were also free parameters in the fits): Orbital period $P_{\mathrm{orb}}$, transit epoch $T_0$, scaled planetary radius $R_\mathrm{p}/R_{\star}$, the stellar density\footnote{The stellar density was mainly chosen for compatibility with the set of input parameters used by \pyan, described in Sect.~\ref{sec:joint}. A pre-processor routine to \utm\  converts the stellar density into the usually used ratio of the semi-major axis versus the stellar radius, $a_\mathrm{p}/R_{\star}$, using e.g. Eq. (31) of  \citet{pyaneti}.} $\rho_{\star}$ and the transit impact parameter $b$. Initial values for these fits were taken from the SPOC's data validation summary for \target. For the stellar limb-darkening (LD), a quadratic LD law was used, albeit for the fitting we used the $q_1$ and $q_2$ coefficients for an optimised sampling introduced by \citet{Kipping2013}.  An absolute offset in flux values was a further free parameter in our fits, in order to account for potential errors in the normalisation of the flux described in Sect.~\ref{sec:tessphot}. The orbital eccentricity was kept to zero. 

Initial fits were performed with the Amoeba algorithm, leading to an intermediate transit-model that was used for the initial parameters for an MCMC sequence. First efforts without constraints on the input parameters showed significant correlation between the impact parameter, the stellar density, and the planet radius. Also, the limb-darkening parameters could only be poorly constrained from the fits. We therefore choose to impose Gaussian priors on the stellar density, taken from the adopted value in Table~\ref{table: comparison stellar parameters} and on the limb-darkening. For the later, we used the tabulation of LD coefficients for the TESS satellite bandpass by \citet[][Table 25 for the  ATLAS model with plane-parallel geometry]{2017A&A...600A..30C} and an interpolation to the adopted stellar parameters from Tables~\ref{table: spectroscopic parameters} and  \ref{table: comparison stellar parameters}. The obtained values for a square LD law as defined in Eq. (2) of  \citet{2017A&A...600A..30C} were $u_1 = 0.4545 \pm 0.05$ and $u_2= 0.1880 \pm 0.05$, which were converted into priors for the $q_1$ and $q_2$ coefficients, given in Table~\ref{table:final_param_model}.

\begin{figure*}
\includegraphics[width=\linewidth]{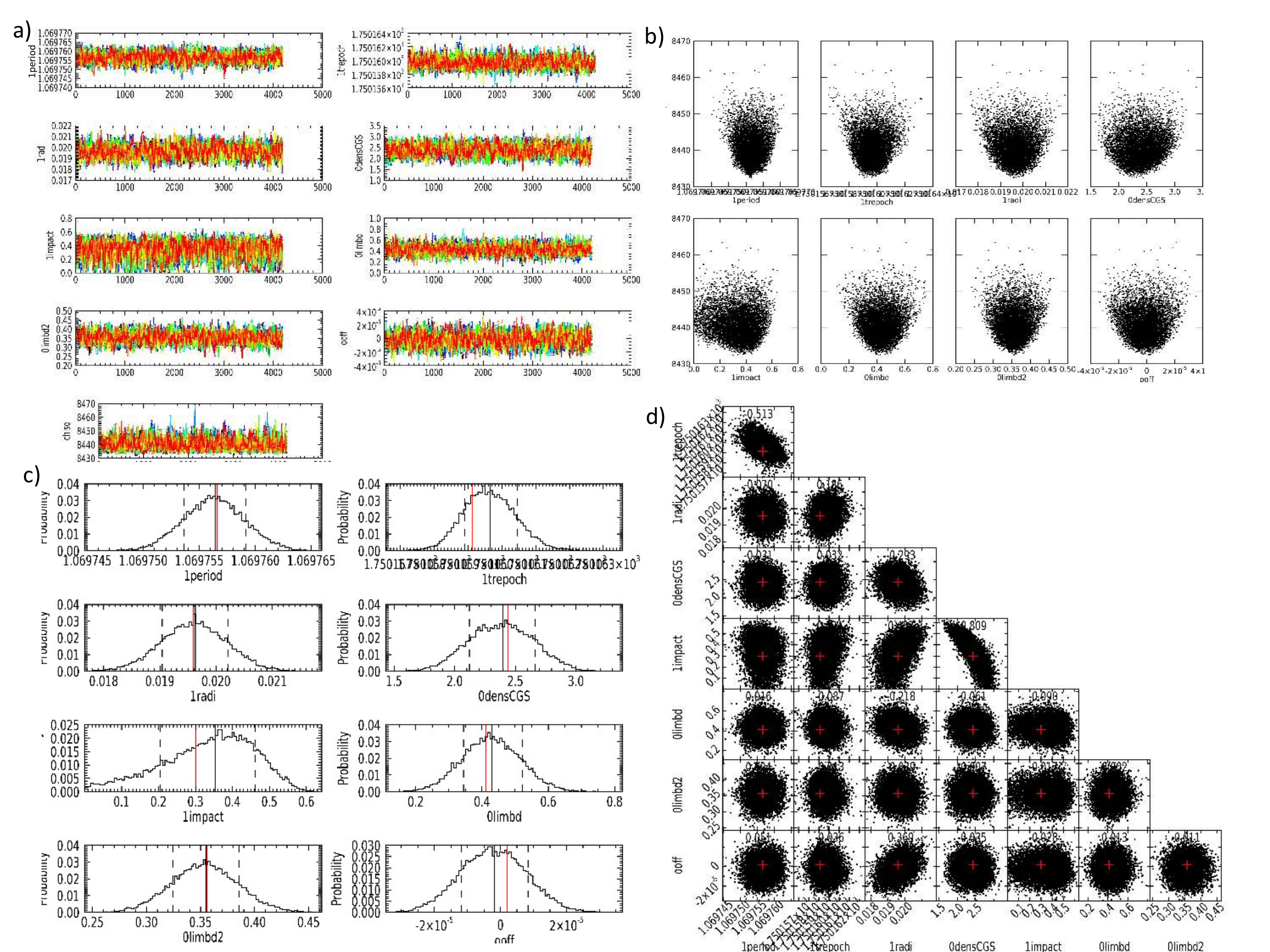}
\caption{Graphical output from MCMC sequence performed by \ufit\ that led to the results reported in Table~\ref{table:final_param_model}.  In all panels, the parameters are indicated by the keywords used in \ufit: 1period: planet period, 1trepoch: epoch of transit, 1radi: relative planet radius, 0densCGS: stellar density in CGS units, 1impact: impact parameter, 0limbd and 0limbd2: Limdarkening coefficients $q_1$ and $q2$,  ooff: off-transit flux-offset against zero. The sub-figures are: a) Values of the parameters against link-number of the MCMC sequence, excluding burn-in. Each MCMC chain is shown by a different color. The lowest panel shows the evaluation of the $\chi^2$ values. b) Scatter plot of parameters versus the $\chi^2$ value. c) Histograms of parameter distributions. The median value is shown by the vertical black line; the dashed lines delimit the 68.3\% credible interval and the red line gives the value of the best fit. d) Cornerplot  of the correlations among parameters. The red crosses give the values of the best fit.
\label{fig:ufit_dens3}}
\end{figure*}

 The final MCMC sequence consisted of 16 parallel chains that were iterated until a sufficient mixing of parameters was achieved, based on the Gelman-Rubin statistics following the precepts of \citet{2006ApJ...642..505F} and \citet{2013PASP..125...83E}. The resultant values (included in Table~\ref{table:final_param_model}) were then derived from the posterior distributions of the parameters, based on 4280 steps,  after a burn-in period of $\approx$ 1000 steps. These distributors were in all cases close to Gaussian shapes; see Fig.~\ref{fig:ufit_dens3} for this and several further diagnostic plots from the MCMC sequence. The best-fit transit model against the phase-folded input lightcurve is also included in Fig.~\ref{fig:wrap_lc}. 

\section{Detection of the transiting planet in RV data by FCO analysis} \label{app:fco}
The FCO (Floating Chunk Offset) method, developed by \citet{2010A&A...520A..93H} and \citet{ 2014A&A...568A..84H}, is best suited for the determination of RV amplitudes of short-periodic planets whose nightly RV variations are expected to be larger than the individual RV measures' uncertainties. In short, sets of nightly RV data -- with at least two well separated data-points per night -- are treated as independent chunks of data with unknown (free) RV offsets. Systematics (both instrumental and effects from other planets or stellar activity) on time-scales larger than a single night are therefore suppressed by the FCO method. RV offsets for each nightly set of RVs are then fitted against an RV model and the parameters of the best fit are obtained. 

Only the RVs from \hn\ were used in this analysis. The FCO method could not be applied to the data from the other telescopes, because all their RVs are single data-points in a given night (with the exception of two nights from APF, where RVs spaced about 20 min apart were obtained, which is too short a separation to be suitable for the FCO analysis). In the \hn\ data, there are 28 nights in which two or more RVs were obtained, which enabled the use of 77 out of the 93 RVs from \hn. For the fitting of the RVs, we used the same \ufit fitter as described in Appendix~\ref{app:utm}, but with a modelling module \texttt{ufit\_rvcurve} for the generation of a Keplerian RV model from any number of RV data-sets, each with its own RV-offset $\gamma_i$.  For the FCO method, each 'chunk' with a nightly set of two to six RVs is considered an independent set of data. 

\begin{figure}
\centering
	\includegraphics[width=1\linewidth]{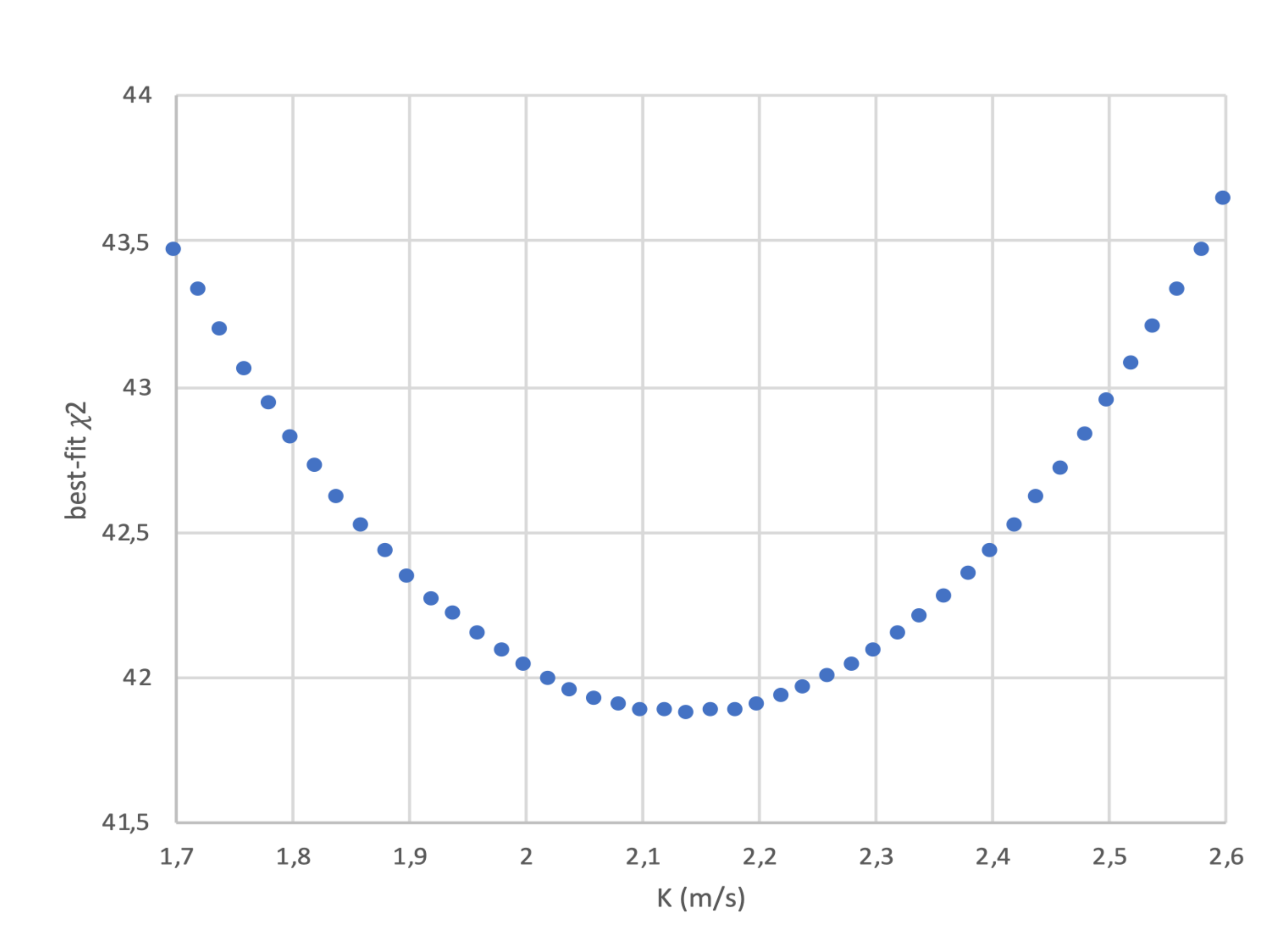}	
    \caption{Best-fit $\chi^2$ from FCO fits of \hn\ data against RV models of TOI-1416b with fixed RV-amplitudes $K$.}  \label{fig:FCOchisq} 
\end{figure}

\begin{figure}
\centering
	\includegraphics[width=1\linewidth]{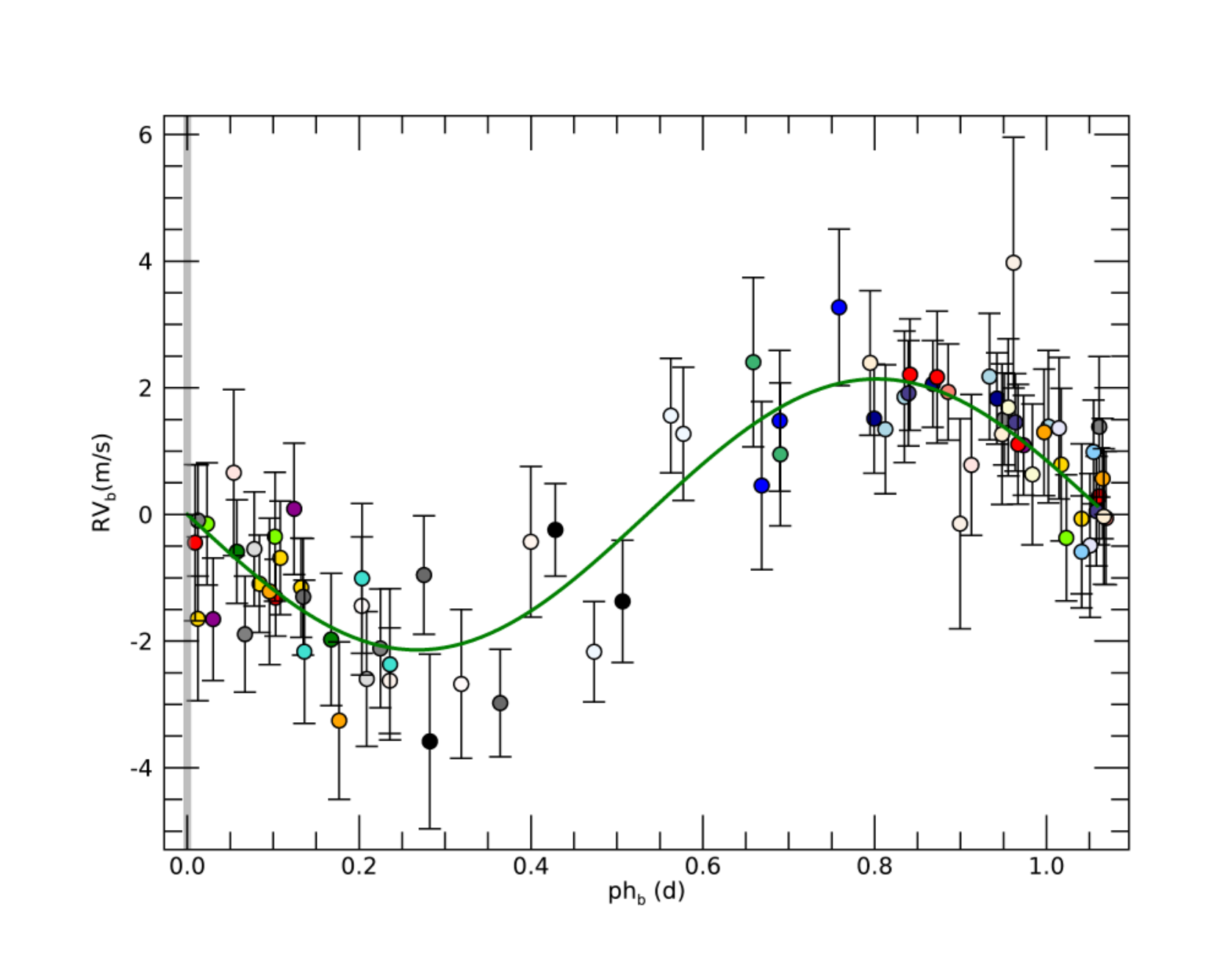}
	\includegraphics[width=1\linewidth]{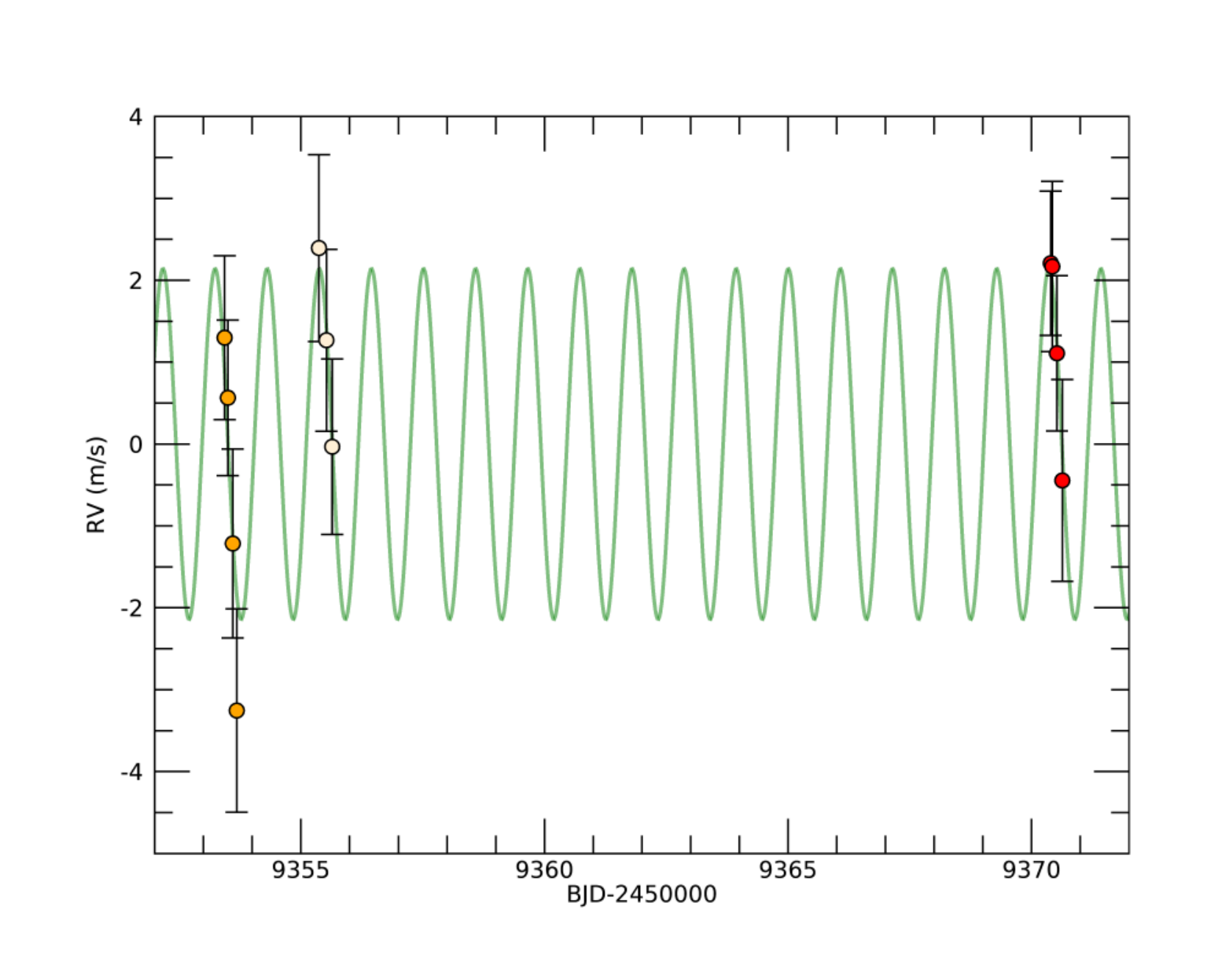}
    \caption{Upper panel: The phase-folded RV model (green line) of TOI-1416 b which corresponds to the best fit from the FCO method, obtained by vertically offsetting nightly chunks of RV data against the model. RV points from the same nights have identical colours. Lower panel: A small section of the RV model plotted against time, with RVs from three different nights.}   
\label{fig:FCOrv}
\end{figure}

A fit using the FCO method for a candidate with an ephemeris known from transits and assuming a circular orbit is in principle very simple, as it contains as free parameters only the RV amplitude $K$ and the nightly RV offsets $\gamma_i$, with $i = 1,...,28 $ indexing the individual nights. However, when using \ufit\ with both the AMOEBA or the MCMC fitter,  resultant RV amplitudes tended to be stuck close to the amplitude's initial value. This behaviour was caused by the large number of 28 nights, where each one corresponds to a free parameter $\gamma_i$. Due to this, either fitter found it difficult to vary the RV amplitude, since any improvement in the fit requires that most of the nightly RV offsets are changed \emph{simultaneously} by the correct amounts. This is difficult to achieve for any fitting algorithm, and is an expected behaviour when free parameters are strongly correlated. We therefore kept the RV amplitude -- and hence the entire RV model -- fixed and fitted only for the RV offsets $\gamma_i$, which reduces the fitting task to a set of 28 simple linear fits. The input RV amplitude was then stepped through a series of suitable values and the $\chi^2$ of each corresponding fit was logged, which led to the curve shown in Fig~\ref{fig:FCOchisq}. The minimum of this curve, and the range where $\chi^2$ increases by 1, indicate an amplitude of $K_b = 2.14\pm0.35 $ m s$^{-1}$. For the best-fit model with  $K_b = 2.14$ m s$^{-1}$, the reduced $\chi^2$ is 0.87 and the residuals of the RVs against the model have an {\it rms} of 0.82 m s$^{-1}$.  Fig.~\ref{fig:FCOrv} shows the RVs of the nightly chunks against the best RV model, with a zoomed-out section across three nights of the RV time-series showing the excellent fit in that range.  
We note that a further FCO fit of the \hn RVs with the open source code \pyan\footnote{Available at \url{https://github.com/oscaribv/pyaneti}} \citep[][see also Sect.~\ref{sec:joint}]{pyaneti,pyaneti2} gave a nearly identical result, of \IGkbfco.

\section{The Neptune desert in radius or mass versus insolation or effective temperature} 
\label{neptune_desert}

\begin{figure}[h]
\centering
	\includegraphics[width=1\linewidth]{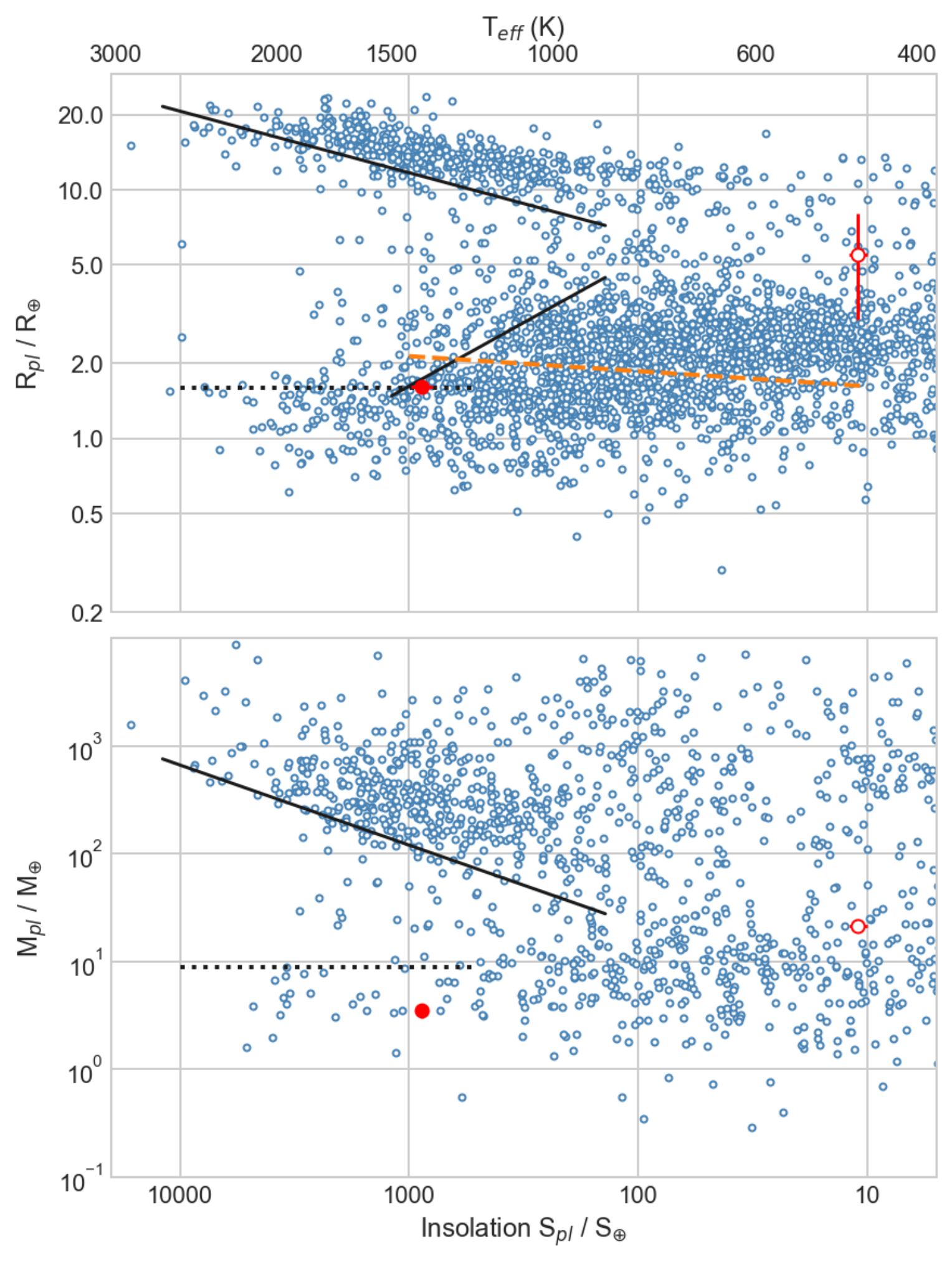}	
   \caption{Similar to Fig.~\ref{fig:radiusperiod}, but with planet radii (top panel) and masses (bottom panel) plotted against the planets' insolation (lower X-axis) {and their effective temperature (upper X-axis)}. The solid black lines show the delineation of the Neptune Desert against insolation resp. $T_\mathrm{eff}$, whereas the horizontal dotted black lines show the same lower limits to the Neptune Desert as those proposed for periods of $P \lesssim 2 d$. The dashed orange line in the upper panel indicates the radius valley against insolation from \citet[][]{2022AJ....163..179P}. \pname\ is indicated by the filled red circle and $c$ by the unfilled one.}   
\label{fig:insolation}
\end{figure}

In Fig.~\ref{fig:insolation} we show plots similar to Fig.~\ref{fig:radiusperiod}, but plotting the planets' radii and masses against the incident bolometric flux or insolation, instead of orbital period (Fig.~\ref{fig:radiusperiod}). The insolation was calculated from first principles from values obtained from the NASA Exoplanet Archive. It is of note that the upper boundary of the Neptune desert is significantly sharper than in the plots against orbital period, whereas the lower boundary remains diffuse; in particular for the plot of planet masses. Given the sharper upper boundary, we propose an upper radius-limit of the Neptune desert against insolation as:
\begin{equation}
\log R_{hi}/R_{\oplus} = 0.248 \log S + 0.33\ ,\hspace{0.5cm} S\gtrsim 150,
\end{equation}
where $S$ is the insolation relative to the Earth's insolation. The corresponding lower limit is then 
\begin{equation}
 \log R_{lo}/R_{\oplus}=\left\{
 	\begin{array}{rl}
    -0.51 \log S + 1.74\ , & 150 \lesssim S \lesssim 1000\\
	  0.20\ ,& \text{$S \gtrsim 1000$},
  \end{array}
  \right.
\end{equation} 
where the same limit of  $R_{lo}=1.60 R_{\oplus}$ as given in Sect.~\ref{sec:results} for very short orbital periods applies also to the strongest insolations. For the limit of the desert against mass, there are much fewer planets with mass measurements, and we only derive an upper limit of 
\begin{equation}
\log M_{hi}/M_{Jup} = 0.74 \log S - 2.35\ ,\hspace{0.5cm} S\gtrsim 150,
\end{equation}
whereas a lower limit cannot be discerned with reliability, given the small sample of known short-period low-mass planets, that is furthermore suffering a strong selection effect against detectability towards smaller masses. In Fig.\ref{fig:insolation} we hence indicate only the same lower mass limit that is given in Sect.~\ref{sec:results} for very short orbital periods, namely  $M_{lo}=8.9 M_{\oplus}$, resp. $\log M_{lo}/M_\mathrm{jup}= -1.55$. In above limits for both radius and mass, we maintain the gradients of the limits against period by \citet{2016A&A...589A..75M}, after multiplication of $\log P$ with a factor of  -4/3, which arises from the dependency of $S$ on the period. 

{For convenience, we also provide the same limits against the planets' effective temperature $T_\mathrm{eff}$, using the conversion $T_\mathrm{eff} = (S)^{1/4}\ 255$ K
, with the 255 K corresponding to the effective temperature of the Earth. We then obtain for the limits of radius against $T_\mathrm{eff}$:
\begin{equation}
\log R_{hi}/R_{\oplus} = 0.99 \log T_\mathrm{eff} -2.72\ ,\hspace{0.7cm} T_\mathrm{eff} \gtrsim 900 K\\
\end{equation}
\begin{equation}
 \log R_{lo}/R_{\oplus}=\left\{
 	\begin{array}{rl}
    -2.04 \log T_\mathrm{eff}  -3.17\ , & 900K \lesssim T_\mathrm{eff} \lesssim 1450 K\\
	  0.20\ ,& \text{$T_\mathrm{eff} \gtrsim 1450 K$}
  \end{array}
  \right.
\end{equation} 
and for the upper mass limit of the desert:
\begin{equation}
\log M_{hi}/M_{Jup} = 3.0 \log T_\mathrm{eff} - 9.5\ ,\hspace{0.7cm} T_\mathrm{eff} \gtrsim 900 K,
\end{equation}
with the same lower limit as indicated previously against insolation. With  $log M/M_{\oplus} = 2.50 \log M/M_{Jup}$, we may easily convert the desert mass limits to the units of Earth masses used in Fig~\ref{fig:insolation}}.

\FloatBarrier

\section{Further figures mentioned in main text}
\label{app:morefigs}

\begin{figure*}
\includegraphics[width=\linewidth]{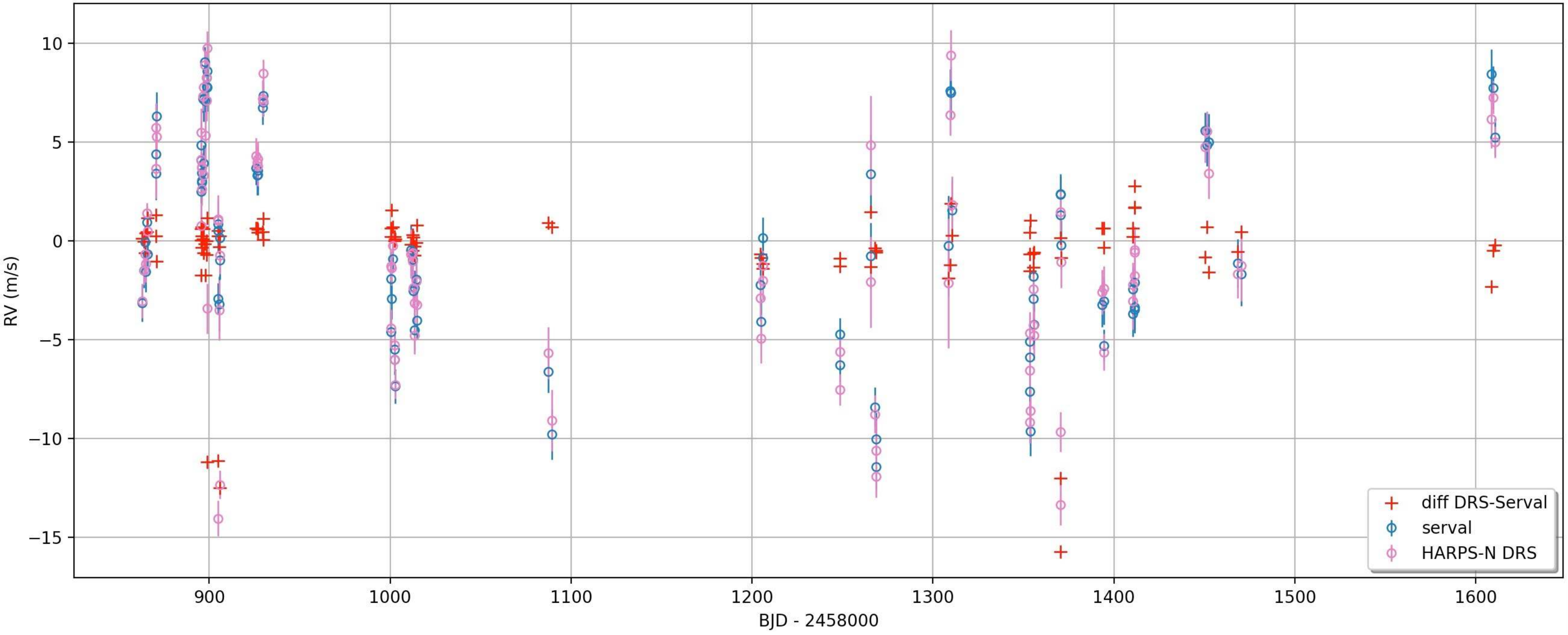}

\caption{Comparison between \hn\ RVs measured with the \hn\ DRS from CCFs (open red circles) and with {\tt serval}  (open green circles). The red crosses show the difference between the two data-sets. The five points in which this difference is significantly negative correspond to exposures that were prematurely terminated, and which are not correctly processed by the DRS. Both data-sets have been averaged to zero without considering these five points. The difference between the two data-sets has a standard deviation of 0.92 m s$^{-1}$ (excluding again these five points).}
\label{fig:HNdrssrv}
\end{figure*}

\begin{figure*}
  \centering
\begin{subfigure}{1\textwidth}
  \includegraphics[width=1\linewidth]{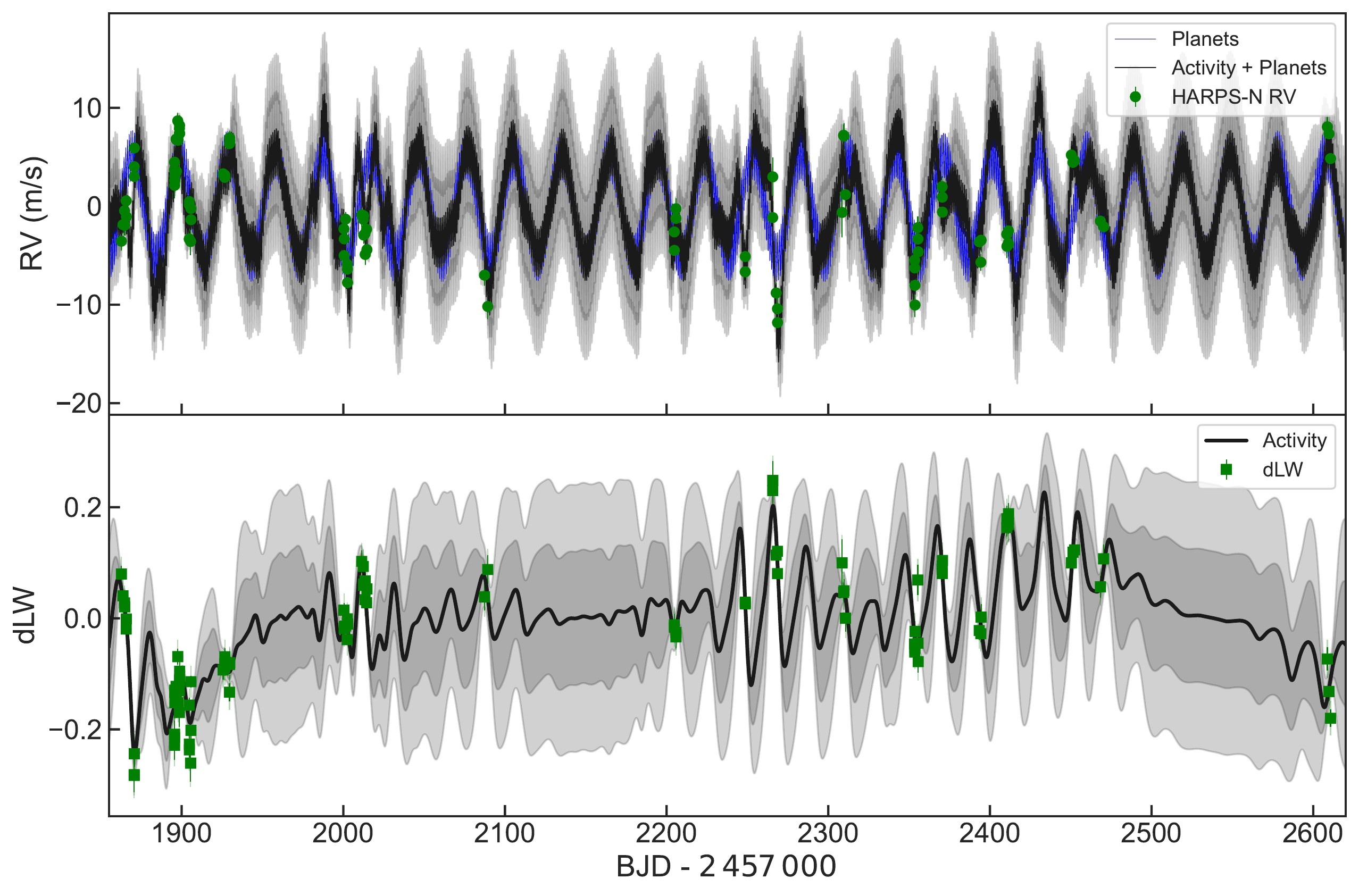} 
  \label{fig:first}
\end{subfigure}
\begin{subfigure}{.49\textwidth}
  \includegraphics[width=\textwidth]{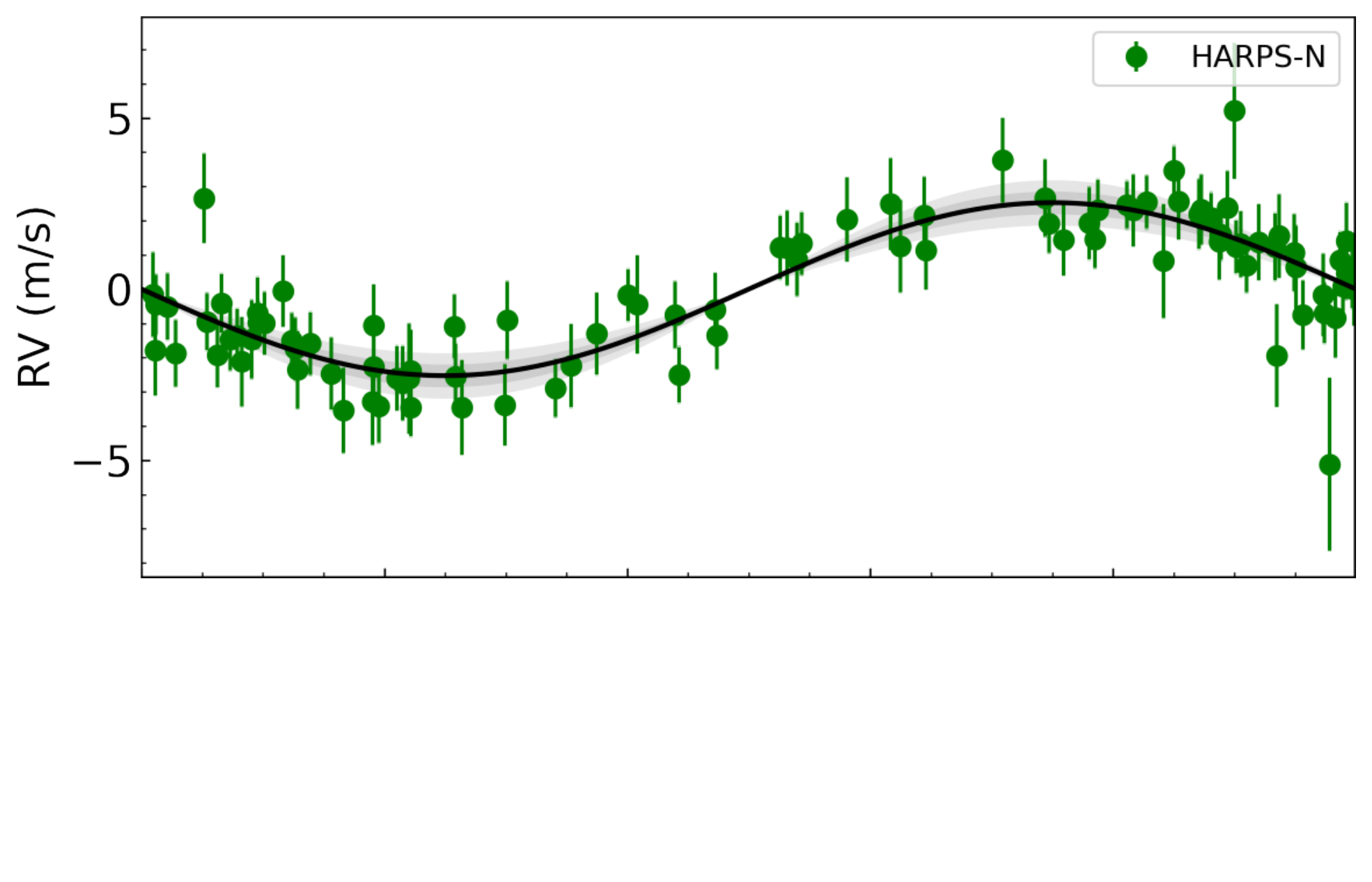} 
  \label{fig:second}
\end{subfigure}
\begin{subfigure}{.49\textwidth}
  \includegraphics[width=\textwidth]{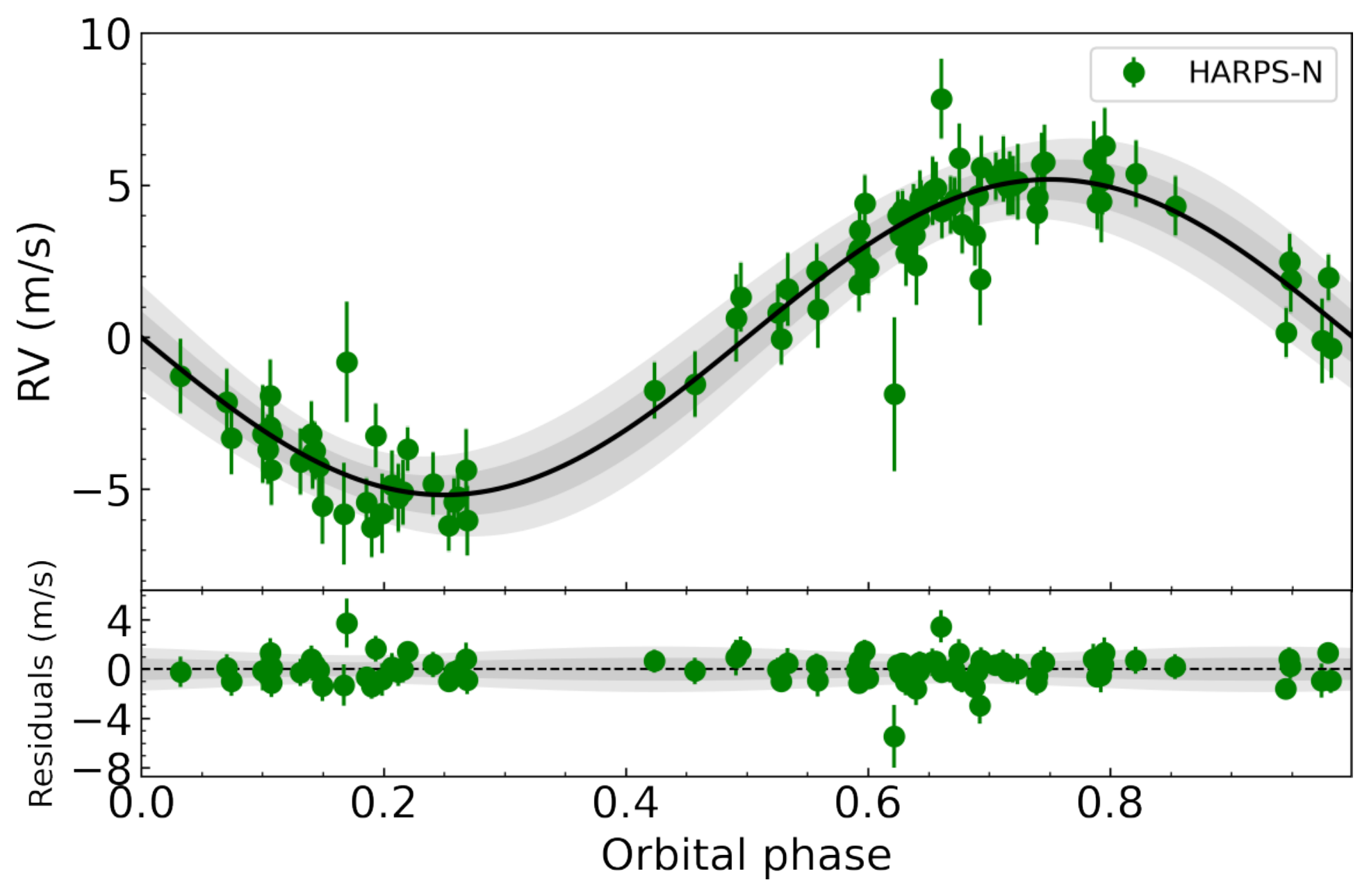}
  \label{fig:third}
\end{subfigure}

\caption{Like Fig.~\ref{fig:RVGP-1pl}, but for Model 3 with the additional fit for a Keplerian signal with the lunar synodic period (29.53d). The lower left panel is again for the transiting planet $b$, while the lower right panel shows the RVs folded over the period of the additional signal. A similar plot for Model 2 does not show any relevant differences to the shown one.}

\label{fig:RVGP-2pl}
\end{figure*}

\begin{itemize}

\item Fig.~\ref{fig:HNdrssrv} gives a comparison of the  RVs measured with \hn\ using the  DRS and the {\tt serval} pipelines.

\item Fig.~\ref{fig:RVGP-2pl}  shows the output of the \pyan\ joint-fit for Model 3, with two planets $b$ and $c$. 

\end{itemize}

\end{appendix}
\listofobjects

\end{document}